\documentclass[preprint,12pt]{elsarticle}
\journal{Nucl. Instr. Meth. A}

\usepackage{lineno}

\usepackage{amssymb}
\usepackage{amsmath}

\usepackage{units}
\usepackage{booktabs}
\usepackage{xspace}
\usepackage{hyperref}
\usepackage[binary-units=true]{siunitx}
\usepackage{subcaption}
\usepackage{rotating}
\usepackage[markup=default]{changes}
\definechangesauthor[color=red]{MS}  

\usepackage{doi}
\DeclareSIUnit \lightspeed {\text{\ensuremath {c}}}
\usepackage{hyperref}
\usepackage{makecell}

\begin{document}


\newcommand{\desyii}{{\mbox{DESY II}}\xspace}
\newcommand{\desyiii}{{\mbox{DESY III}}\xspace}
\newcommand{\petraiii}{{\mbox{PETRA III}}\xspace}
\newcommand{\petraiv}{{\mbox{PETRA IV}}\xspace}
\newcommand{\petra}{{PETRA}\xspace}
\newcommand{\doris}{{DORIS}\xspace}
\newcommand{\pia}{{PIA}\xspace}
\newcommand{\linacii}{{\mbox{LINAC II}}\xspace}

\newcommand{\diitbf}{{\desyii Test Beam Facility}\xspace}

\newcommand{\geant}{\textsc{Geant4}\xspace}
\newcommand{\slic}{\textsc{SLIC}\xspace}

\begin{frontmatter}



\title{The \diitbf \tnoteref{t1}}
\tnotetext[t1]{Paper dedicated to the memory of our colleague Ulrich~K\"otz}

\author[desy]{R.~Diener}
\author[desy]{J.~Dreyling-Eschweiler\corref{cor}}
\ead{jan-dreyling-eschweiler@desy.de}
\cortext[cor]{Corresponding author}
\author[desy]{H.~Ehrlichmann}
\author[desy]{I.~M.~Gregor}
\author[desy]{U.~K\"otz}
\author[desy]{U.~Kr\"amer}
\author[desy]{N.~Meyners}
\author[desy]{N.~Potylitsina-Kube}
\author[desy]{A.~Sch\"utz}
\author[desy]{P.~Sch\"utze}
\author[desy]{M.~Stanitzki}
\address[desy]{DESY, Notkestrasse 85, D-22607 Hamburg}

\begin{abstract}
DESY Hamburg operates a test beam facility with three independent beam lines at the \desyii synchrotron.
It is world-wide one of very few facilities providing test beams in the GeV range.
To this end, it offers electron/positron beams with user-selectable momenta from 1-\SI{6}{GeV/c}. 
The available infrastructure for the users is unique, including a high field solenoidal magnet and permanently installed high-precision pixel beam telescopes. 
This publication gives a detailed description of the facility, the available infrastructure, and the simulated and measured performance.
\end{abstract}


\end{frontmatter}





\newpage

\section{Introduction}\label{sec:intro}

\begin{figure}[htb]
\centering
\includegraphics[width=0.95\textwidth]{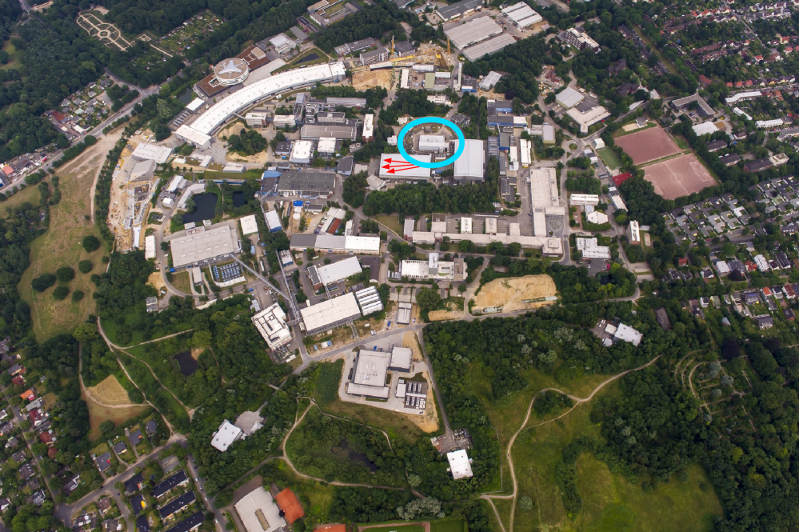}
    \caption{(Colour online) Aerial view of the DESY Campus at Hamburg-Bahrenfeld with the \desyii synchrotron (blue) and the location of the test beam lines (red) in Hall~2. }
\label{fig:intro:desyoverview}
\end{figure}

DESY (Deutsches Elektronen-Synchrotron)\footnote{\url{http://www.desy.de}} operates a test beam facility at its campus at Hamburg-Bahrenfeld (Fig.~\ref{fig:intro:desyoverview}). 
The facility offers three independent beam lines with electron or positron particles with selectable momenta from 1-\SI{6}{\GeV/c} and is located in building 27 (``Halle 2''), one of the experimental halls at DESY. 
The beam lines are attached to the \desyii synchrotron, which typically runs electron beams with an oscillating energy from 0.45-\SI{6.3}{\GeV}. 
This Test Beam Facility is one of the few worldwide that offers users access to multi-GeV beams.
It has essential infrastructure for the development and testing of nuclear and particle physics detectors and generic detector R\&D. 

After the start-up of the original DESY electron synchrotron in 1964 the developments and studies of new detector components were done by using 
the primary beam when not used for the recognized experiments. 
The place behind the pair spectrometer in beam line 24 was a popular location which also provided an energy measurement of the electrons. 
With the installation of \desyii this possibility ceased to exist and the current facility started serving the test beam needs of the community.

Since its inception and start of operation simultaneously to the start of \desyii in 1987, the usage of the \diitbf has continuously increased. 
At the same time, there were continuous investments into this facility and more infrastructure added, including a high-field solenoid and two high-precision pixel beam telescopes. 
All of these are available to all users of the facility. 

The EU has supported both access and enhancements to the \diitbf within the FP6-EUDET \cite{EUDET}, the FP7-AIDA \cite{AIDA} and the Horizon 2020-AIDA2020 \cite{AIDA2020} grants.
The transnational access offers travel support for users from outside Germany thus enabling further groups to use the facility. 

This paper is organized as follows:
first an overview of the \desyii synchrotron is given (Sec.~\ref{sec:desy2}), 
then the test beam generation (Sec.~\ref{sec:desy2beamgen}),
the beam line instrumentation (Sec.~\ref{sec:beaminstr}),
and the individual beam areas (Sec.~\ref{sec:tbareas})
are explained. 
All additional test beam infrastructure, like magnets and beam telescopes, is then described in detail (Sec.~\ref{sec:addinf}).
The performance of the \diitbf is presented by 
the results of different measurements (Sec.~\ref{sec:performance}) and 
is compared to simulations of the test beam (Sec.~\ref{sec:simulations}),
It is concluded with 
a report on the user community (Sec.~\ref{sec:userstats}) 
and a summary and an outlook on future improvements (Sec.~\ref{sec:summaryoutlook}). 

\newpage

\section{The \desyii Synchrotron}\label{sec:desy2}

More than thirty years ago, on the 22$^{\rm nd}$ of March 1985, the first electron beam in \desyii was circulated \cite{Hemmie:1983et,Hemmie:1985uw}. 
After the final connection of all transport lines and magnet circuits, from 
spring 1987 on, \desyii has delivered electron or positron beams with high 
stability and reliability up to \SI{7}{\GeV} beam 
energy to \doris, \petra, HERA and the \diitbf. 

\begin{figure}[htbp]
\centering
\includegraphics[width=0.7\textwidth]{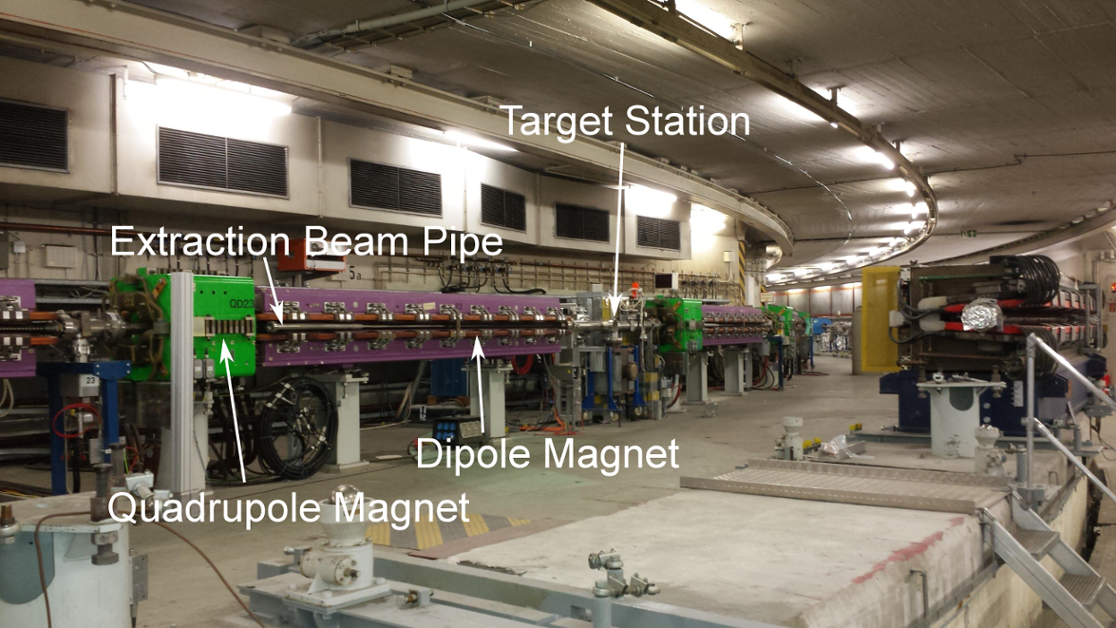}
\caption{(Colour online) View of a section of \desyii  on the left side of the DESY tunnel, the magnets on the right side are dipoles from the \desyiii proton synchrotron}
\label{fig:desy2:desytunnel}
\end{figure}

\desyii is installed in the DESY tunnel. It is \SI{292.8}{\m} long and has an average radius of \SI{46.6}{\m} (Fig.~\ref{fig:desy2:desytunnel}).
The tunnel houses the \desyii synchrotron and the \desyiii proton synchrotron, 
which served as an \SI{8}{\GeV} proton injector for HERA  and was decommissioned in 2007. 
As \desyiii has been re-using most infrastructure from the original DESY synchrotron including the dipole magnets, 
\desyii was designed from scratch.

\begin{figure}[htbp]
\centering
    
    \includegraphics[width=0.9\textwidth]{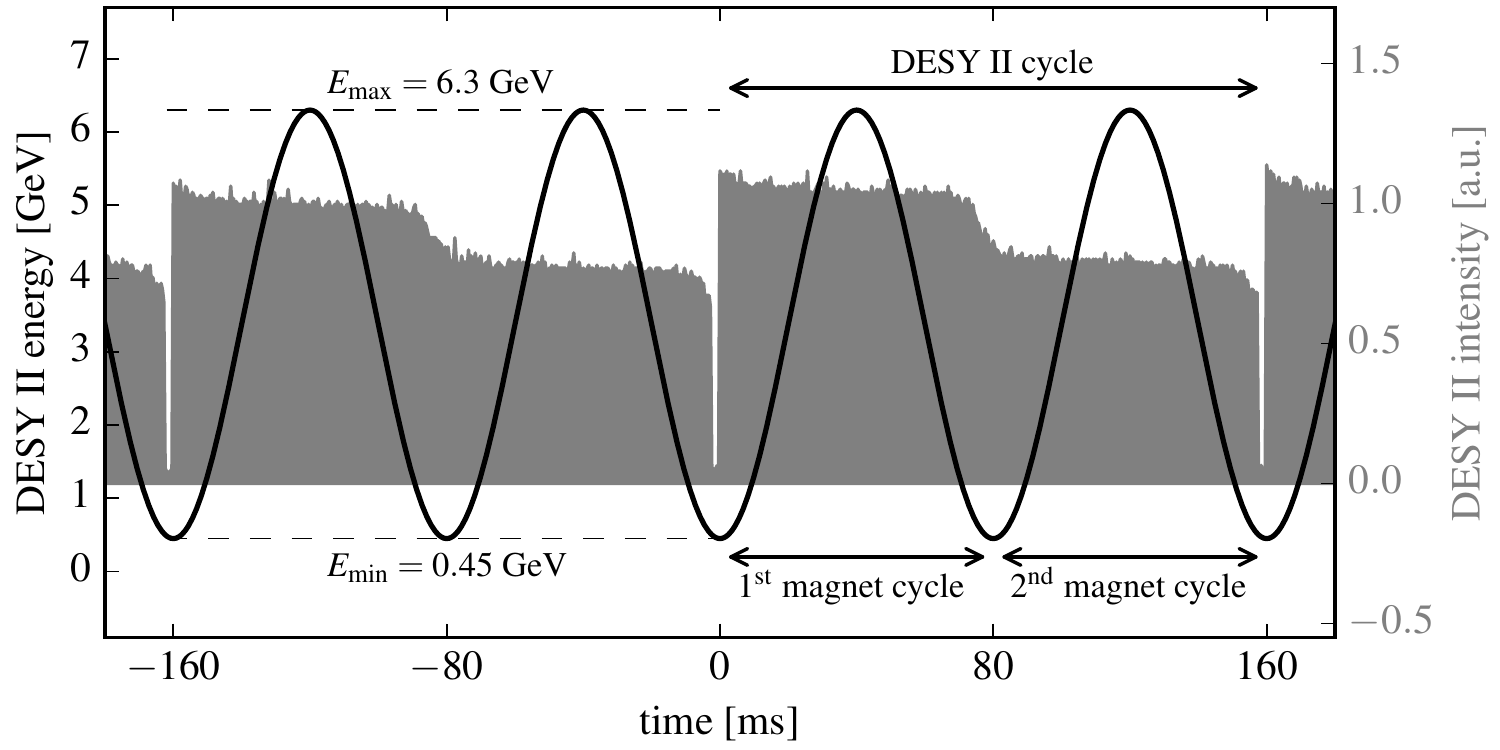}
\caption{The exclusive \desyii operation without beam extraction:
A schematic of the sinusoidal \desyii  beam energy (left $y$-axis, black) and 
a typical scope picture of the \desyii beam intensity monitor (right $y$-axis, grey). 
One \SI{160}{ms} \desyii cycle starts with the injection of a bunch which is stored for two \SI{80}{\milli\second} magnet cycles before being dumped.
The beam intensity of the second magnet cycle is expected to be smaller than the first due to beam losses during traversing $E_{\rm min}$.
}
\label{fig:desy2:cycle}
\end{figure}

The magnet system of \desyii consists of five circuits: 
the dipole circuit, two quadrupole, and two sextupole circuits. 
All of the circuits operate synchronously with a sinusoidally oscillating current with a frequency of \SI{12.5}{\Hz} which corresponds to a \SI{80}{ms} magnet cycle (Fig.~\ref{fig:desy2:cycle}). 
24 horizontal and 24 vertical DC corrector coils allow orbit manipulation at lower beam energies. 
The optical lattice is formed by 8 $\times$ 3 FODO
cells. 
Two independent HF systems ---each consisting of one klystron and four \SI{500}{\mega\hertz} 7-cell \petra-type cavities--- can be operated 
either individually in single mode or both together for reaching higher beam energies. 

The \desyii standard operation without an extraction stores a bunch of electrons for two magnet cycles which defines the \SI{160}{\milli\s} \desyii cycle (Fig.~\ref{fig:desy2:cycle}):
One bunch of about 10$^{10}$ electrons or positrons is injected on-axis at $E_{\rm min} = \SI{0.45}{\GeV}$ from the linear accelerator \linacii (LINear ACcelerator) via the small storage ring \pia (Positron Intensity Accumulator) shortly after the time of minimal fields, 
and is accelerated to typically $E_{\rm max} = \SI{6.3}{\GeV}$, with a possible maximum of \SI{7}{\GeV}. 
The beam is typically stored for two magnet cycles or one \desyii cycle and is dumped about \SI{160}{\milli\s} after injection, just before the next injection (Fig.~\ref{fig:desy2:cycle}). 
It is important to note, that the beam get accelerated twice to the maximum energy of \SI{6.3}{\GeV} with one deceleration to \SI{0.45}{\GeV} in between
which causes beam losses for the second magnet cycle (Fig.~\ref{fig:desy2:cycle}).
The life-time at low energies is additionally affected by the number of test beam targets in the primary beam (Sec.~\ref{sec:desy2beamgen:primarytarget}),
and other machine effects like the tunes and the HF settings.

Nowadays \desyii is the injector for \petraiii and the  
the \desyii beam can be completely extracted during the acceleration cycle. 
When \petraiii runs in standard operation with the constant current or so-called ``top-up'' mode, 
\desyii has to deliver beam every few minutes (Fig.~\ref{fig:performance:p3topup} in Sec.~\ref{sec:performance:p3topup}).
When reaching the beam energy of \SI{6.0}{\GeV} within the first magnet cycle the beam is extracted and \desyii stays empty until the next injection.
Further and as a consequence, the \desyii magnet circuits have to run continuously, consuming roughly \SI{600}{\kilo\watt} electrical power. 
The \desyii HF power of about \SI{400}{\kilo\watt} for the default operation is only switched on if beam has been requested by \petraiii or the \diitbf.

For the last several years, \desyii has been running in a completely automated operation mode without any manual actions by the operators. 
Beam requests from all users of \petraiii and the \diitbf are detected to set all necessary parameters for the requested operation mode.
\desyii has an availability of close to \SI{99}{\percent} during this operation.

\newpage

\section{Test Beam Generation}\label{sec:desy2beamgen}

\begin{figure}[htbp]
\centering
\includegraphics[width=0.925\textwidth]{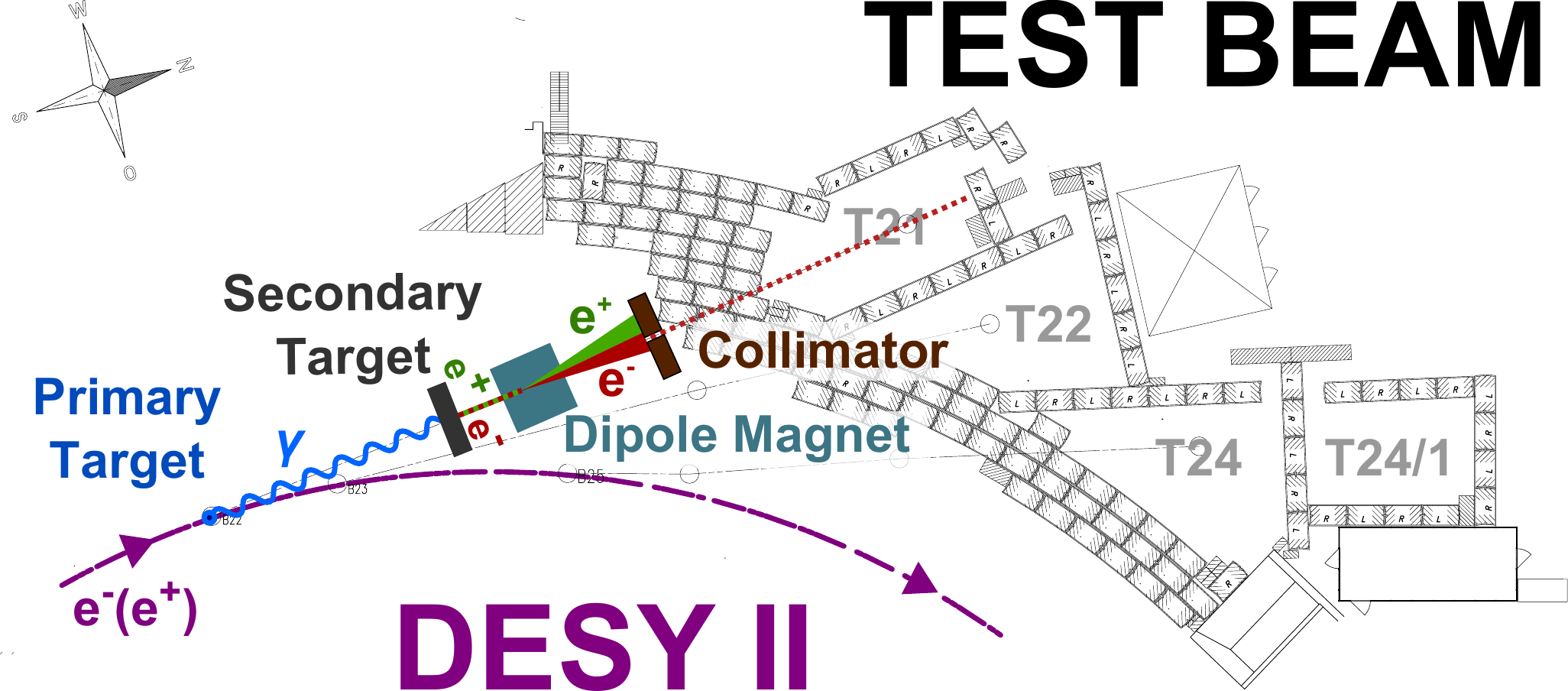}
\caption{(Colour online) A schematic view of the test beam generation at the \diitbf, here for beam line TB21. 
Bremsstrahlung photons generated in the primary target travel through the tunnel and hit the secondary target generating electron/positron pairs.
The dipole magnet selects particles according to their momentum and charge and the particle beam can be further collimated before entering the test beam areas.}
\label{fig:desy2beamgen:schematic}
\end{figure}

At the \diitbf, the test beams are generated by a double conversion instead of using a direct extraction of the primary beam in \desyii.
Initially bremsstrahlung photons are generated by a fiber target positioned in the \desyii beam orbit.
These photons hit the secondary target generating electron/positron pairs. 
Depending on the polarity and strength of the magnetic field of the following dipole magnet, the test beam particles reaching the test beam areas are electrons or positrons with a certain momentum.
There are three independent beam lines, called TB21, TB22 and TB24, 
named after the positions of the primary targets located behind the quadrupoles QF21, QF22 and QF24 respectively. 
A schematic view of the beam generation and the three beam lines is shown in Figure~\ref{fig:desy2beamgen:schematic}.

Before describing all individual elements of one beam line in detail, 
an overview of the beam generation is given. 
The primary target is a several \SI{}{\micro\m} thick fiber (Sec.~\ref{sec:desy2beamgen:primarytarget})
which can be positioned at the main \desyii orbit in order to produce the highest rate of bremsstrahlung photons. 
Since the fiber is fixed on the orbit it intercepts the beam over the full \desyii magnet cycle (Sec.~\ref{sec:desy2}):
photons are generated by electrons from the injection energy of \SI{0.45}{\GeV} up to full energy of \SI{6.3}{\GeV} (see Fig.~\ref{fig:bremsstrahlung_spectrum} in Sec.\ref{sec:simulations}).
These bremsstrahlung photons travel along an extraction beam pipe tangentially to the \desyii orbit and then leave the \desyii vacuum through a \SI{500}{\micro\m} thick aluminum exit window. 
They travel through the air for up to \SI{22}{\m} and then hit the secondary targets (Sec.~\ref{sec:desy2beamgen:secondary}), where electrons and positrons are generated through the pair production process. 
A high vacuum beam line starts right after the secondary target. 
The electrons and positrons then pass through a dipole magnet (Sec.~\ref{sec:desy2beamgen:magnets}), 
which allows selection of the particle flavor and momentum 
depending on the polarity and strength of the field. 
The selected particles then travel in an evacuated beam pipe and can be collimated by the controllable primary collimator (Sec.~\ref{sec:desy2beamgen:primarycoll}). 
After passing the beam shutter (Sec.~\ref{sec:desy2beamgen:shutter}) the test beam particles enter the test beam areas located in Hall~2 (Sec.~\ref{sec:tbareas}). 
Inside the beam areas an exchangeable fixed-size secondary collimator is located (Sec.~\ref{sec:tbareas:secondarycoll}),
along with shielding elements acting as beam dumps (Sec.~\ref{sec:tbareas:shieldingdump}).

\subsection{Primary Targets}\label{sec:desy2beamgen:primarytarget}

\begin{figure}[htbp]
    \center
    \includegraphics[width=0.6\textwidth]{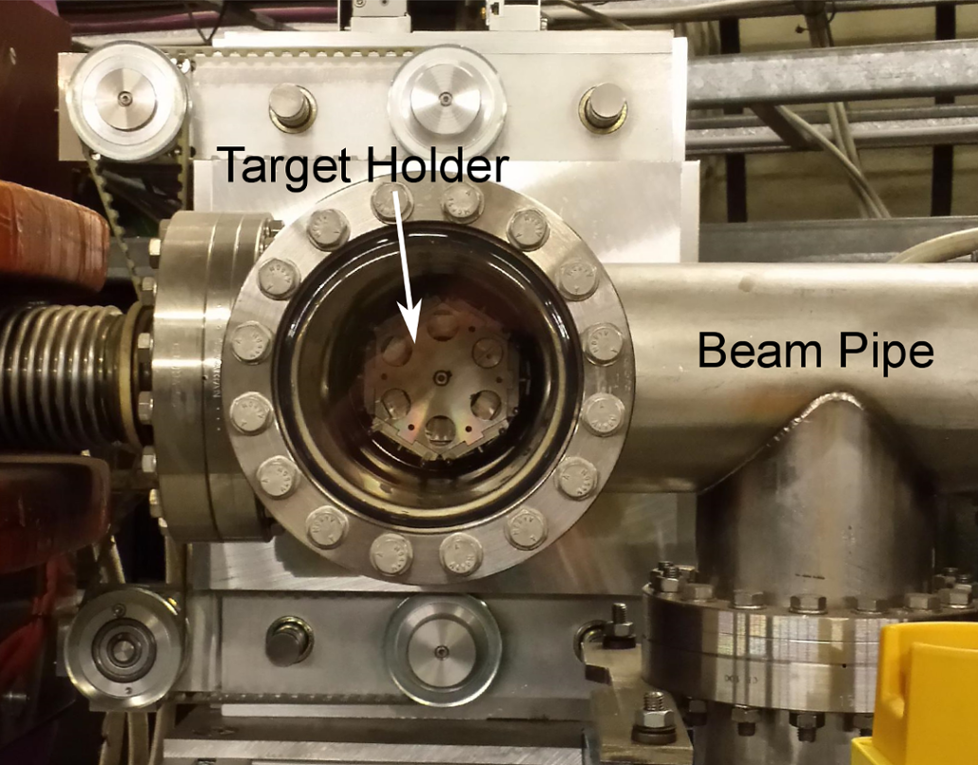}
    \caption{(Colour online) The primary target station with the hexagonal target holder. 
    }
\label{fig:desy2beamgen:targetphotos}
\end{figure}

For each beam line a primary target station is installed inside the vacuum pipe.
A station consists of up to six carbon fibers.
The fibers are each about \SI{7}{\micro\m} thick and \SI{30}{\mm} long and each fiber is glued between the two prongs of a metal fork. 
Three long forks and three short forks are mounted on a hexagonal target holder, the so-called ``revolver'' (Fig.~\ref{fig:desy2beamgen:targetphotos}). 
The revolver can be rotated to select one of the six fibers. 
The selected fiber can be horizontally moved in the \desyii beam orbit to generate bremsstrahlung.

\begin{sidewaysfigure}[htbp]
\centering
\includegraphics[width=\textwidth]{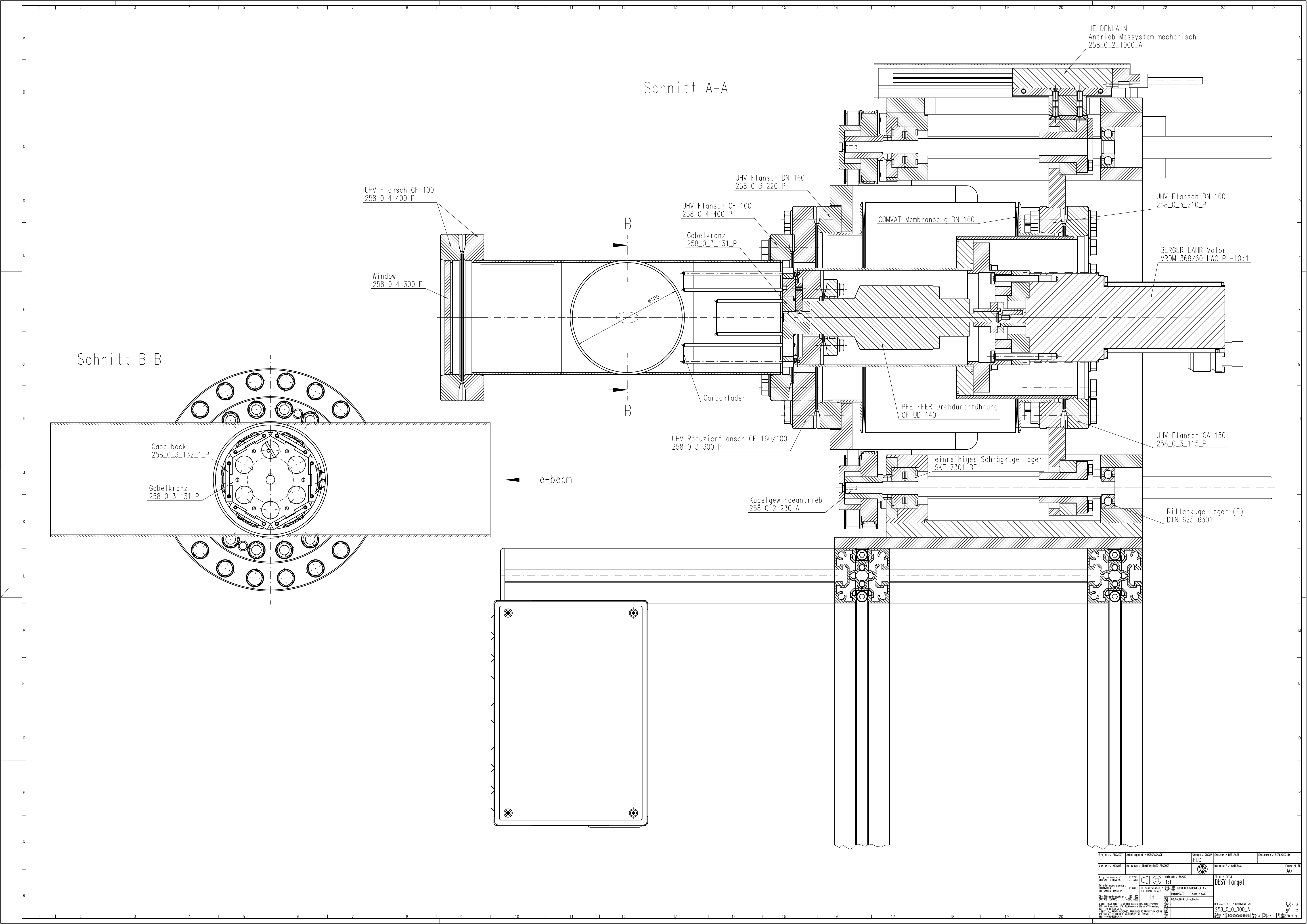}
\caption{Technical drawing of the primary target control station.
   }
\label{fig:desy2beamgen:targetdetails}
\end{sidewaysfigure}

A detailed drawing of an entire target station is shown in Figure~\ref{fig:desy2beamgen:targetdetails}.
The target stations can be remotely controlled via a CAN bus interface.
They are fully integrated in the accelerator control software and can be controlled using a simple GUI.
The selected fiber usually remains in the \desyii beam orbit, 
unless there are special beam tests performed or there is maintenance during the shutdown. 
While the status is visible to the test beam users, the operation of the primary target control is restricted to the shift crews in the DESY accelerator control room (BKR).

\subsection{Secondary Targets}\label{sec:desy2beamgen:secondary}

The secondary targets are located inside the \desyii tunnel right in front of the test beam dipole magnets. 
The bremsstrahlung photons hit this target and produce electron-positron pairs as well as a non-negligible neutron background.
Each station has up to eight different targets allowing the user to adjust the particle rates and yields to their needs. 
The available target options for each beam line are listed in Table~\ref{tab:desy2beamgen:secondarytargets:specs}. 

\begin{table}[htbp]
  \begin{center}
    \begin{tabular}{ c r r r} 
    \textbf{Target No.} 	& \textbf{TB21} & \textbf{TB22} & \textbf{TB24} \\\toprule 
    1 						&Cu \SI{5}{\mm}	&Cu \SI{5}{\mm}	&Cu Wire \SI{1}{\mm} \\
    2 						&Cu \SI{4}{\mm}	&Cu \SI{1}{\mm}	&Cu \SI{4}{\mm}		\\
    3 						&Cu \SI{3}{\mm}	&Cu \SI{3}{\mm}	&Cu \SI{3}{\mm}		\\
    4 						&Al \SI{3}{\mm}	&Al \SI{4}{\mm}	&Al \SI{3}{\mm}		\\
    5 						&Al \SI{2}{\mm}	&Al \SI{3}{\mm}	&Al \SI{2}{\mm}		\\
    6 						&Al \SI{1}{\mm}	&Al \SI{1}{\mm}	&Al \SI{1}{\mm}		\\
    7 						&Cu Wire \SI{1}{\mm} &Cu \SI{10}{\mm}	&Cu Wire \SI{1}{\mm}		\\
    \bottomrule
    \end{tabular} 
   \end{center}  
\caption{ \label{tab:desy2beamgen:secondarytargets:specs}
The available secondary targets of each beam line. The plates have a dimension of \SI[product-units=repeat]{45x60}{\mm}.}   
\end{table}

\begin{figure}[htbp]
\centering
\includegraphics[width=0.7\textwidth]{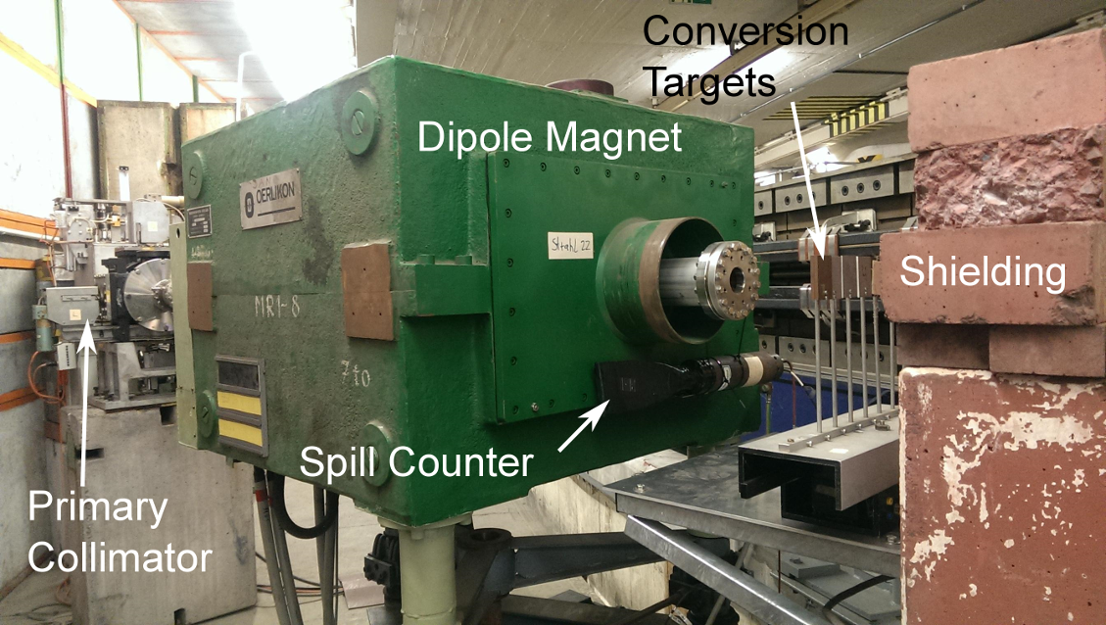}
\caption{(Colour online) The test beam magnet (Type MR) of TB22. 
The secondary conversion target system is also visible in front of the magnets.
    }
\label{fig:desy2beamgen:primarymagnets}
\end{figure}

The secondary target stations for each beam line can be remotely controlled by the users from the test beam huts. 
One of the secondary target stations is shown in Figure~\ref{fig:desy2beamgen:primarymagnets}.

\subsection{Test Beam Magnets}\label{sec:desy2beamgen:magnets}

The test beam magnets are dipole magnets (``DESY Typ MR''), located \SI{60}{\cm} behind the secondary target,
and are used to bend particles of the desired momentum and charge.
The integrated length of the magnet is \SI{710}{\mm} and the entry window to the evacuated beam pipe is right in the center of the dipole field, ensuring a very homogeneous magnetic field.
The maximum field of the dipole is \SI{1.38}{\tesla} with a maximum current of \SI{375}{\A}. 

\begin{table}[htbp]
  \begin{center}
    \begin{tabular}{ l c c }\\
                        & \textbf{MR}    &  \textbf{MB}   \\ \toprule 
I$_{max}$ (A)           & 375   & 1500  \\
B at I$_{max}$ (T)      & 1.38  &  2.24 \\
P at I$_{max}$ (kW)     &  25   & 400   \\
integrated length  (mm)&  710  & 1029  \\ 
Weight (t)              &  7    & 7.5   \\\midrule
Usable Aperture &       &       \\   
vertical  (mm)  &  100  &108       \\
horizontal(mm)  &  220  &303    \\
    \bottomrule
    
    \end{tabular} 
   \end{center}  
    \caption{ \label{tab:desy2beamgen:primarymagnet:specs}Key parameters of the primary magnets used in all beam lines (MR) and the additional one in TB24 (MB).}   
\end{table}

\begin{sidewaysfigure}[htbp]
\begin{subfigure}{0.47\hsize}
\centering
\includegraphics[width=0.99\hsize]{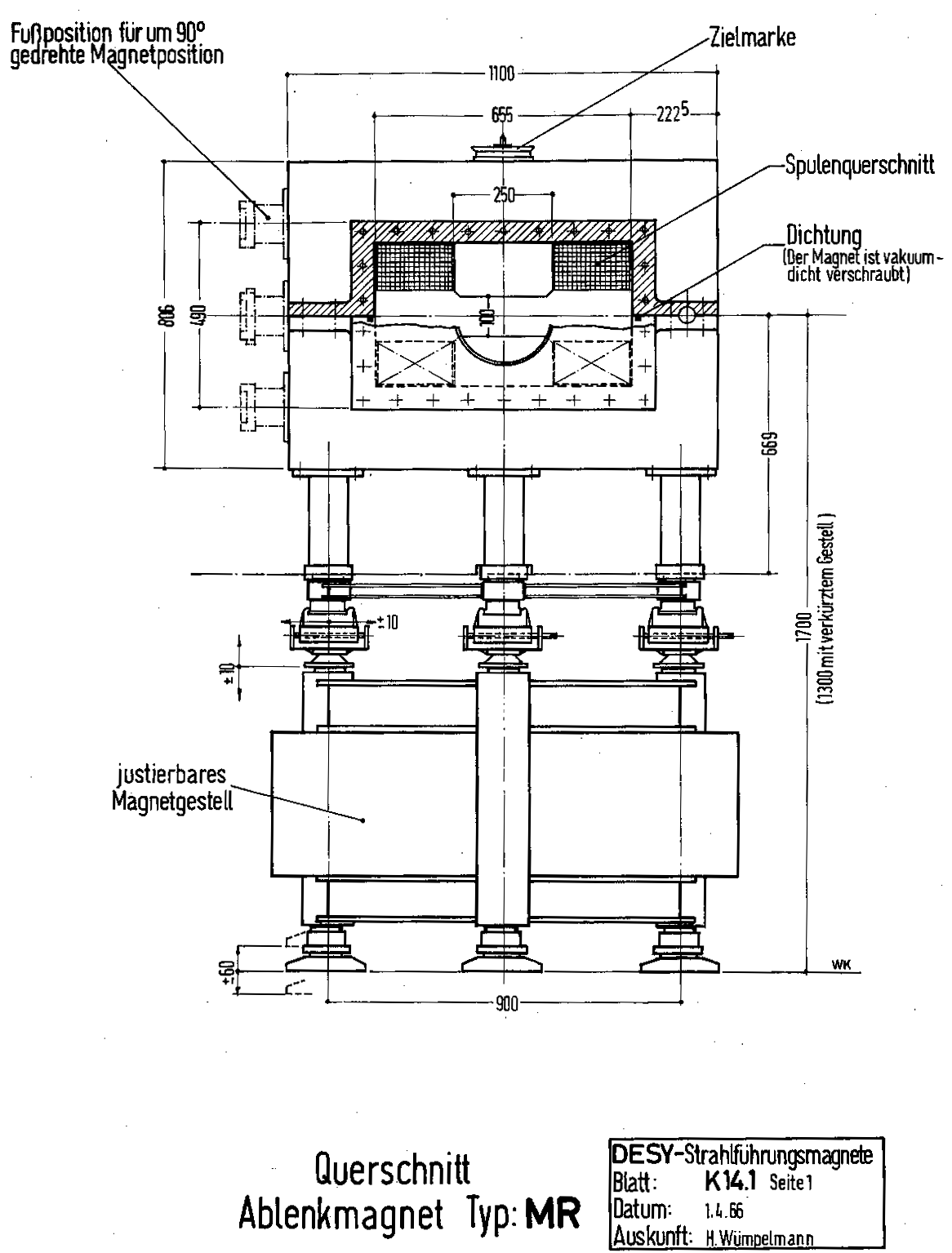}
\caption{MR magnet}
\label{fig:desy2beamgen:primarymagnetsdrawingsmr}
\end{subfigure}\hfill
\begin{subfigure}{0.47\hsize}
\centering
\includegraphics[width=0.99\textwidth]{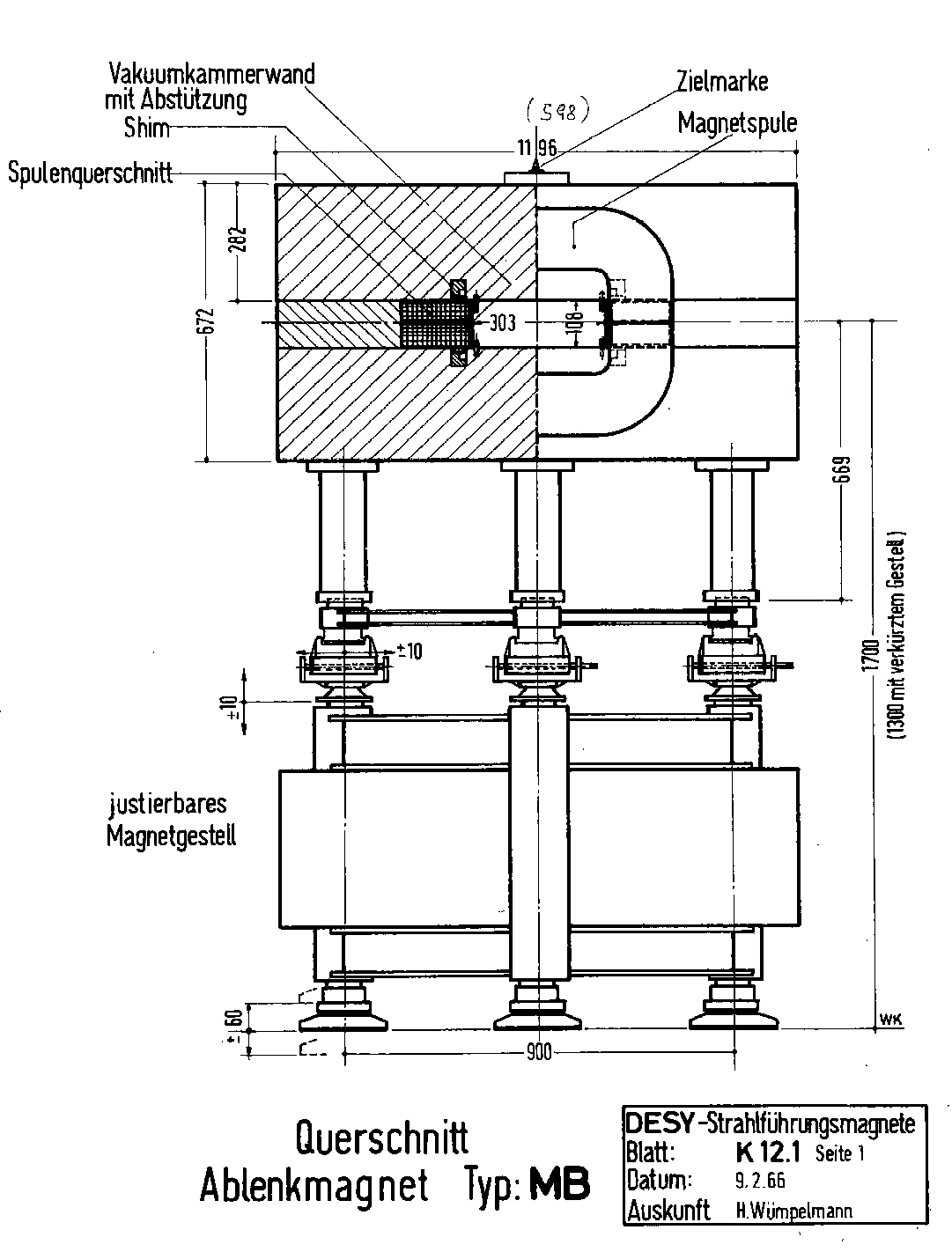}
\caption{MB magnet}
\label{fig:desy2beamgen:primarymagnetsdrawingsmb}
\end{subfigure}
\caption{Technical drawings of test beam magnets for selecting the beam momentum.}
\end{sidewaysfigure}

For TB24, there is an additional magnet (``DESY Typ MB'') in front of the test beam dipole magnet, which is required due to geometrical reasons. 
The particles need to be bent further in order to reach the corresponding beam area. 
The maximum field of the dipole is \SI{2.24}{\tesla} with a maximum current of \SI{1500}{\A}. 
Table~\ref{tab:desy2beamgen:primarymagnet:specs} summarizes the key parameters of both magnets. 
Pictures of a MR magnet and a MB magnet are shown in Figure~\ref{fig:desy2beamgen:primarymagnets}.
The technical drawings of these two magnets are shown in Figure~\ref{fig:desy2beamgen:primarymagnetsdrawingsmr} and \ref{fig:desy2beamgen:primarymagnetsdrawingsmb}.

The user can remotely control the individual magnets from either huts or the accelerator control room and 
select the appropriate particle momentum by selecting the corresponding current from the magnet power supplies (Tab.~\ref{tab:desy2beamgen:magnets:specs}). 

\begin{table}[htbp]
 
  \begin{center}
    \begin{tabular}{ c c c c}\\ 
        \thead{\textbf{Particle}\\ \textbf{Momentum}\\\textbf{(GeV/c)}} &  \thead{\textbf{TB21/TB22} \\ \textbf{ER/400} \\\textbf{(A)}}& \thead{\textbf{TB24} \\ \textbf{ER/400} \\ \textbf{(A)}}& \thead{\textbf{TB24} \\ \textbf{MK1500} \\ \textbf{(A)}} \\ \toprule
	
1.0	 &37.46		 &37.46		&66.92  \\
2.0	 &74.92		 &74.92		&133.84 \\
3.0	 &112.39	 &112.39	&200.76\\
4.0	 &149.85	 &149.85	&267.68\\
5.0	 &187.31	 &187.31	&334.60\\
6.0	 &224.80	 &224.80	&401.51\\
  
   \bottomrule
    \end{tabular} 
   \end{center}  
\caption{\label{tab:desy2beamgen:magnets:specs} The current settings of the power supplies as a function of the required particle momentum for both TB21/22 (ER/400) and TB24 (ER/400+MK1500).}

\end{table}

\subsection{Primary Collimator}\label{sec:desy2beamgen:primarycoll}

\begin{figure}[htbp]
\centering
\includegraphics[width=0.6\textwidth]{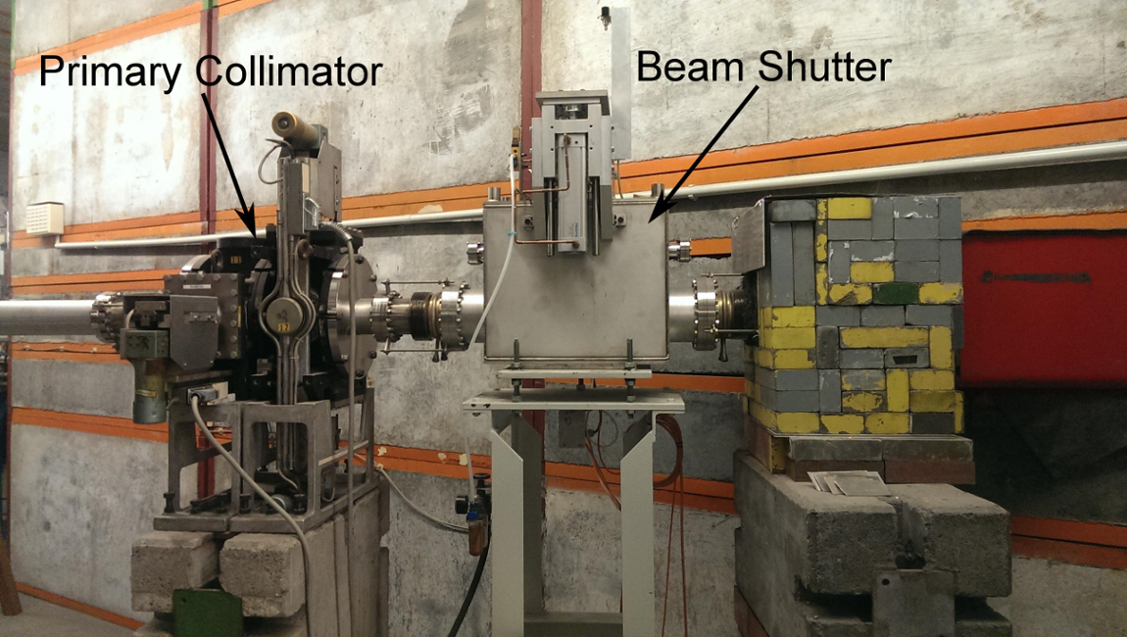}
\caption{(Colour online) The primary collimator and the beam shutter in the \desyii tunnel.
    }
\label{fig:desy2beamgen:collimatorandshutter}
\end{figure}

\begin{sidewaysfigure}[htbp]
\centering
\includegraphics[width=0.95\textwidth]{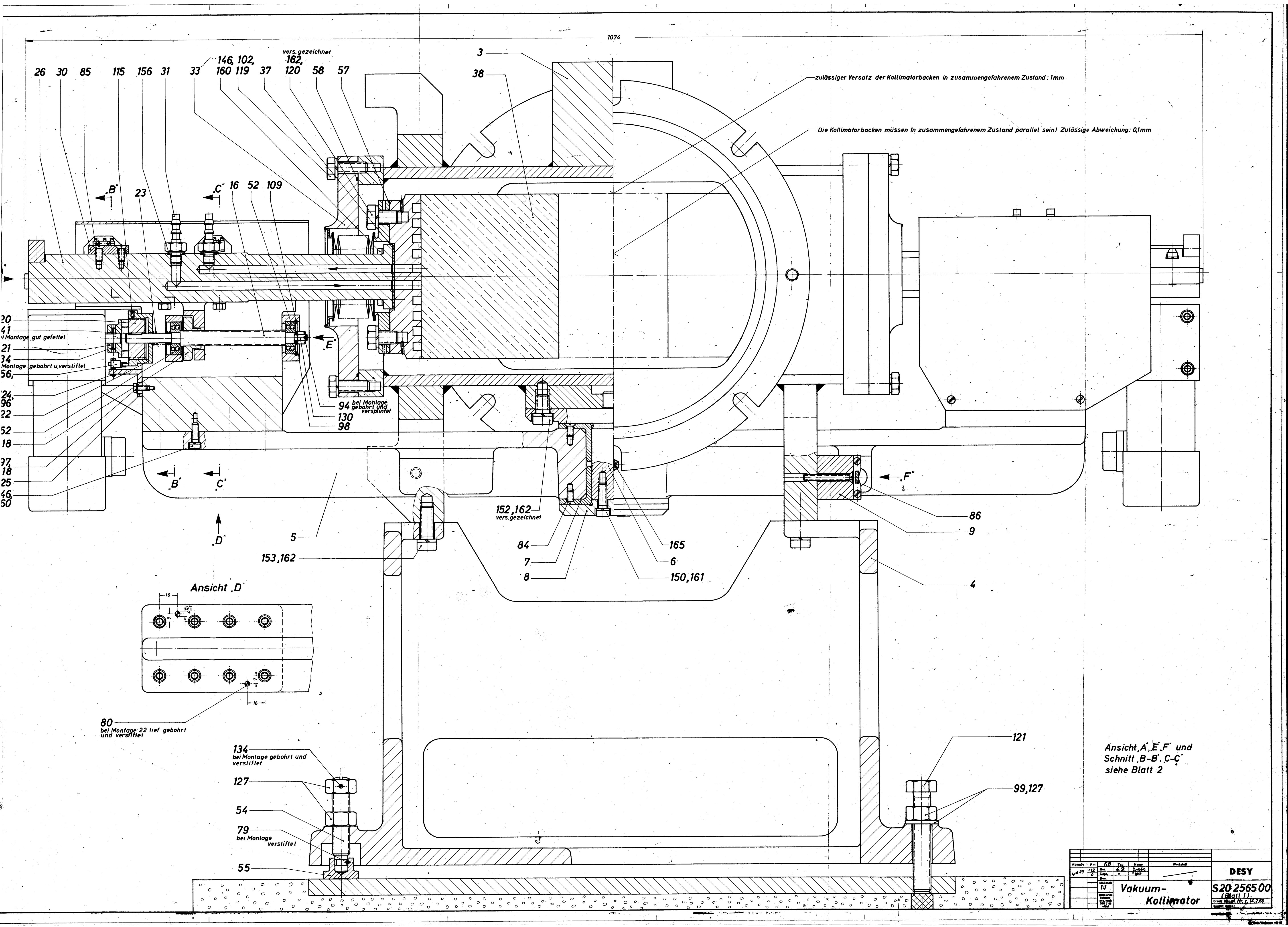}
\caption{Technical drawing of the primary collimator.}
\label{fig:desy2beamgen:collimator}
\end{sidewaysfigure}

The primary collimator is located behind the momentum defining magnet (Fig.~\ref{fig:desy2beamgen:collimatorandshutter}). 
It is a combination of a horizontal and a vertical collimator unit. 
The main part of each unit is a pair of motorized \SI{100}{\mm} thick jaws made out of tungsten (Fig.~\ref{fig:desy2beamgen:collimator}). 
The jaw position can be remotely controlled from the hut. 
The option to cool the jaws with water is not necessary due to the low beam intensities and therefore not being used.

\subsection{Test Beam Shutter}\label{sec:desy2beamgen:shutter}

\begin{sidewaysfigure}[htbp]
\centering
\includegraphics[width=0.95\textwidth]{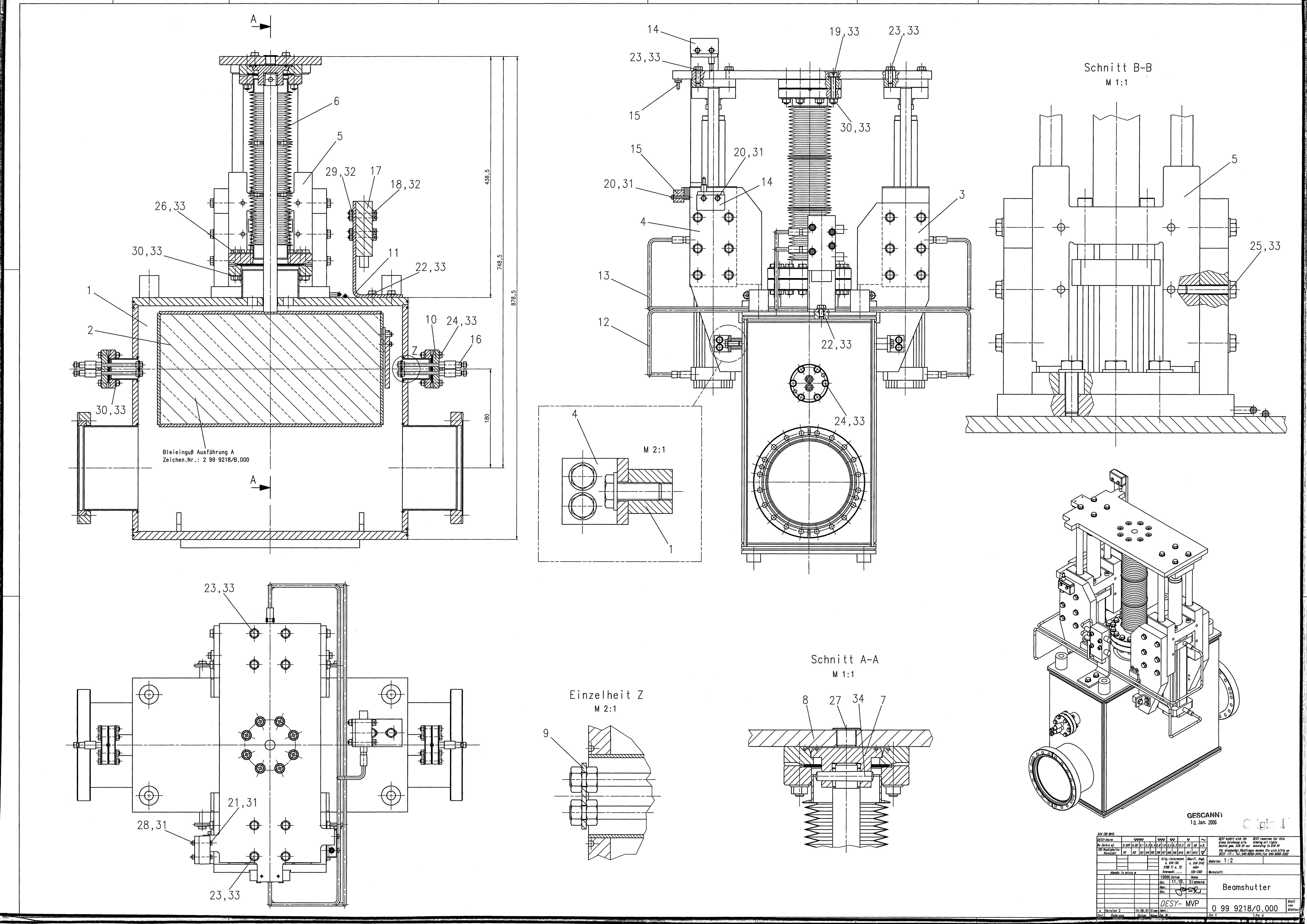}
\caption{Technical drawing of the Test Beam Shutter.}
\label{fig:desy2beamgen:shutter}
\end{sidewaysfigure}

Directly before the opening in the shielding wall of the accelerator a beam shutter is placed in each beam line  (Fig.~\ref{fig:desy2beamgen:collimatorandshutter}) 
to switch off the beam to the corresponding test beam area and to enable safe access to it. 
It is controlled by the safety interlock system. 
Its main part is a \SI{40}{\cm} long lead block with a diameter of \SI{20}{\cm} diameter that is moved by compressed air (Fig.~\ref{fig:desy2beamgen:shutter}).


\newpage

\section{Beam Line Instrumentation}\label{sec:beaminstr}

Each beam line is equipped with various monitoring systems to indicate the beam rate: 
so-called spill counters and 
beam monitors (Sec.~\ref{sec:beaminstr:rate}). 
There is not installed an independent instrumentation to measure the test beam momentum
permanently.
However different methods were used to periodically cross-check the calibration
of the current of the corresponding Test Beam Magnet (see Sec.~\ref{sec:desy2beamgen:magnets}) 
to select a certain test beam momentum (Sec.~\ref{sec:beaminstr:energycalibration}).

\subsection{Monitoring the Beam Rate}\label{sec:beaminstr:rate}
The \diitbf has currently no option for monitoring the beam generation directly inside the beam pipe. 
However, at each beam line at least two monitoring systems serve as guidance for beam operators to maximize the individual test beam rates by positioning fibers properly: 
Spill counters located at the beam line can measure the background photon radiation, 
beam monitors located in the test beam areas behind the vacuum pipe measure the passage of beam particles.

\subsubsection{Spill Counter}\label{sec:beaminstr:rate:spillcounter}
Information about the bremsstrahlung beam is determined from background radiation which accompanies the photon beam after production or after the secondary target (see Sec.~\ref{sec:desy2beamgen:secondary}). 
All beam lines have scintillator counters mounted close to the primary photon and secondary electron beam. 
The background rate associated with the spill from DESY II is measured using these counters. 
An additional spill counter based on Cherenkov light has been successfully operated directly in the beam at beam line 21. 
This counter is positioned after the conversion targets. 
A photomultiplier tube (PMT) in the shadow of heavy concrete shielding detects the Cherenkov light produced by electron positron pairs in a \SI{15}{\cm} long air radiator. 

Since both type of spill counters are located after the targets but before the dipole magnet and the collimators,  
there is no information on the momentum or on the shutter status.
Therefore they indicate that a test beam is produced by the targets, 
but do not indicate whether it propagates all the way through to the beam area.

\subsubsection{Beam Monitor}\label{sec:beaminstr:rate:beammonitor}

\begin{figure}
\begin{center}
\includegraphics[width=0.6\textwidth]{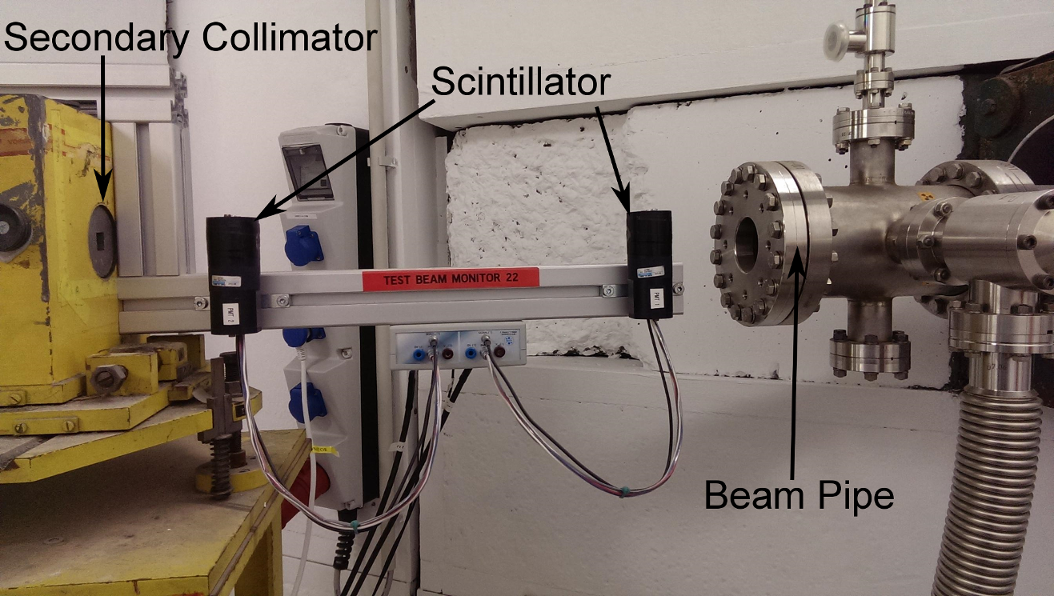}
\caption{(Colour online) A beam monitor consisting of two scintillators installed right at the end of the evacuated beam pipe.
    }
\label{fig:beaminstr:beammonitors}
\end{center}
\end{figure}

A set of two scintillation counters has been installed just in front of the
secondary collimator in each experimental test beam area \cite{bielski2012}.
These counters are $10 \times \SI{10}{\mm}$ large and \SI{3}{\mm} thick and measure in
coincidence the particle rate of the incoming beam right after leaving the evacuated beam pipe (Fig.~\ref{fig:beaminstr:beammonitors}).

The PMT units are connected to inputs of a EUDET Trigger Logic Unit (TLU) \cite{tlu}. The TLU 
provides a coincidence of the PMT signals and transmits all rates to the machine control system of the synchrotron (Sec.~\ref{sec:beaminstr:tine}).

\subsection{Momentum Calibration}\label{sec:beaminstr:energycalibration}
The momentum of the test beam and therefore the current of the corresponding test beam magnet (see Tab.~\ref{tab:desy2beamgen:magnets:specs} in Sec.~\ref{sec:desy2beamgen:magnets}) 
have occasionally been calibrated using a non-permanent lead glass counter setup.
However, the energy resolution of the lead glasses is not sufficient to provide a permanent measurement to determine the momentum of single particles on an event by event basis.
In addition, the pixel telescope together with the large MD magnet in TB21 was used as a spectrometer to confirm the momentum calibration by measuring the particle deflection by the magnet and determining the momentum of test beam particles (see Sec.~\ref{sec:performance:momentum}).

\subsection{Integration in the DESY-wide Machine Control System}\label{sec:beaminstr:tine}
For \desyii and \petraiii several machine parameters are measured and are recorded, time-stamped and archived 
using the TINE Accelerator Control and Monitoring system~\cite{tine-0957-0233-18-8-012}. 
One important parameter for the \diitbf which is used for some of the performance measurements (see Sec.~\ref{sec:performance}) is 
the \desyii beam intensity which is measured by using the \desyii beam monitoring system.
The system consists of monitors at three locations along the \desyii beam pipe, which inductively measure the beam current which is converted in numbers of particles.
Within in one \desyii cycle (see Fig.~\ref{fig:desy2:cycle}) the intensity is measured \SI{10}{\milli\second} after injection and \SI{10}{\milli\second} after reaching $E_{\rm max}$. 

Important \diitbf parameters like target 
positions, beam shutter states, magnet currents and the rates from the beam monitors (Sec.~\ref{sec:beaminstr:rate:beammonitor}) 
are also stored in the system. 
\diitbf users can run the TINE tools being available in the beam huts to access the archives and extract all relevant machine conditions.

\newpage

\section{Test Beam Areas}\label{sec:tbareas}

\begin{figure}[htbp]
\begin{center}
\includegraphics[width=1.0\textwidth]{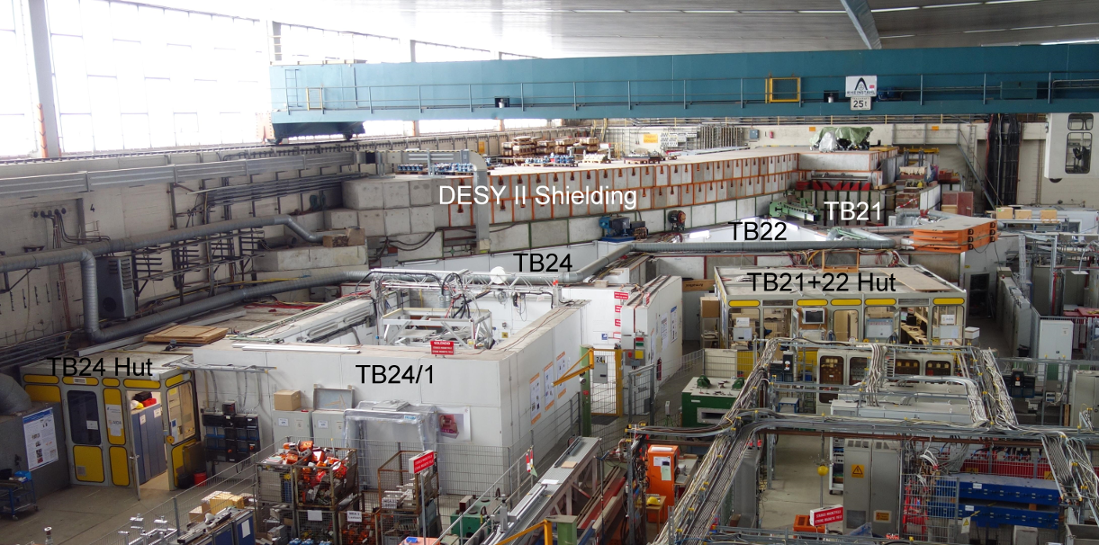}
\caption{(Colour online) View of the \diitbf inside of Hall 2.
    }
\label{fig:tbareas:halle2}
\end{center}
\end{figure}

There are three user areas available at the \diitbf corresponding to the three beam lines TB21, TB22 and TB24. 
The area of TB24 is subdivided into two subsequent areas, TB24 and TB24/1. 
This is done by using a second shutter system which allows working in TB24/1 while beam is available in TB24. 
For each area, there is a hut available, which also houses all the user controls for the corresponding beam line and provides network connectivity. 
A drawing of the areas and the huts is depicted in Figure~\ref{fig:desy2beamgen:schematic}, the corresponding view from the north in Figure~\ref{fig:tbareas:halle2}.

All user areas provide power distribution boxes with both \SI{230}{\volt} two-phase (CEE 7/7) and \SI{400}{\volt} three-phase outlets (IEC 60309 3L+N+PE \SI{16}{\ampere}/\SI{32}{\ampere}).
Furthermore each area has several sets of patch panels that provide connectivity to the huts, minimizing the amount of cabling which needs to be performed by the user. 
There are panels for high-voltage using SHV connectors and Coax-RG58 connector for signals.
For data transmission there are panels with CAT7 Ethernet 
and RJ45 connectors providing a local \SI{1}{\giga\bit\per\second} network 
and panels with both single-mode (OS2) and multi-mode (OM3) fibers and LC connectors.

Remote-controllable stages with a capacity of up to \SI{1}{\tonne} can be installed in all areas allowing users to mount their devices.
All areas also provide a IP-based CCTV system, cooling water and nitrogen gas.
Finally, a laser alignment system is installed in each test beam area using Class 1M line lasers. 
The system provides the users with a tool to align their devices with the beam axis. 
Horizontal and vertical alignment are supported and can be individually enabled. 

The following sections describe two of the last beam line components which are located in each beam area: 
the secondary collimator (Sec.~\ref{sec:tbareas:secondarycoll}) and 
the shielding and beam dumps (Sec.~\ref{sec:tbareas:shieldingdump}). 

\subsection{Secondary Collimator}\label{sec:tbareas:secondarycoll}

\begin{figure}[htb!]
\begin{center}
\includegraphics[width=0.6\textwidth]{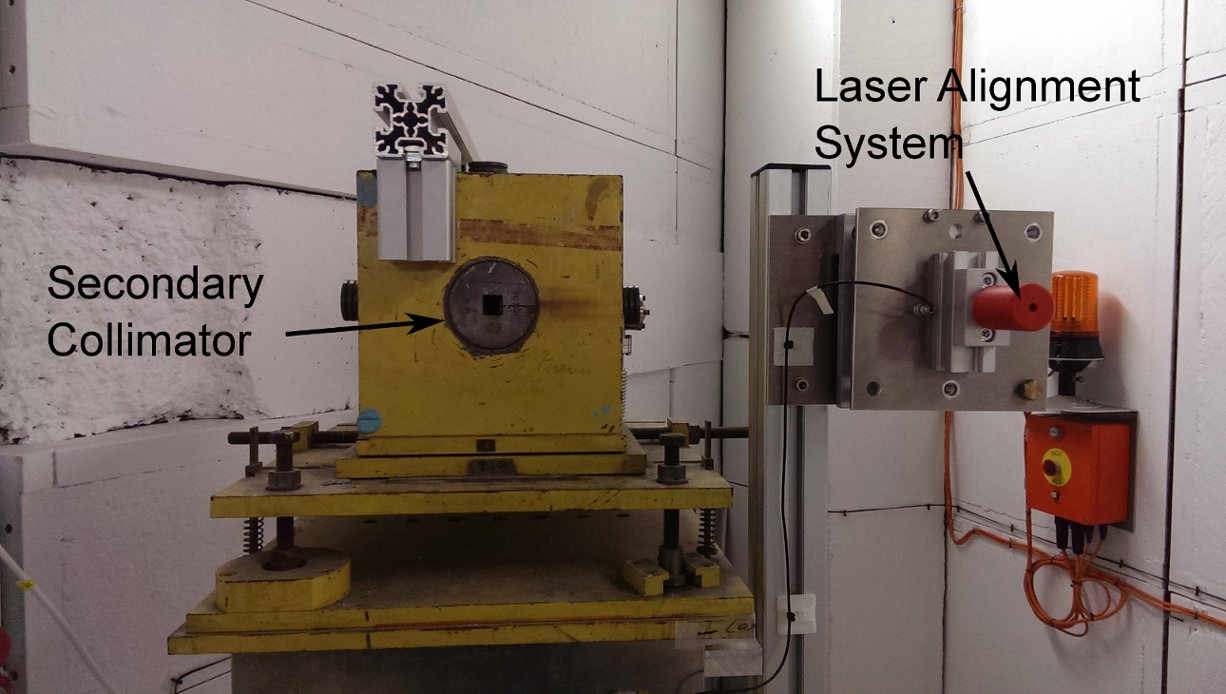}
\caption{(Colour online) The secondary collimator in TB22, the round, gray lead inset is visible in the center of the yellow painted collimator block. 
    A part of the laser alignment system can be seen on the right (red cap).
    }
    \label{fig:tbareas:secondarycollimator}
\end{center}
\end{figure}

In each beam line, a secondary collimator is installed right after the end of the evacuated beam pipe, which can be used to further 
collimate the beam (Fig.~\ref{fig:tbareas:secondarycollimator}). 
It uses lead insets with different bore diameters and shapes, which can be manually selected.
The beam monitors (see Sec.~\ref{sec:beaminstr}) are installed just after the end of the evacuated beam pipe and 
before the secondary collimator. Various lead insets 
with windows sizes ranging from \SI[product-units=repeat]{5x5}{\mm} to \SI[product-units=repeat]{20x20}{\mm} are available. 

\subsection{Shielding and Beam Dumps}\label{sec:tbareas:shieldingdump}


In all test beam areas, the walls are made out of shielding concrete blocks where the electron beams end. 
This is sufficient to limit the doses outside the areas below \SI{1}{\milli\sievert\per a} 
which corresponds to the legal limit for persons not occupationally exposed to radiation. 
To reduce the dose even further, additional beam dumps have been installed in front of the walls.

\begin{figure}[htb!]
\begin{subfigure}[b]{0.56448\textwidth}
\includegraphics[width=\textwidth]{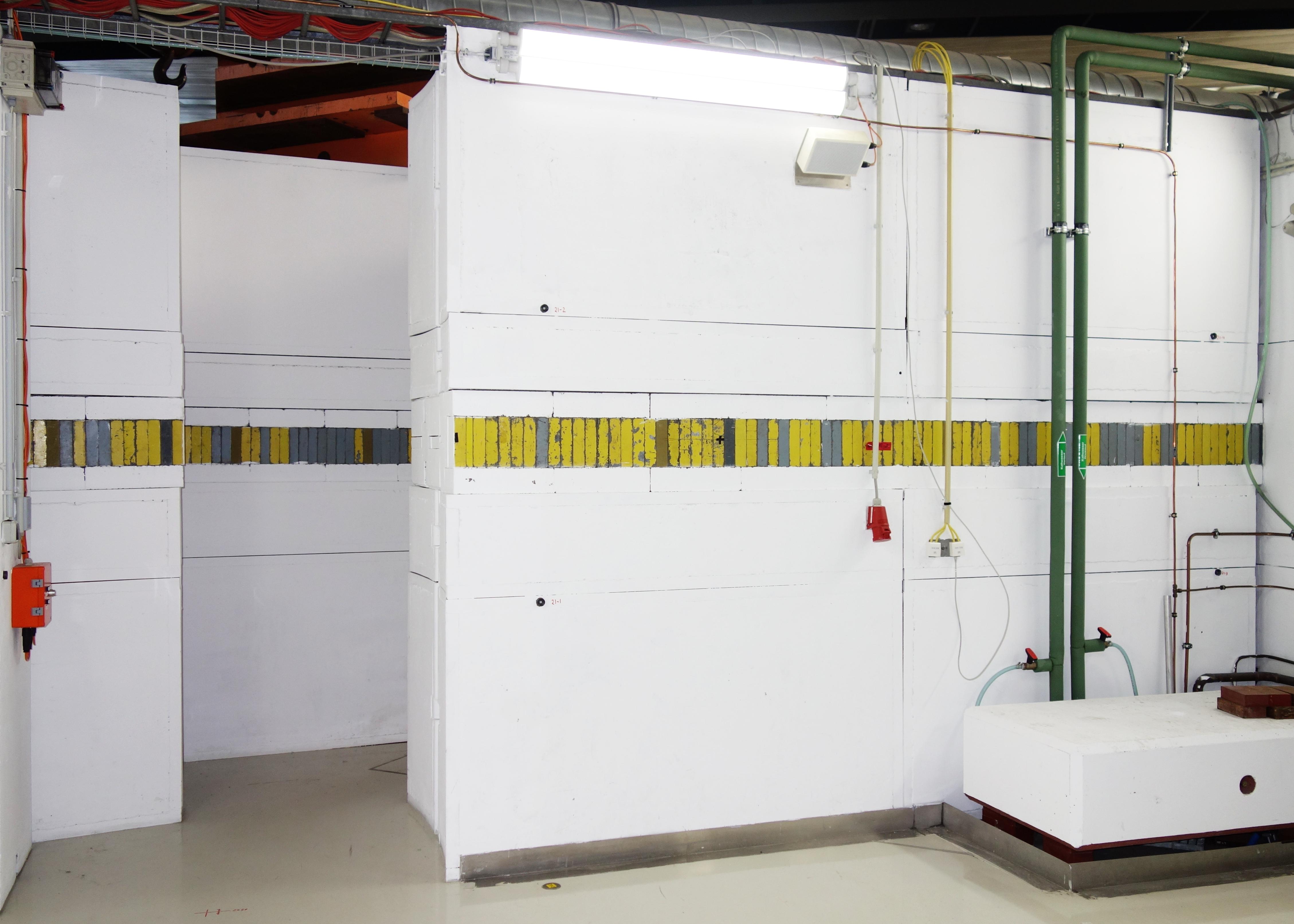}
\caption{TB21}
\label{sfig:beamdumpTB21}
\end{subfigure}
\hfill
\begin{subfigure}[b]{0.288\textwidth}
\includegraphics[width=\textwidth]{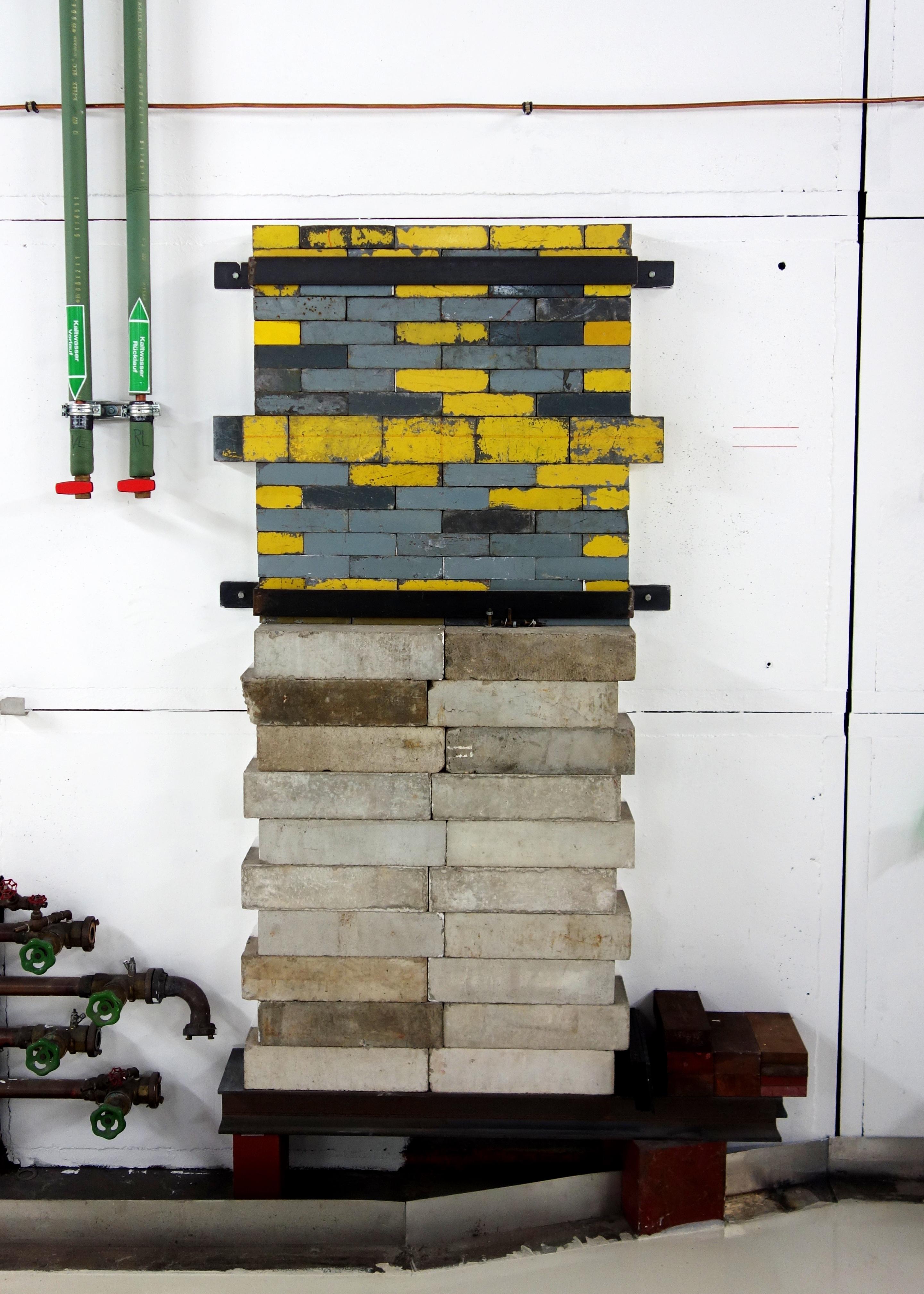}
\caption{TB22}
\label{sfig:beamdumpTB22}
\end{subfigure}
\begin{subfigure}[b]{0.56448\textwidth}
\includegraphics[width=\textwidth]{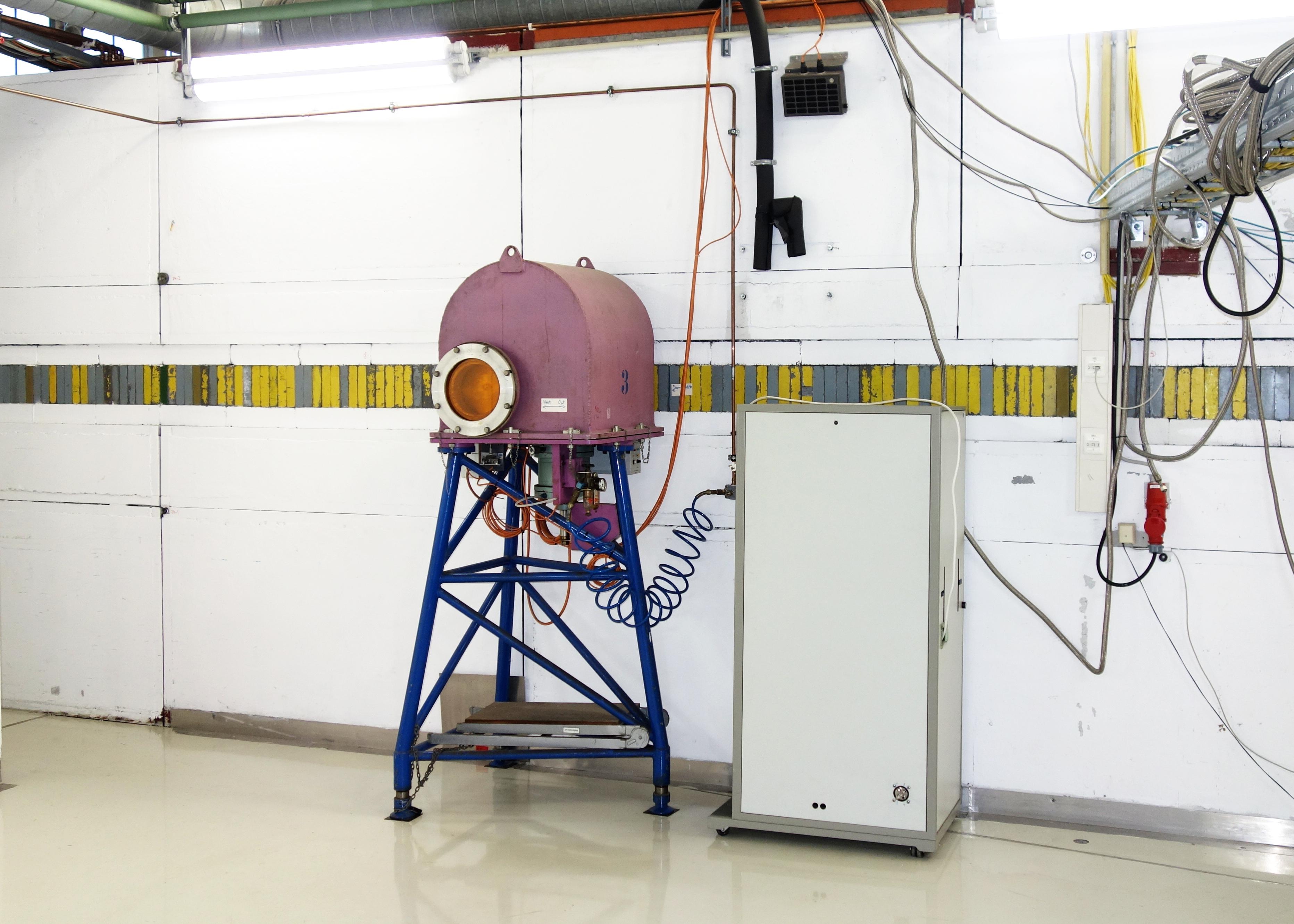}
\caption{TB24}
\label{sfig:beamdumpTB24}
\end{subfigure}
\hfill
\begin{subfigure}[b]{0.288\textwidth}
\includegraphics[width=\textwidth]{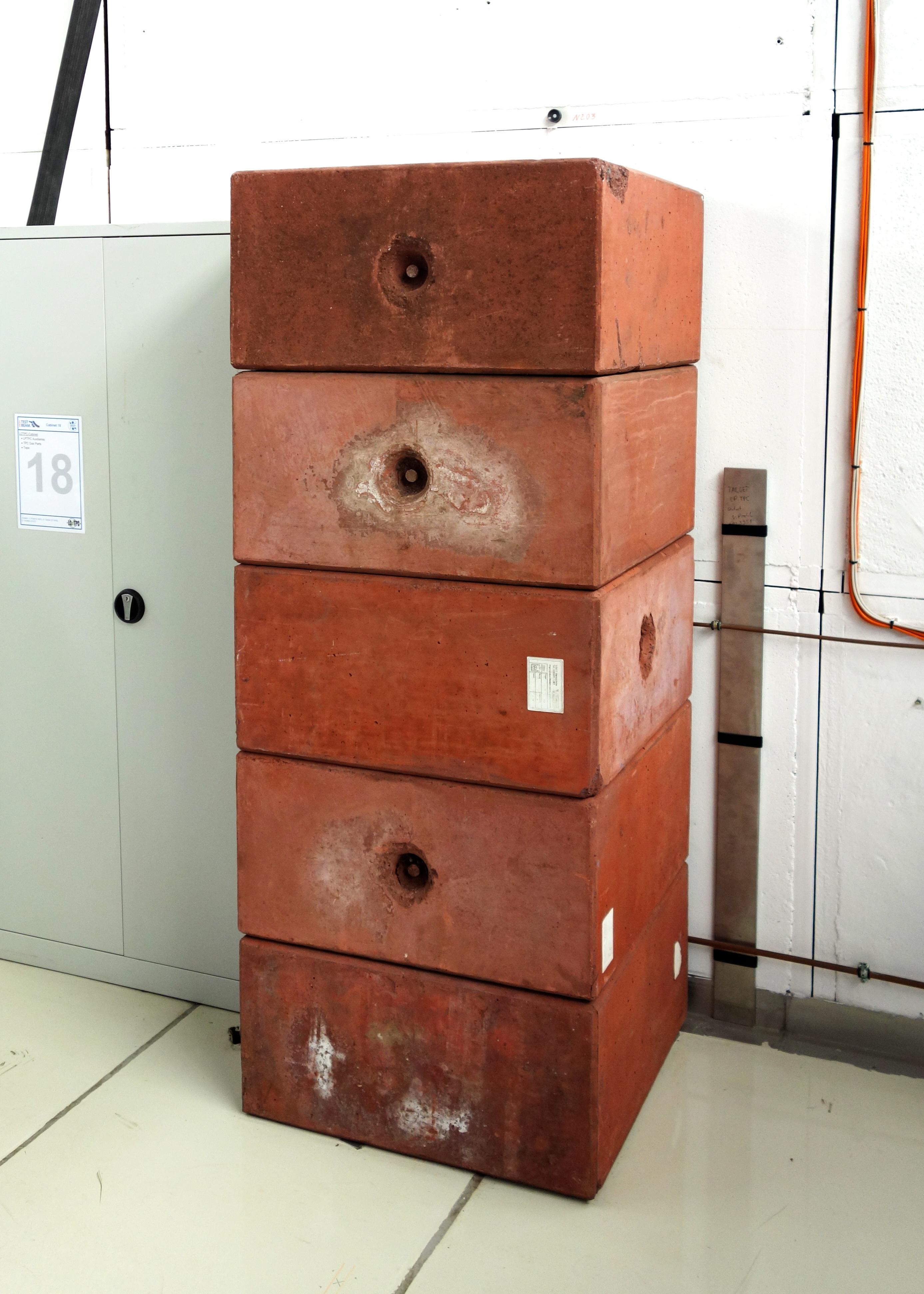}
\caption{TB24/1}
\label{sfig:beamdumpTB24-1}
\end{subfigure}
\caption{(Colour online) Beam dumps in the test beam areas: A horizontal row of (yellow and gray) lead blocks in the areas TB21 and TB24, \protect\subref{sfig:beamdumpTB21}) 
and \protect\subref{sfig:beamdumpTB24}), an array of lead blocks in TB22, \protect\subref{sfig:beamdumpTB22}) and a pillar of (red) shielding 
concrete blocks in TB24/1, \protect\subref{sfig:beamdumpTB24-1}).}
\label{fig:beamdumps}
\end{figure}

In area TB21, a horizontal line of lead blocks with the same thickness has been 
installed along the whole wall, since the installed dipole magnet could deflect 
the beam in this direction (Fig.~\ref{sfig:beamdumpTB21}). 
A similar approach has been used in area TB24.
In area TB22, the beam dump consists of an array of lead blocks with a 
thickness of \SI{10}{\cm} at the wall (Fig.~\ref{sfig:beamdumpTB22}). 
In area TB24/1, the beam dump has 
been realized with a set of additional shielding concrete blocks of \SI{80}{\cm} 
thickness that cover the entire range to which the beam can be deflected by the PCMAG 
solenoid (Fig.~\ref{sfig:beamdumpTB24-1}). 
Since at this beam line only 
electrons are available and the solenoid polarization is chosen such, 
that they are deflected downwards, there is no need to extend the shielding 
significantly above the beam height of about \SI{1.7}{\m}. 
The results of a \geant~\cite{Agostinelli:2002hh,Allison:2006ve,Allison:2016lfl} simulation have been used to determine the heights 
at which the beam hits the area wall after passing the PCMAG with a \SI{1}{\tesla} field.


The resulting radiation doses of a FLUKA~\cite{BOHLEN2014211,Ferrari_fluka:a} simulation, 
 for the example of a \SI{10}{\cm} thick lead block array in front of a test beam area wall as in TB22 are presented in the following. 
Two beam momenta were chosen for the simulation, \SI{2.8}{\GeV/c} for the maximum rate ---a particle rate of \SI{40}{\kilo\hertz} is 
assumed, according to rate tests with fully open collimators and a high current 
in DESY II--- and \SI{6}{\GeV/c} for the 
maximum momentum for which a rate of \SI{1}{\kilo\hertz} is assumed. 
The simulations were performed using FLUKA version \emph{Fluka2011.2x}
simulating 10 runs with 10,000 primary histories, each. 

\begin{figure}[htb!]
\begin{subfigure}[b]{0.5\textwidth}
\includegraphics[width=\textwidth]{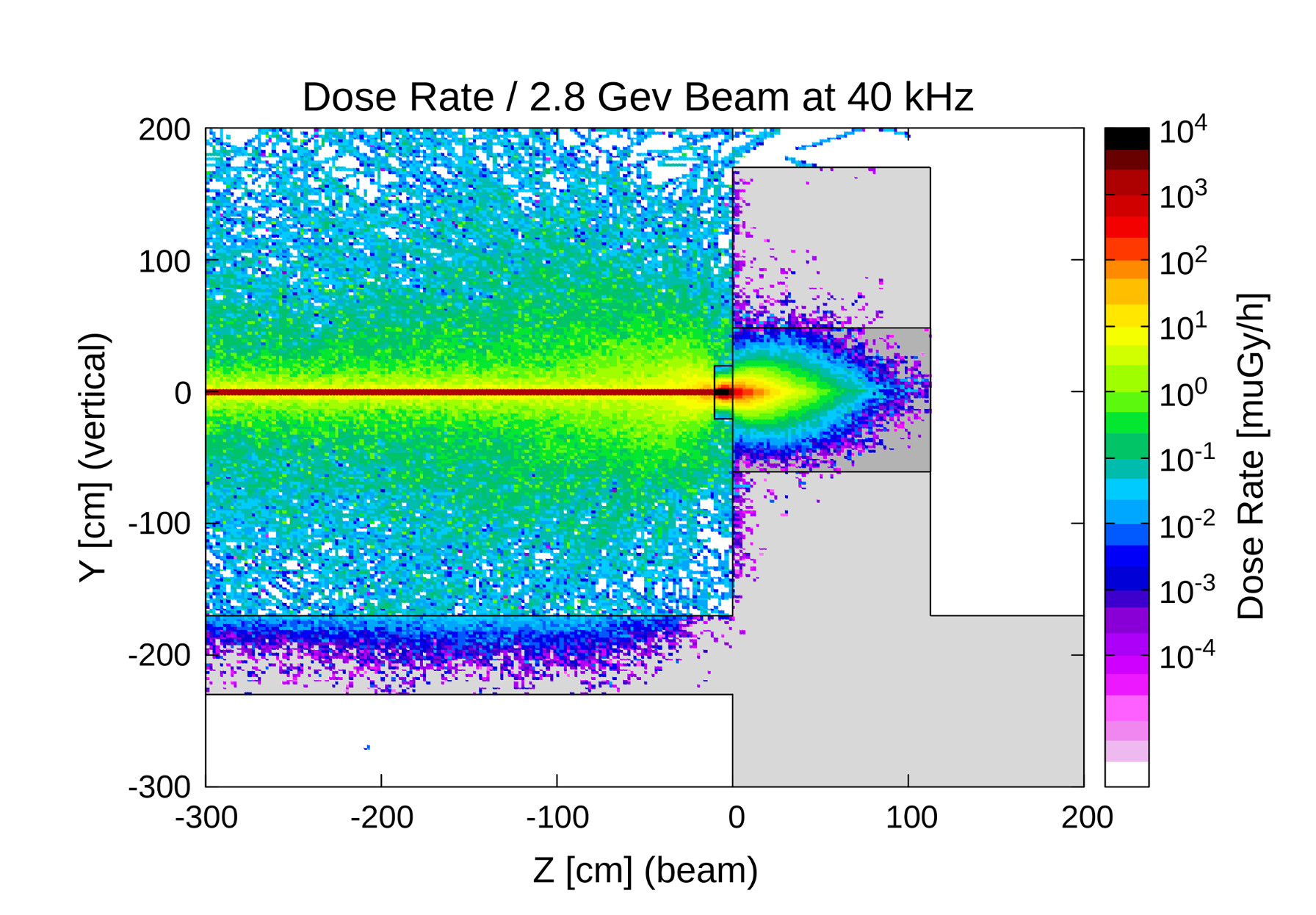}
\caption{}
\label{sfig:doseratebeamdumpTB24_2.8GeVLead}
\end{subfigure}
\hfill
\begin{subfigure}[b]{0.5\textwidth}
\includegraphics[width=\textwidth]{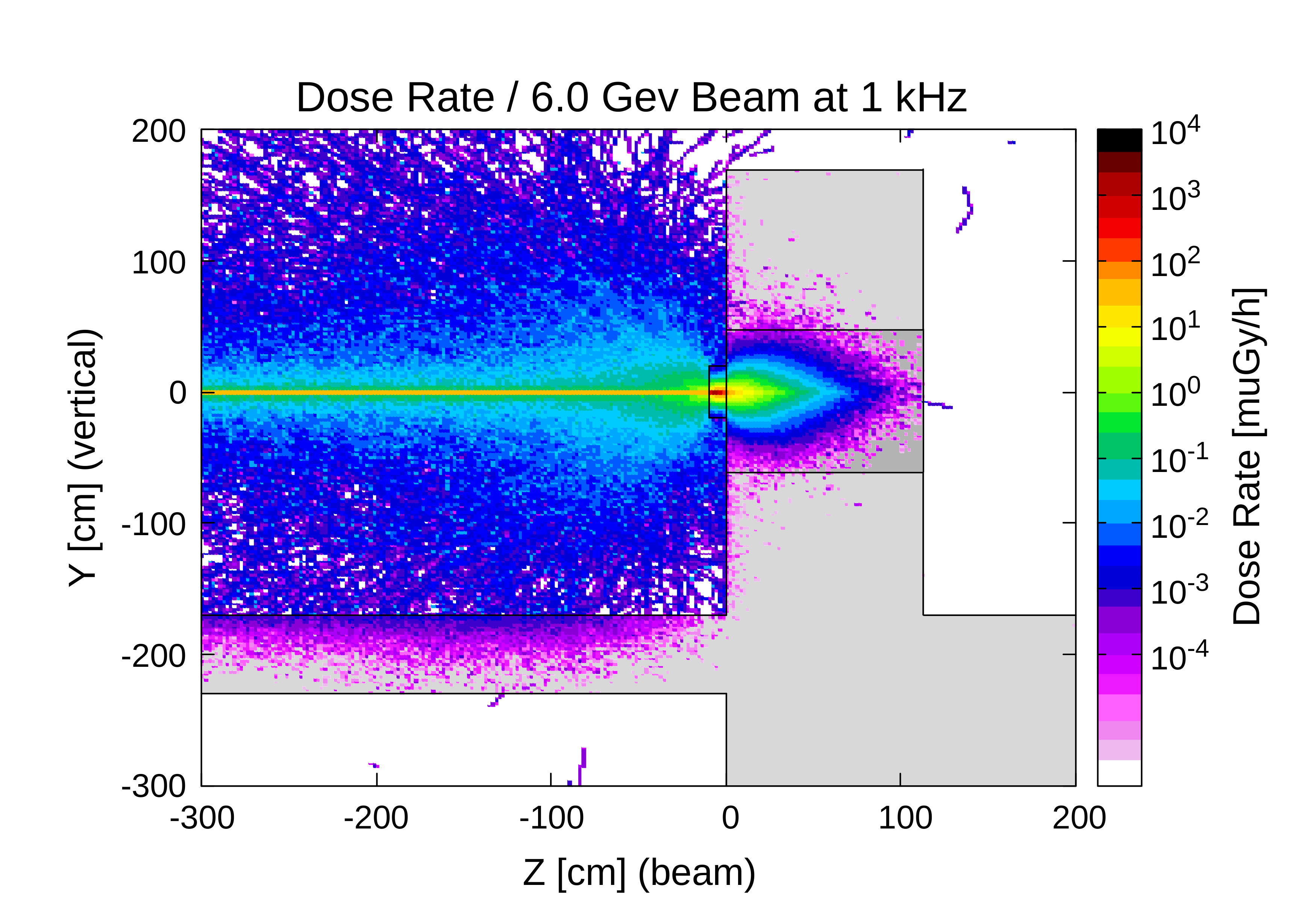}
\caption{}
\label{sfig:doseratebeamdumpTB24_6GeVLead}
\end{subfigure}
\caption{(Colour online) Dose rates in \SI{}{\micro\Gray\per\hour} for \protect\subref{sfig:doseratebeamdumpTB24_2.8GeVLead}) a \SI{2.8}{\GeV/c} (\SI{40}{\kilo\hertz}) and \protect\subref{sfig:doseratebeamdumpTB24_6GeVLead}) a \SI{6}{\GeV/c} (\SI{1}{\kilo\hertz}) beam entering from the left side.}
\label{fig:doseratebeamdumplead}
\end{figure}

\begin{figure}[htb!]
\begin{subfigure}[b]{0.5\textwidth}
\includegraphics[width=\textwidth]{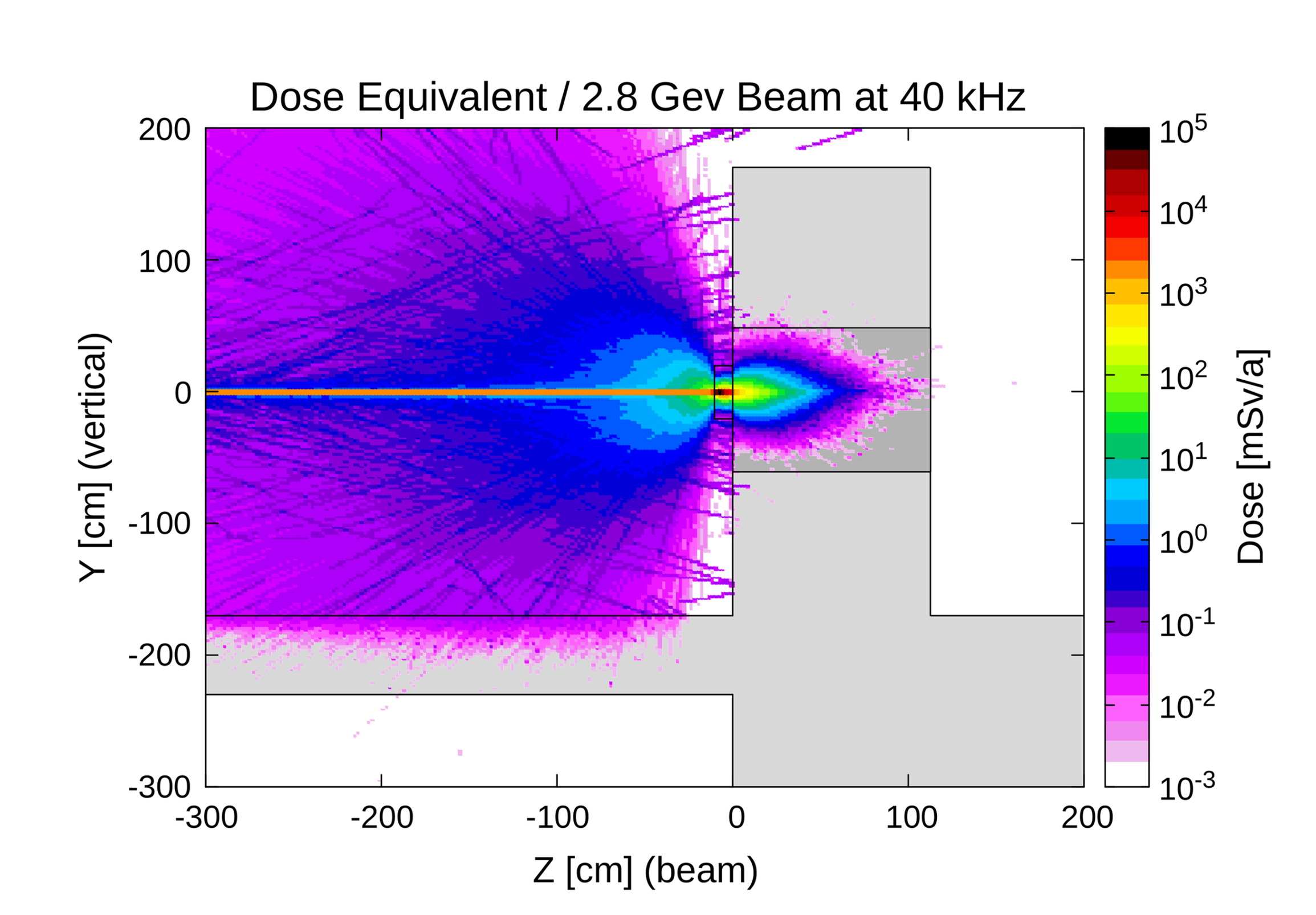}
\caption{}
\label{sfig:doseequivalentbeamdumpTB24_2.8GeVLead}
\end{subfigure}
\hfill
\begin{subfigure}[b]{0.5\textwidth}
\includegraphics[width=\textwidth]{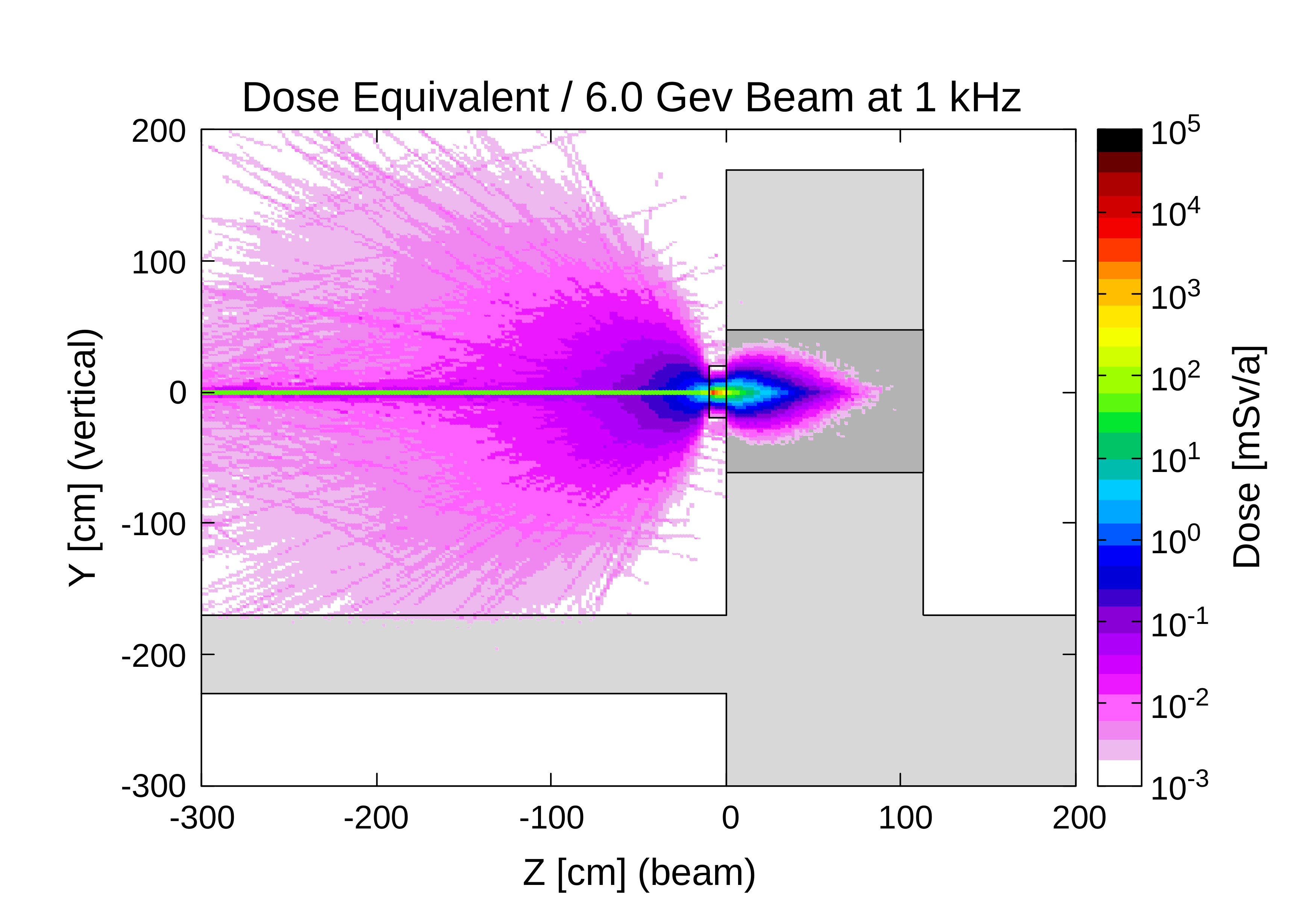}
\caption{}
\label{sfig:doseequivalentbeamdumpTB24_6GeVLead}
\end{subfigure}
\caption{(Colour online) Dose equivalents in \SI{}{\milli\sievert\per a} for \protect\subref{sfig:doseequivalentbeamdumpTB24_2.8GeVLead}) a \SI{2.8}{\GeV/c} (\SI{40}{\kilo\hertz}) and \protect\subref{sfig:doseequivalentbeamdumpTB24_6GeVLead}) a \SI{6}{\GeV/c} (\SI{1}{\kilo\hertz}) beam entering from the left side.}
\label{fig:doseequivalentbeamdumplead}
\end{figure}

Figure~\ref{fig:doseratebeamdumplead} shows the resulting dose rate in \SI{}{\micro\Gray\per\hour}. 
Figure~\ref{fig:doseequivalentbeamdumplead} shows the resulting equivalent dose in \SI{}{\milli\sievert\per a}, assuming 2000 working hours per year which corresponds to 50 working weeks per year with 40 working hours per week. 
The plots show averaging projections from \SI{-2.5}{\cm} to \SI{+2.5}{\cm} of the beam position in the plane looking from the side on the beam. 
The beam height is at zero on the vertical axis. 
The simulation includes the \SI{60}{\cm} thick floor under the beam area to judge the doses in the tunnels below some parts of the test beam areas. 
At beam height, the wall consists of a \SI{112.8}{\cm} thick block made out of shielding concrete with magnetite 
aggregate \cite{DESYHandbuch} (dark gray). The rest of the wall and the floor are made out of normal concrete (light gray).

As can be seen, the resulting doses outside the test beam area are very low for both beam momenta and rates.
The dose rate stays under \SI[parse-numbers=false]{10^{-3}}{\micro\gray\per\hour}, mostly significantly under \SI[parse-numbers=false]{10^{-4}}{\micro\gray\per\hour}.
The equivalent dose stays in both cases below \SI[parse-numbers=false]{10^{-2}}{\milli\sievert\per a}, which is well below the legal limit.

\newpage

\section{Additional Infrastructures}\label{sec:addinf}

In addition to the common infrastructure available in all test beam areas at the \diitbf, 
certain beam lines offer additional and unique infrastructure like 
gas systems, slow control units, large bore test magnets and pixel beam telescopes. 

Fully-fledged gas systems are available to the users in TB22 and TB24 
including a gas warning system suitable for handling flammable gas and gas cabinets for storing gas bottles. 
Premixed gases can be provided by the central gas group of DESY.

In the framework of the AIDA2020 project~\cite{AIDA2020} a generic slow control system~\cite {Wu:2290758} has been installed at the \diitbf
which allows to record the environmental conditions during beam tests.
The system provides a standard suite of sensors for measuring temperature, pressure or humidity and can be easily extended with other sensors. 
As an additional feature, users can record the data and embed it into the data stream of the EUDAQ data acquisition framework described in Section~\ref{sec:addinf:telescopes}. 
Two identical systems have been installed on two movable racks which allow to move them into the test beam area, where they have been required.

In the following sections, unique and important infrastructure is described which is only available for certain beam lines and is located in the corresponding test beam area: 
\begin{itemize}
    \item {\bfseries in TB21:} Big dipole magnet (Sec.~\ref{sec:addinf:brm}) and DATURA, one of the two permanently installed EUDET-type Pixel Beam Telescopes (Sec.~\ref{sec:addinf:telescopes}).
    \item {\bfseries in TB22:} DURANTA, one of the two permanently installed EUDET-type Pixel Beam Telescopes (Sec.~\ref{sec:addinf:telescopes}).
    \item {\bfseries in TB24/1:} Superconducting solenoid PCMAG (Sec.~\ref{sec:addinf:pcmag}).
\end{itemize}

\subsection{Big Red Magnet -- MD Dipole}\label{sec:addinf:brm}

\begin{figure}[htb]
\centering
\includegraphics[width=0.45\textwidth]{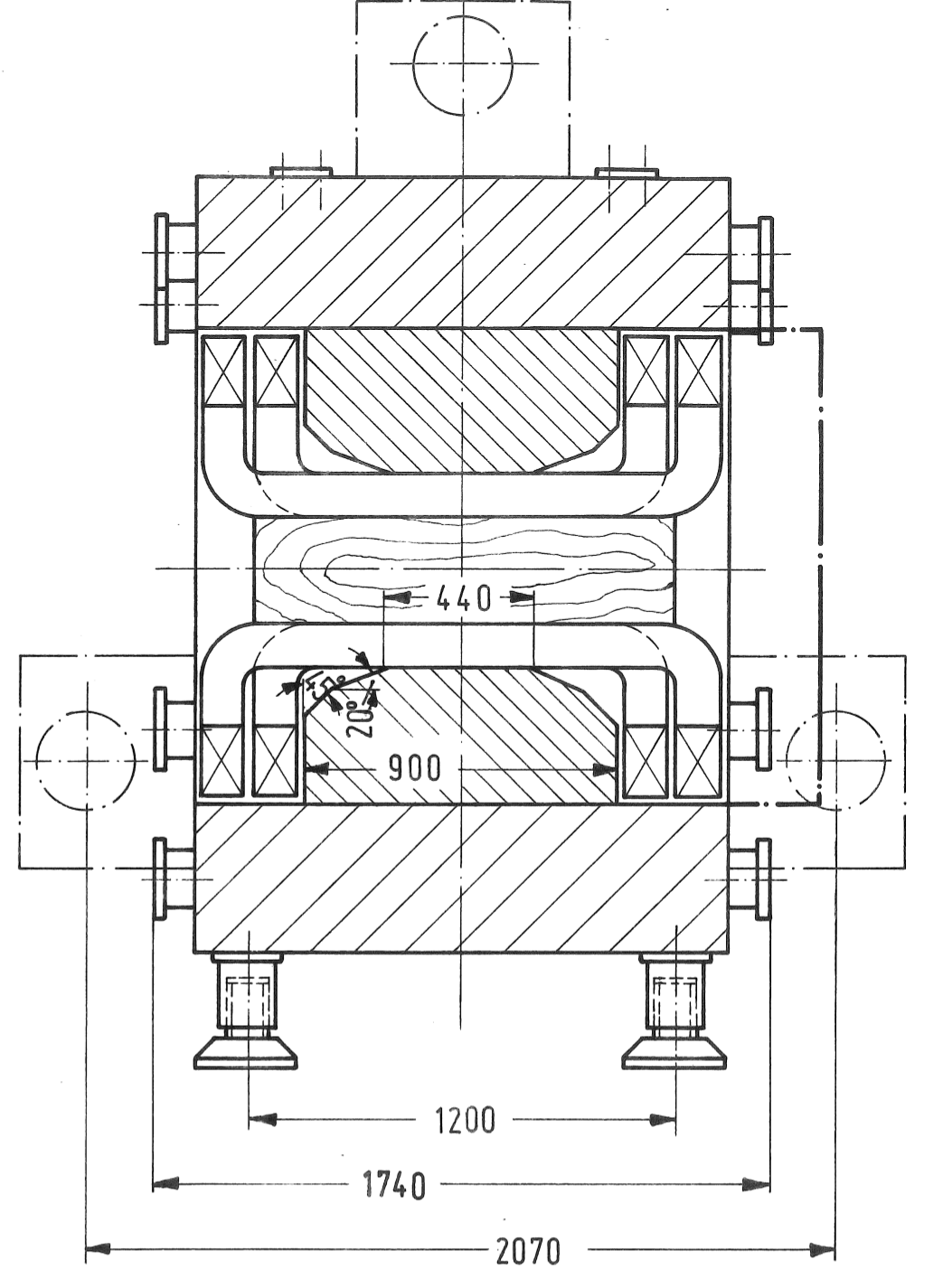}
\caption{Technical drawing of the Big Red Magnet with measures given in \SI{}{\cm}.}
\label{fig:addinf:brmdrawing}
\end{figure}

\begin{figure}[htb]
\centering
\includegraphics[width=0.7\textwidth]{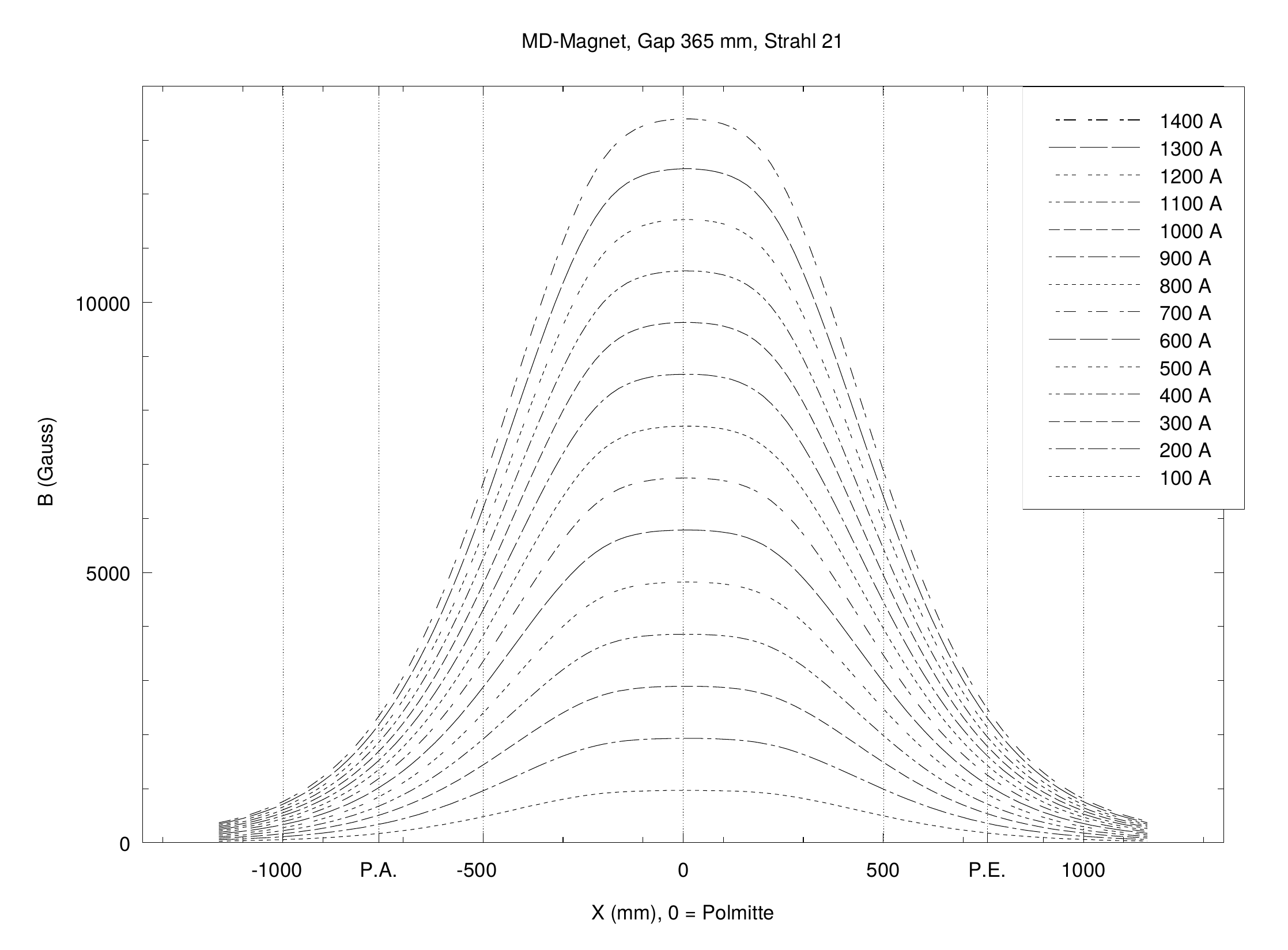}
\caption{The Big Red Magnet: measurement of the dipole field.}
\label{fig:addinf:brmfieldmap}
\end{figure}

Directly after the secondary collimator in TB21, a large window dipole magnet ---DESY type MD, spark chamber magnet--- is installed (see in the background of Fig.~\ref{fig:addinf:tscope_pic:tb21}).
It has an integrated length of about \SI{1}{\m} and an opening, which is about \SI{1.50}{\m} wide and \SI{0.35}{\m} high (Fig.~\ref{fig:addinf:brmdrawing}).
The maximum field is \SI{1.35}{\tesla} at \SI{1400}{\A} which has been measured in the middle of the field volume (Fig.~\ref{fig:addinf:brmfieldmap}).

This magnet has been used as a particle spectrometer:  
By measuring the deflecting angle of the particle track using a pixel telescope 
the momentum of the incoming beam can be precisely determined 
(see Sec.~\ref{sec:performance:primarycollimator}).
Furthermore it has been a crucial component for user groups performing 
gamma-electron separation in tests in calorimeters including an additional 
bremsstrahlung or Lorentz angle measurements in Silicon sensors. 
Finally, it has also been used to qualify electronics for the usage in high 
magnetic fields.

\subsection{PCMAG}\label{sec:addinf:pcmag}

\begin{figure}[htbp]
\centering
\includegraphics[width=0.50\textwidth]{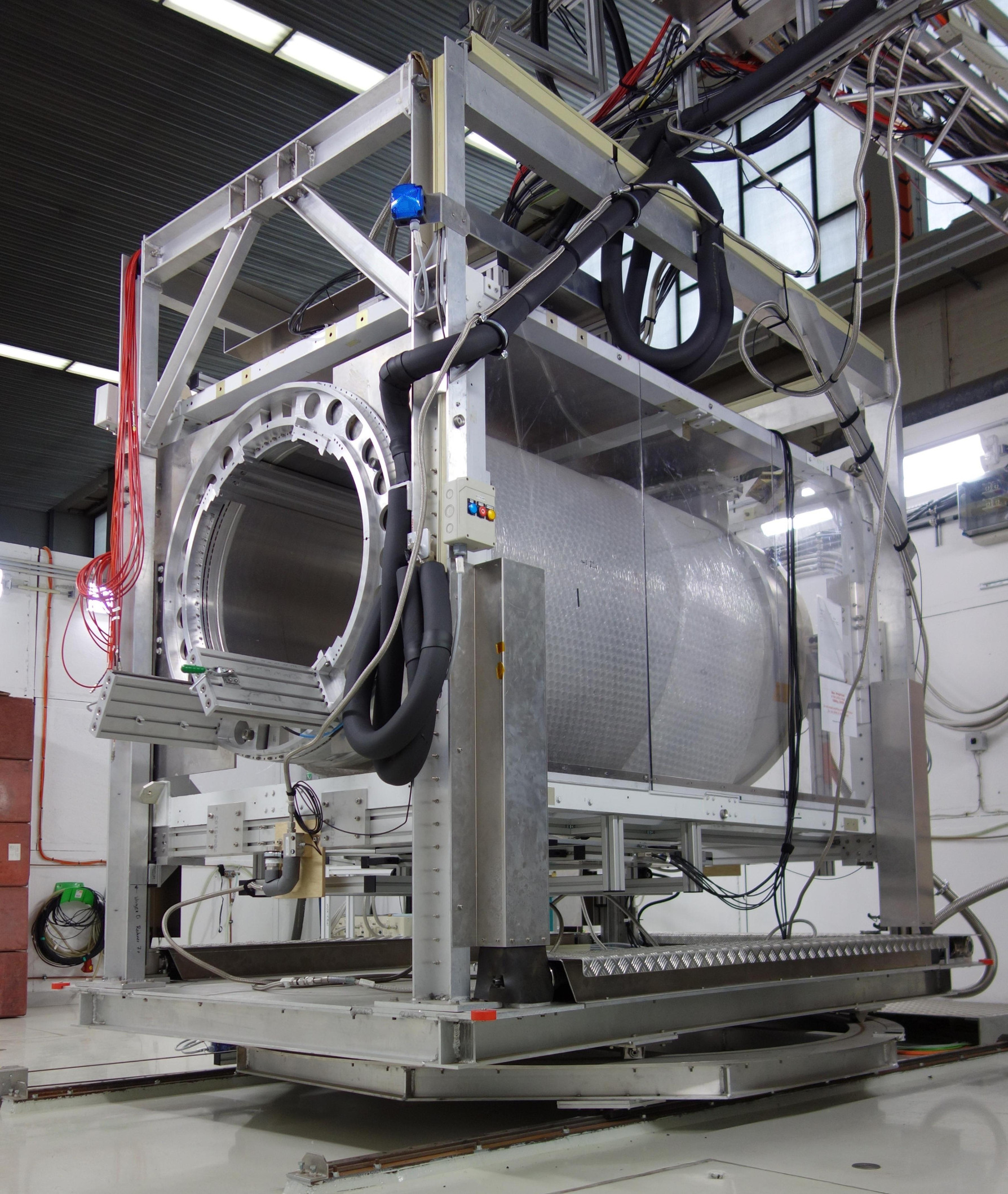}
\caption{(Colour online) TB24/1 test beam area at DESY: movable lifting stage with the PCMAG \SI{1}{\tesla} solenoid.}
\label{fig:addinf:pcmag:t241setup}
\end{figure}

In test beam area TB24/1, a large-bore superconducting solenoid is installed, called PCMAG (Fig.~\ref{fig:addinf:pcmag:t241setup}). 
This magnet has been provided by KEK and has been installed in 2006 at DESY, initially funded by the EU-FP6 EUDET project \cite{EUDET}. 
The PCMAG magnet has a superconducting coil which can produce fields of up to \SI{1.25}{\tesla}. 
At its operational current of \SI{438}{\A} the field in the center has a strength of \SI{1}{\tesla}.
Originally, the PCMAG was  designed for airborne experiments \cite{pcmag:yamamoto}. 
It has therefore been designed to be very lightweight and without a return yoke. This results in quite a large stray field. 
Table~\ref{tab:addinf:pcmag:pcmagvalues} lists the important parameters of the solenoid.

Due to its small radiation length of the coil and the comparably large bore, 
the PCMAG is well suited to be used in the test beam setup. 
Inside the PCMAG, a support structure has been installed. 
On this, detectors can be mounted on two rails and be moved along the $z$ direction of the magnet which is perpendicular to the test beam axis. 
This support structure is also necessary, since the inner wall of the magnet is not very robust 
and cannot hold heavy weights. 
The whole magnet is mounted in a remote controlled movable stage which provides horizontal and 
vertical movement perpendicular to the beam axis as well as rotations up to $\pm \SI{45}{\degree}$ in the horizontal plane.

\begin{table}[htbp]

  \begin{center}
    \begin{tabular}{ l c} \toprule 
      Central Magnetic Field (operational) & \SI{1.0}{\tesla} \\ 
      Central Magnetic Field (maximum) &  \SI{1.25}{\tesla} \\ 
      Operational Current at \SI{1}{\tesla} Field &  \SI{438}{\A} \\ \midrule
      Coil Length &  \SI{1.3}{m} \\ 
      Coil Diameter &  \SI{1.0}{m} \\ \midrule
      Radiation Length (wall including coil) & $\SI{0.2}{X_0}$ \\ 
      Interaction Length & $\SI{0.04}{\lambda_i}$ \\  
       & $\SI{4}{\g\per\cm^2}$ \\ \midrule
      Warm Bore Aperture &  \O~\SI{0.85}{\m} $\times$ length \SI{1.0}{\m} \\ 
      Weight &  \SI{430}{\kg}\\ \bottomrule
    \end{tabular}
    \caption{  \label{tab:addinf:pcmag:pcmagvalues}Main parameters of the PCMAG solenoid magnet. Original values from \cite{pcmag:yamamoto}, updated to current modification status where appropriate.}
  \end{center}
\end{table}

\begin{figure}[htb!]
\begin{subfigure}[b]{0.48\textwidth}
\includegraphics[width=\textwidth]{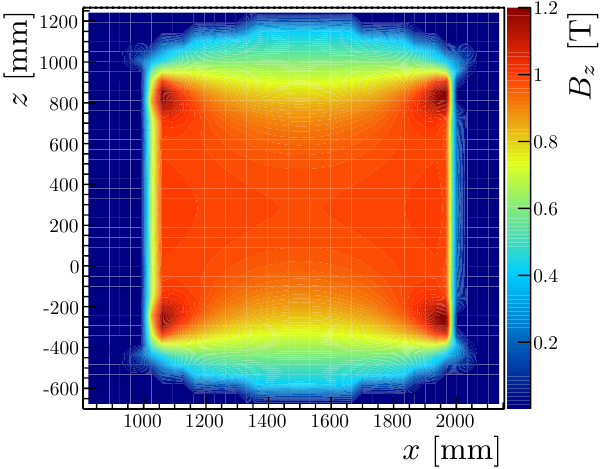}
\caption{$z$-component of $B$}
\label{fig:addinf:pcmag:pcmagfieldZ}
\end{subfigure}
\begin{subfigure}[b]{0.48\textwidth}
\includegraphics[width=\textwidth]{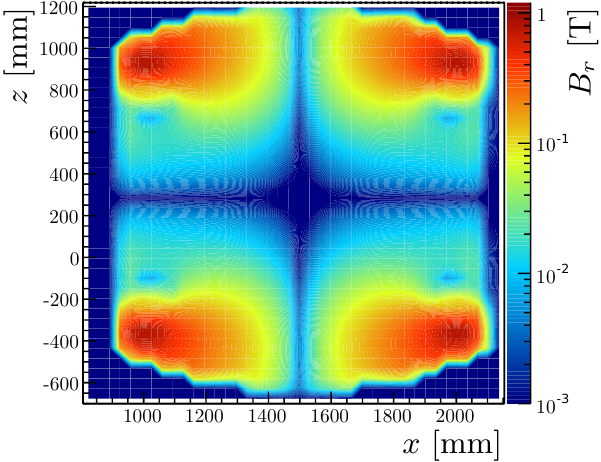}
    \caption{$r$-component of $B$}
\label{fig:addinf:pcmag:pcmagfieldR}
\end{subfigure}
\caption{(Colour online) Field maps of the measured magnetic field components $B_i$ of the PCMAG~\cite{pcmag:fieldana,ZenkerPHD}. 
    Radial direction $r$ is rotated around and perpendicular to the $z$ axis.
    }
\label{fig:addinf:pcmag:pcmagfield}
\end{figure}

The field of the magnet was measured with support from CERN to a precision of a few $10^{-4}$ \cite{pcmag:fieldmeas}. 
These measurements resulted in a field map that can be used in correction and reconstruction software (Fig.~\ref{fig:addinf:pcmag:pcmagfield} and \cite{pcmag:fieldana}). 

The cooling of PCMAG has been upgraded within the AIDA programme \cite{AIDA} from liquid Helium fillings to a commercial, closed circuit Helium compressor cooling system. 
This modification makes long, uninterrupted running periods possible, since it avoids internal pipe blocking by air introduced during liquid Helium re-filling. 
In addition, the PCMAG was modified to work with a continuously connected power supply instead of the previously used persistent current mode. 
This has increased the safety of the setup. 
Further, user groups gained more control over the setup, since they steer the magnet current by themselves.


\subsection{EUDET-type Pixel Beam Telescopes}\label{sec:addinf:telescopes}

The \diitbf is equipped with EUDET-type pixel beam telescopes which allow to track the test beam particles \cite{Jansen:2016bkd}. 
These kinds of test beam trackers were originally developed within the EUDET project \cite{EUDET} in order to meet most user requirements in terms 
of easy integration of the device under test (DUT), precise spatial resolution and suitable event rates. 

\begin{figure}[htbp]
\centering
\begin{subfigure}[b]{0.48\textwidth}
\includegraphics[width=\textwidth]{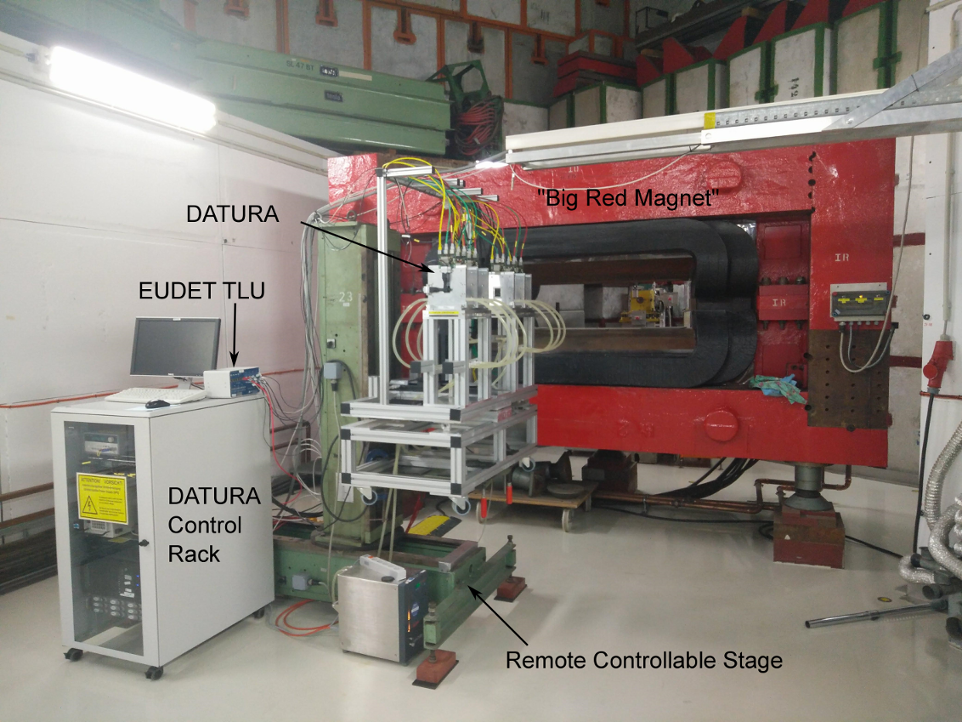}
\subcaption{DATURA in TB21}
\label{fig:addinf:tscope_pic:tb21}
\end{subfigure}
\hfill
\begin{subfigure}[b]{0.48\textwidth}
    \includegraphics[width=\textwidth]{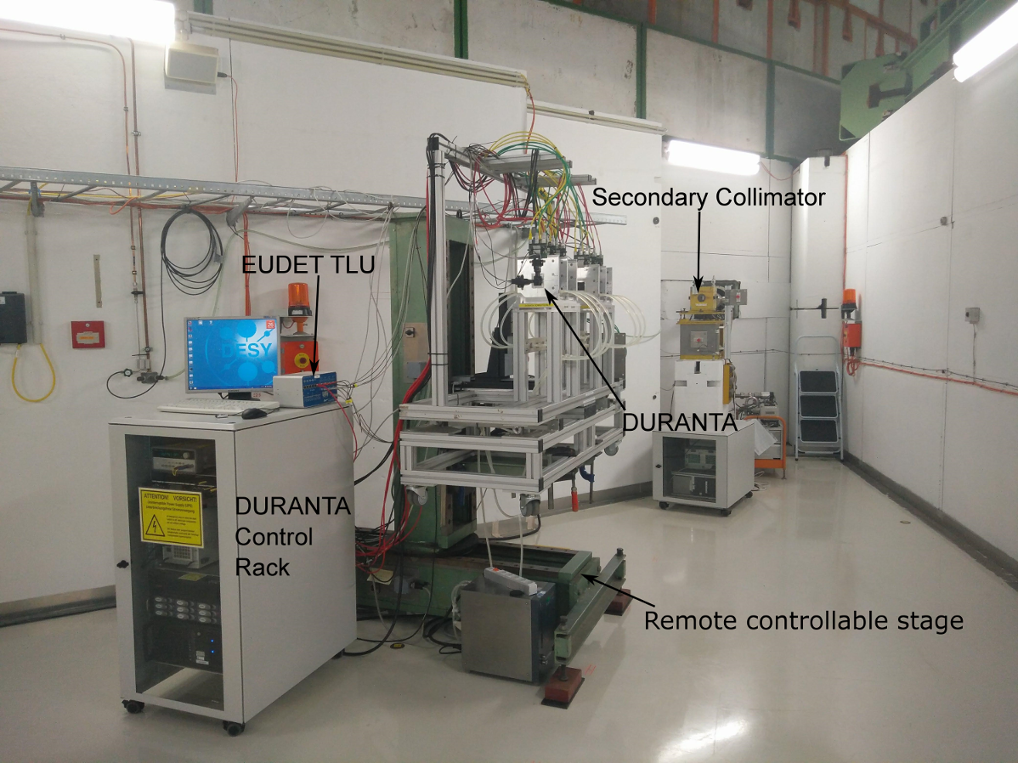}
\subcaption{DURANTA in TB22}
\label{fig:addinf:tscope_pic:tb22}
\end{subfigure}
\caption{(Colour online) 
    EUDET-type pixel beam telescopes in the test beam areas. 
}
\label{fig:addinf:tscope_pic}
\end{figure}

At the moment  there are seven replicas worldwide located at CERN, ELSA, SLAC and DESY \cite{www:telescope}.
At the \diitbf, the so-called DATURA telescope is installed in TB21 and the DURANTA telescope in TB22 (Fig.~\ref{fig:addinf:tscope_pic}).
In addition, a non-magnetic support frame is available to install one of the telescopes inside the PCMAG in TB24/1 (Sec.~\ref{sec:addinf:pcmag}).

Each beam telescope is composed of two telescope arms incorporating each three planes.
The standard telescope configuration is having the DUT incorporated by three telescope planes upstream and three telescope planes downstream. 
Each plane consists of a MIMOSA26 monolithic active pixel silicon sensor \cite{HuGuo:2010zz}. 
The pitch size is \SI[product-units=repeat]{18.4x18.4}{\micro\m} and pixels are arranged in 
1152 columns and 576 rows, which results in an active area of about \SI[product-units=repeat]{21.2x10.6}{\mm}. 
Pixel states are continuously read out in a rolling shutter by buffering line by line, the on-chip digitization provides a binary pixel information, and the output data stream is zero-suppressed. 
Therefore, the integration time is \SI{115.2}{\micro\s} per frame. 

\begin{figure}[htbp]
\centering
\begin{subfigure}[b]{0.58\textwidth}
\includegraphics[width=\textwidth]{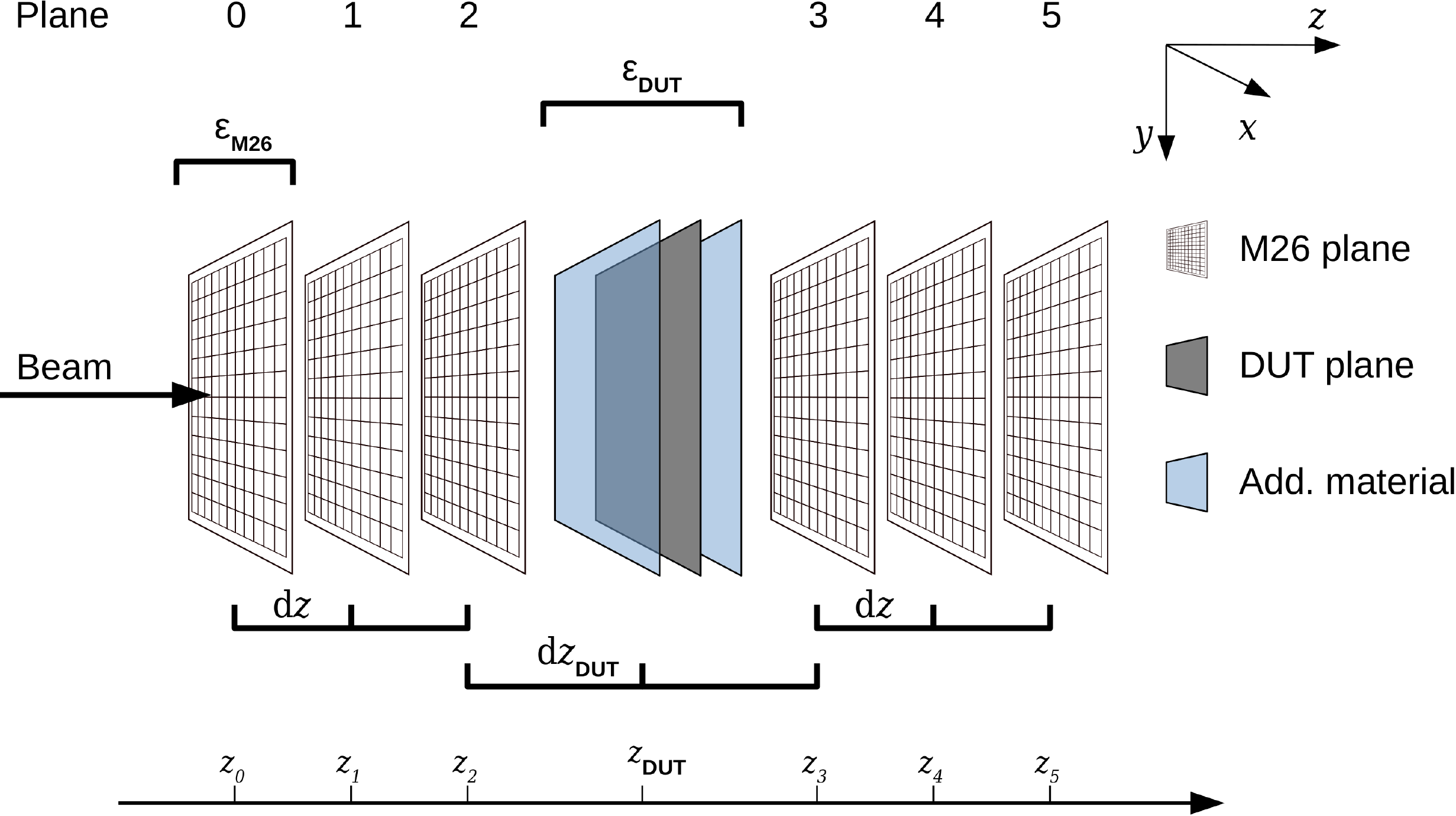}
\subcaption{Configuration}
\label{fig:addinf:tscope:sketch}
\end{subfigure}
\hfill
\begin{subfigure}[b]{0.38\textwidth}
    \includegraphics[width=\textwidth]{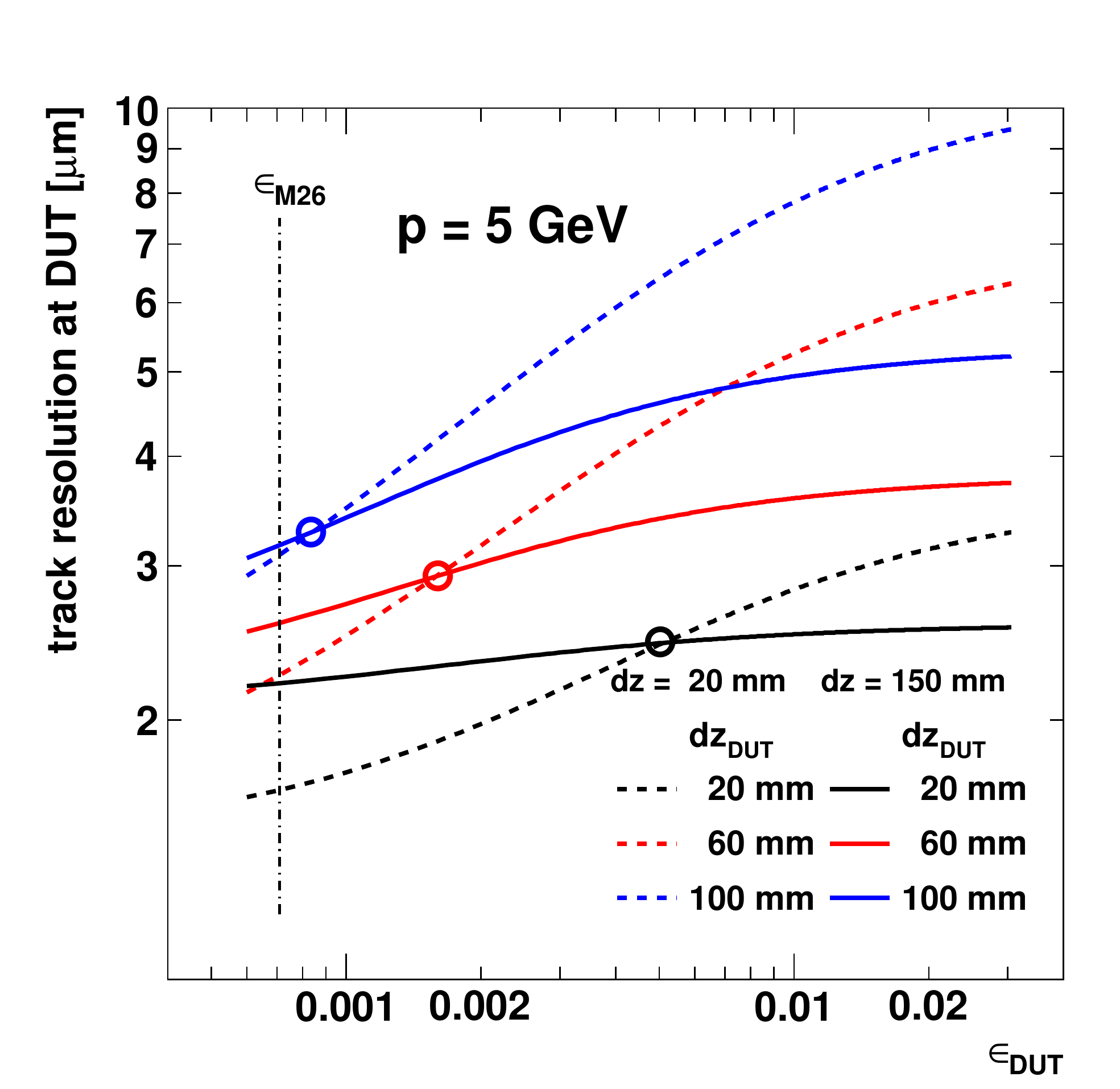}
\subcaption{Resolution}
\label{fig:addinf:tscope:res}
\end{subfigure}
    \caption{(Colour online) 
(\subref{fig:addinf:tscope:sketch}): Sketch of the standard telescope geometry and definition of important parameters. 
(\subref{fig:addinf:tscope:res}): The calculated track resolutions at the DUT for two geometries are shown at a particle momentum of \SI{5}{\GeV/c}.
$\epsilon_{\rm DUT}$ is defined as the DUT thickness normalized to its radiation length. 
Figures from \cite{Jansen:2016bkd}.
}
\label{fig:addinf:tscope}
\end{figure}

Each MIMOSA26 sensor has a thickness of \SI{50}{\micro\m} silicon and is shielded from environmental light using \SI{25}{\micro\m} thick Kapton foil on each side.
This keeps the material budget as low as possible in order to achieve a high track resolution at 1-6~GeV/c at the \diitbf.
The intrinsic resolution of a sensor was measured to be \SI[separate-uncertainty = true]{3.24(9)}{\micro\m} \cite{Jansen:2016bkd}. 
The best track resolution is estimated to \SI[separate-uncertainty = true]{1.83(3)}{\micro\m} using an equidistant plane spacing of \SI{20}{\mm} at a \SI{5}{\GeV/c} test beam. 
The realistic track resolution depends on the beam momentum, the plane spacing and the material budget of the DUT (Fig.~\ref{fig:addinf:tscope:res}).  

A EUDET Trigger Logic Unit (TLU) \cite{tlu} provides timestamp information on a particle passage through four trigger devices in coincidence.
Two trigger devices are located in front of the first telescope plane and two of them behind the last plane. 
Each trigger device is built up by a \SI{3}{\mm} thick and \SI[product-units=repeat]{2x1}{\cm} scintillator matching the 
MIMOSA26 sensor area and attached to a photomultiplier tube. 
The TLU and the DAQ of the MIMOSA26 sensors communicate in a handshake mode, so that if the TLU asserts a trigger, the MIMOSA26 DAQ raises a busy signal during readout of the frame.  
Additional sensors or DUTs can be integrated in the same way or in a no-handshake mode only accepting triggers.

The entire hardware is integrated in the EUDAQ data acquisition framework which merges data streams of all components as event-based data.
EUDAQ version~1 works for synchronous DAQ systems requiring one event per trigger. 
Thus, the trigger rate is limited by the slowest device.
Running only the telescope without any DUT results in an event rate of \SI{2.0}{\kilo\Hz} at a test beam particle rate of about \SI{10}{\kilo\Hz}~\cite{izzatullah2015}. 
Users can integrate the DAQ of their DUT by writing a EUDAQ component which matches a defined interface. 

DUTs are mechanically integrated between the two telescope arms at a $x$-, $y$-, $\phi$-stage system providing a \micron-precision.
This allows a geometrical scan of the DUT response in respect of the particle tracks, which is larger than the
 \SI[product-units=repeat]{2x1}{\cm} active area of the telescope. 
The DAQ PCs can be accessed from the huts via a local network connection
and the data acquisition can be monitored with the EUDAQ online monitor.

\newpage

\section{Performance Measurements}\label{sec:performance}

The performance of the \diitbf is determined by various measurements considering 
the time structure of the test beam particles (Sec.~\ref{sec:performance:timing}), 
the various rate dependencies (Sec.~\ref{sec:performance:rate}) and 
the particle momentum calibration and spread (Sec.~\ref{sec:performance:momentum}).
The results are dependent on the operation mode and the performance of the \desyii synchrotron 
(see Sec.~\ref{sec:desy2}) 
as well as the settings of the beam line components 
(see Sec.~\ref{sec:desy2beamgen}).
The results are in a good agreement with the results performed by simulations 
(see Sec.~\ref{sec:simulations:momentum_dist}).
All measurements were performed when \desyii was operated at $E_{\rm max}=\SI{6.3}{\GeV}$ 
unless indicated otherwise.

\subsection{Timing Structure}\label{sec:performance:timing}

The timing structure of the test beam depends on the performance and the operation mode of the \desyii synchrotron.
Three characteristic cycles have to be considered: 
\begin{itemize}
    \item the \petraiii top-up ranging from every \SI{30}{\s} to a few minutes (Sec. \ref{sec:performance:p3topup})
    \item the \desyii magnet or injection cycle every \SI{80}{\ms} or \SI{160}{\ms} (Sec.~\ref{sec:performance:d2cycle})
    \item the \desyii bunch cycle every \SI{0.976}{\micro\s} (Sec.~\ref{sec:performance:d2bunch})
\end{itemize}

\subsubsection{\petraiii Top-up}\label{sec:performance:p3topup}

\begin{figure}[htbp]
\begin{center}
    \includegraphics[width=0.6\textwidth]{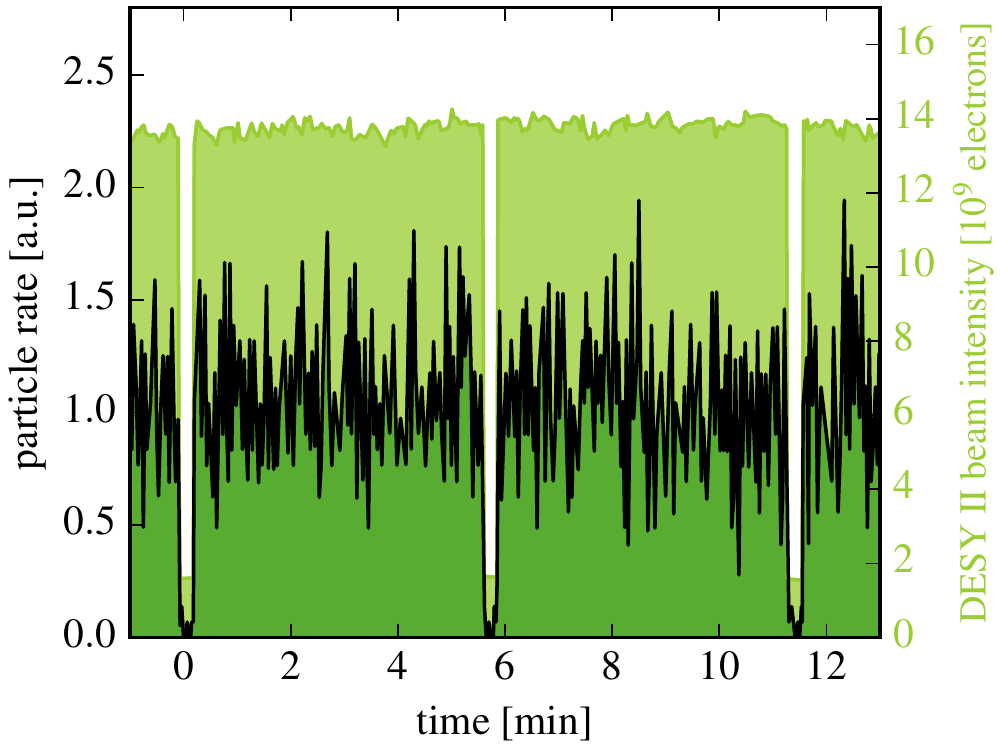}
\caption{(Colour online) 
\petraiii top-up: The stable level of the \desyii intensity in units of 10$^9$ electrons (bright green, right axis) is interrupted every $\sim$ \SI{5}{min} here.
Accordingly the test beam particle rate drops to zero for few tens of seconds (dark green, left axis).
}
\label{fig:performance:p3topup}
\end{center}
\end{figure}

Every few minutes the \desyii synchrotron provides beam to the \petraiii storage ring if the beam lines at \petraiii are operated. 
During that \petraiii top-up, the beam intensity inside \desyii and therefore the particle rate of the test beam drop significantly for few seconds (Fig.~\ref{fig:performance:p3topup}).
The exact timing of this cycle depends on the running mode of \petraiii.

The particle rate shown in Figure~\ref{fig:performance:p3topup} was set to a corresponding magnet current for \SI{5.6}{GeV/c} test beam particles and was measured by the test beam monitor (see Sec.~\ref{sec:beaminstr:rate:beammonitor}),
the \desyii beam intensity using the \desyii beam monitoring system (see Sec.~\ref{sec:beaminstr:tine}).

\subsubsection{\desyii Cycle}\label{sec:performance:d2cycle}

The \desyii synchrotron continuously cycles its magnets using a sinusoidal curve with \SI{12.5}{\Hz} (see Sec.~\ref{sec:desy2} and Fig.~\ref{fig:desy2:cycle}),
which affects the particle momentum and the timing structure of the test beam.
Only for times when the \desyii energy is higher than the selected particle momentum, 
test beam particles with the momentum selected by the users can reach the area. Hence the actual \desyii energy is the upper 
cycle-dependent limit for the momentum of the test beam particles available for the users.

\begin{figure}[htbp]
\begin{subfigure}[t]{0.49\textwidth}
    \includegraphics[width=1.0\textwidth]{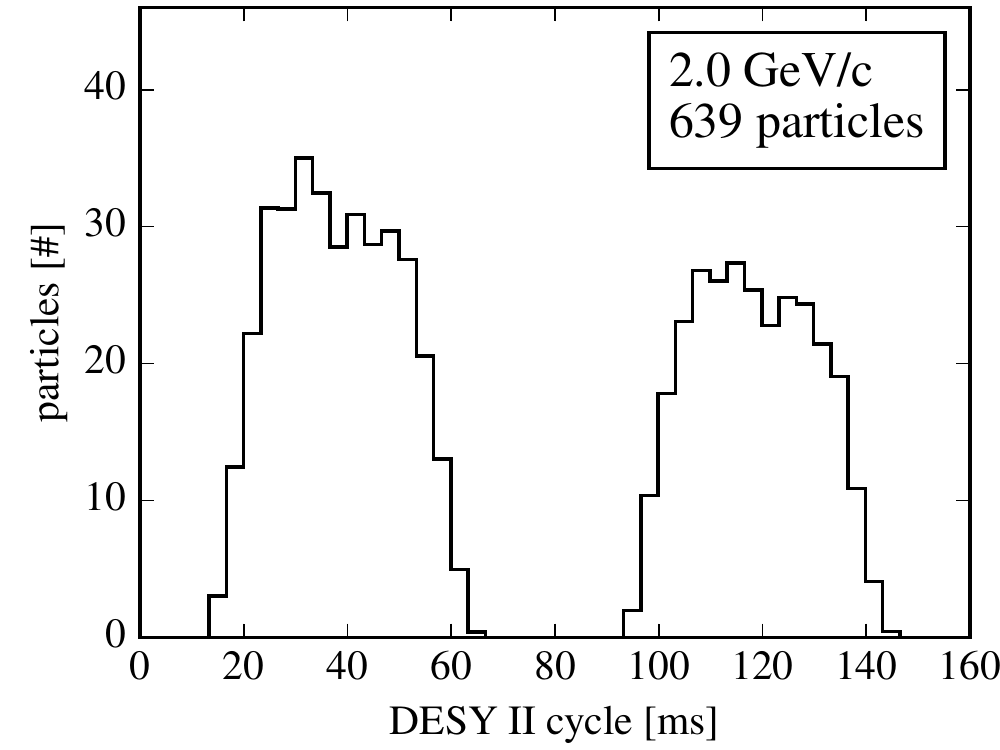}
    \subcaption{at \SI{2}{GeV/c}}
    \label{fig:performance:desy2energy:hist2gev}
\end{subfigure}
\hfill
\begin{subfigure}[t]{0.49\textwidth}
    \includegraphics[width=1.0\textwidth]{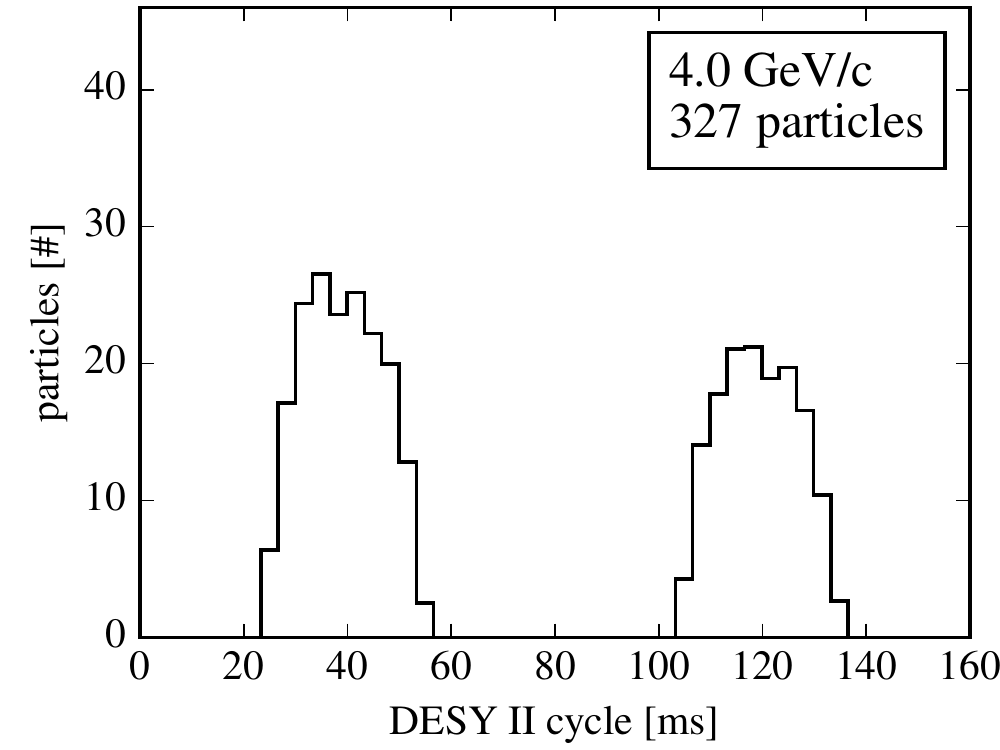}
    \subcaption{at \SI{4}{GeV/c}}
    \label{fig:performance:desy2energy:hist4gev}
\end{subfigure}
\begin{subfigure}[t]{0.49\textwidth}
    \includegraphics[width=1.0\textwidth]{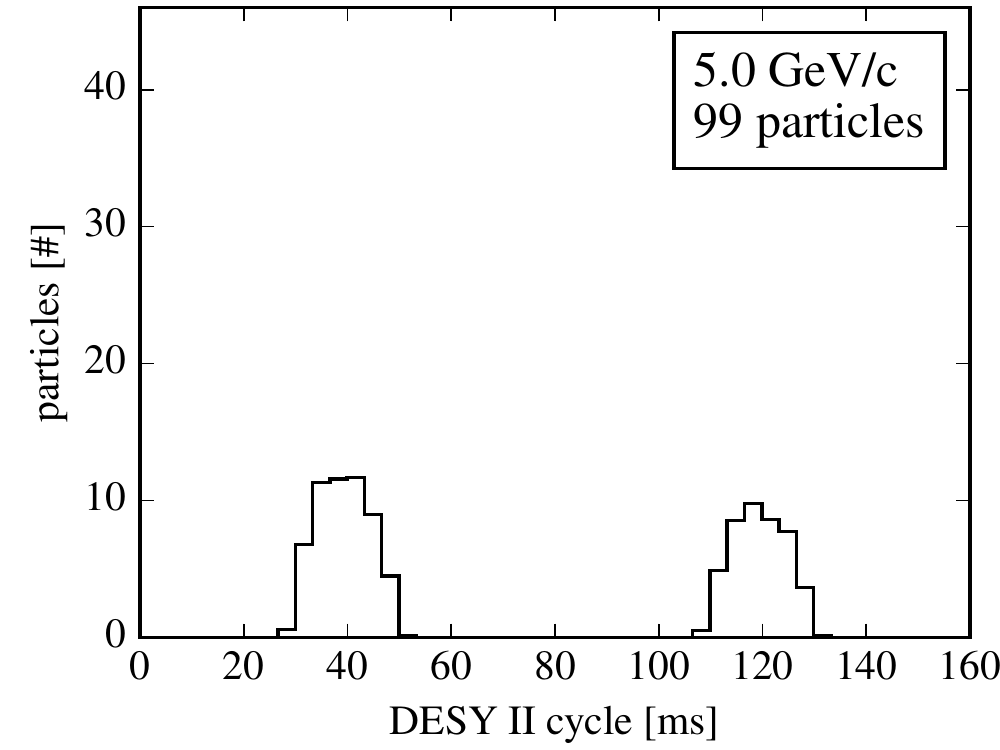}
    \subcaption{at \SI{5}{GeV/c}}
    \label{fig:performance:desy2energy:hist5gev}
\end{subfigure}
\hfill
\begin{subfigure}[t]{0.49\textwidth}
    \includegraphics[width=1.0\textwidth]{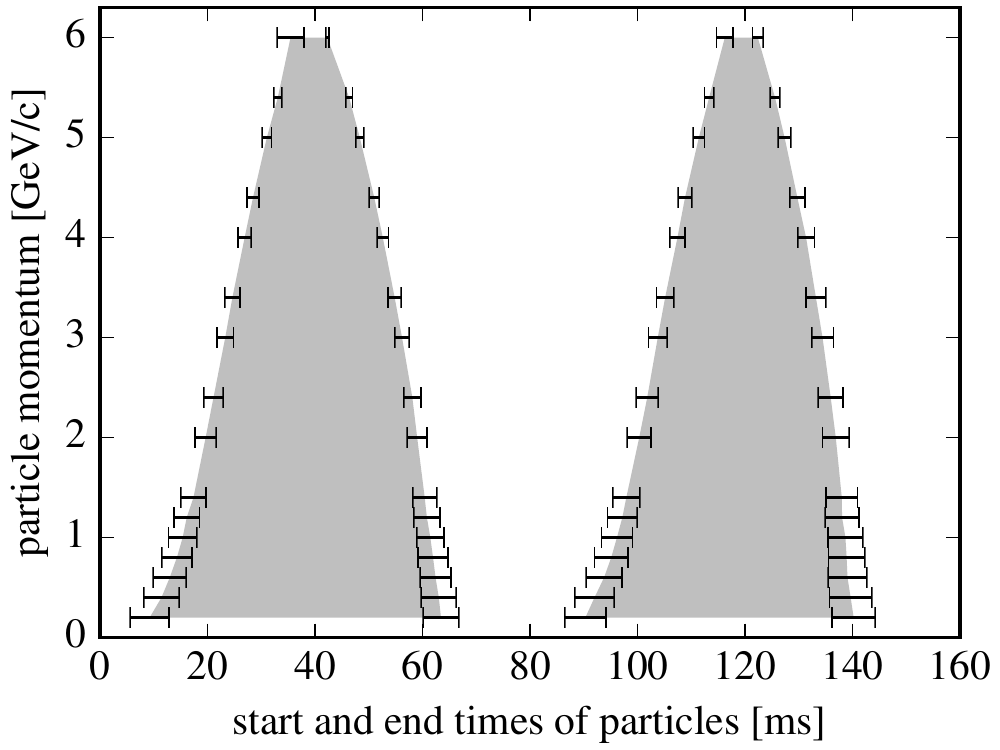}
    \subcaption{momentum vs. time}
    \label{fig:performance:desy2energy:times}
\end{subfigure}
\caption{
Time structure and energy dependence of test beam particles within a \desyii cycle.
(\subref{fig:performance:desy2energy:hist2gev}) to (\subref{fig:performance:desy2energy:hist5gev}): 
Exemplary test beam particle distributions within a \SI{160}{\milli\s} \desyii cycle at selected particle momenta of 2, 4 and \SI{5}{\GeV}.
(\subref{fig:performance:desy2energy:times}):
Momentum dependence of start and end times of test beam particles within a \SI{160}{\milli\s} \desyii cycle. 
    }
\label{fig:performance:desy2energy}
\end{figure}

Thus, within one \desyii cycle or two magnet cycles (\SI{160}{\milli\s}), 
there are dedicated start and end times: 
Within these periods test beam particles are available (Fig.~\ref{fig:performance:desy2energy:hist2gev} to \subref{fig:performance:desy2energy:hist5gev}) 
and the period is the shorter the higher the selected particle momentum (Fig.~\ref{fig:performance:desy2energy:times}).
In addition, due to beam losses in \desyii during traversing $E_{\rm min}$ at \SI{80}{ms} the particle occurrence is lower 
within the second magnet cycle than within the first (see Fig.~\ref{fig:desy2:cycle}).

The measurements were performed by using the \desyii injection signal as the trigger input.
For each selected momentum, multiple \SI{160}{ms} time series were recorded with an oscilloscope and the distributions were extracted by using the internal histogram function.
The total particle numbers given in Fig.~\ref{fig:performance:desy2energy:hist2gev} to \subref{fig:performance:desy2energy:hist5gev}
were normalized to the absolute effective rate at \SI{2}{GeV/c} which was measured at the same time (see Fig.~\ref{subfig:rate_momentum_T24}).
The start and end times as well as their uncertainty were extracted by calculating the weighted mean and RMS of the first or last \SI{5}{\percent} 
of the total events of one single measurement.


\subsubsection{\desyii Bunch Cycle}\label{sec:performance:d2bunch}

\begin{figure} [htbp]
\begin{subfigure}[t]{0.49\textwidth}
    \includegraphics[width=1.0\textwidth]{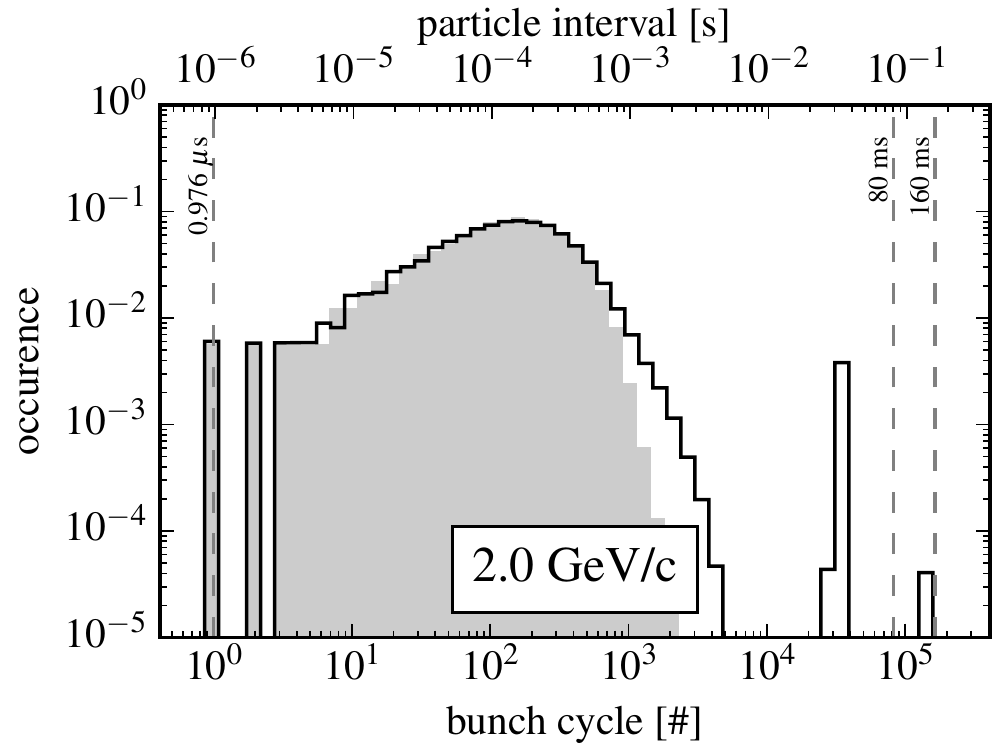}
    \subcaption{at \SI{2}{\GeV/c}} 
    \label{fig:performance:desy2bunch:dist2gev}
\end{subfigure}
\hfill
\begin{subfigure}[t]{0.49\textwidth}
    \includegraphics[width=1.0\textwidth]{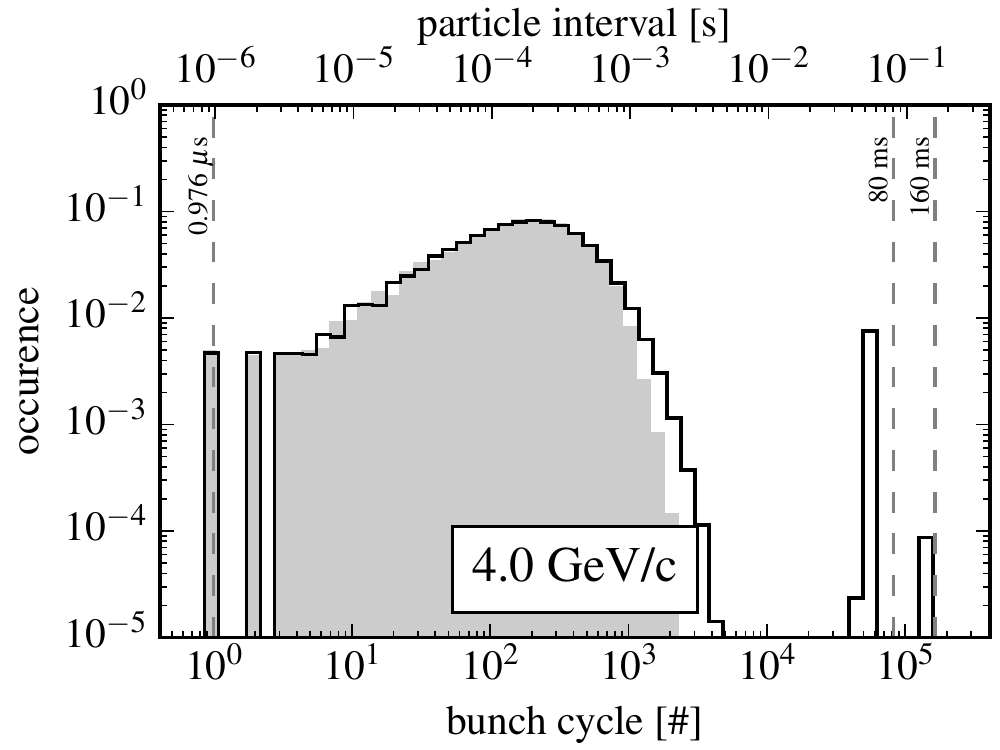}
    \subcaption{at \SI{4}{\GeV/c}}
    \label{fig:performance:desy2bunch:dist4gev}
\end{subfigure}
\begin{subfigure}[t]{0.49\textwidth}
    \includegraphics[width=1.0\textwidth]{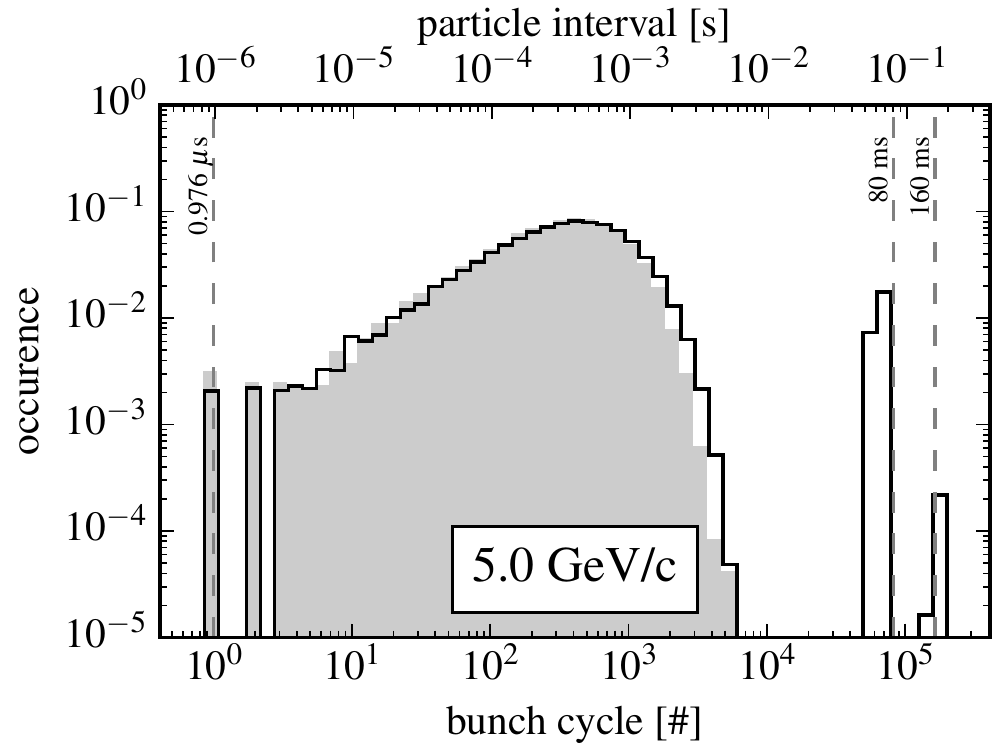}
    \subcaption{at \SI{5}{\GeV/c}}
    \label{fig:performance:desy2bunch:dist5gev}
\end{subfigure}
\hfill
\begin{subfigure}[t]{0.49\textwidth}
    \includegraphics[width=1.0\textwidth]{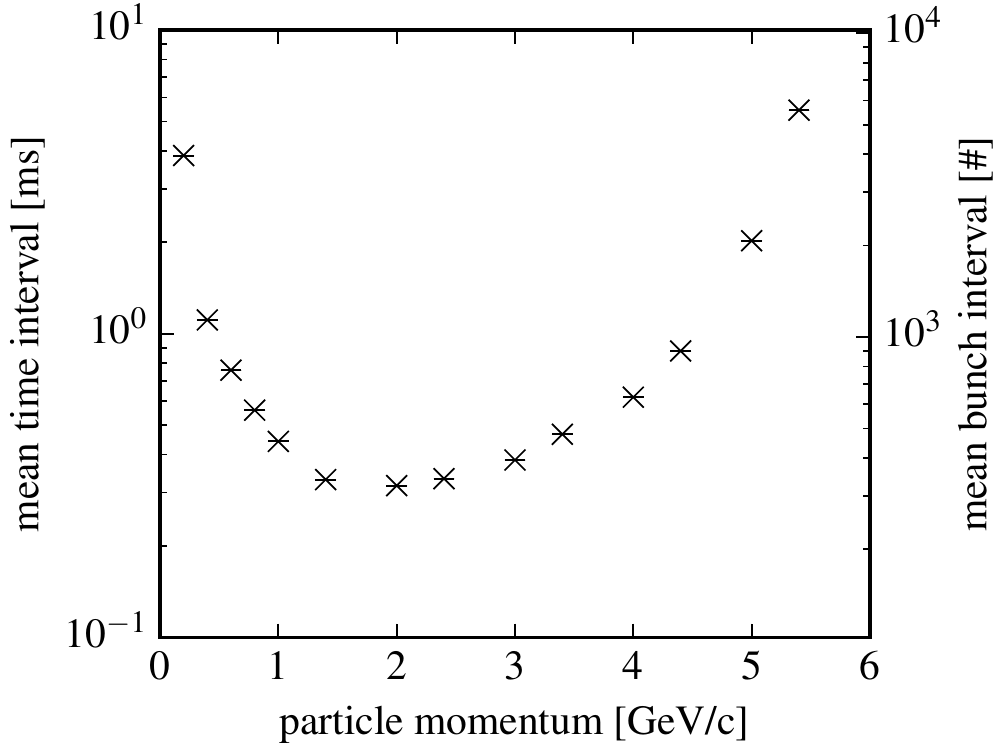}
    \subcaption{mean interval vs. momentum}
    \label{fig:performance:desy2bunch:interval_mean}
\end{subfigure}
\caption{Test beam particle occurrence: 
(\subref{fig:performance:desy2bunch:dist2gev}) to (\subref{fig:performance:desy2bunch:dist5gev}):
    The measured distributions (solid line) for different particle momenta show a similar broad structure  
    peaking around 50 \desyii bunch cycles.
    The distribution below the maximum is well described by a model based on Poisson statistics expressing the probability for multiple test beam particles per bunch (grayish area).
(\subref{fig:performance:desy2bunch:interval_mean}):
The mean interval shows a minimum at \SI{2}{\GeV/c} and increases with lower and higher momenta, 
    which is in accordance with the effective rate measurements in Fig.~\ref{fig:performance:rate_momentum}.
}
\label{fig:performance:desy2bunch}
\end{figure}

A \desyii bunch hits the primary target every $L_{\rm DESY\, II} / c = \SI{0.976}{\micro\s}$.
This time structure can be resolved by measuring the time or bunch intervals between consecutive test beam particles:
The distributions (Figs.~\ref{fig:performance:desy2bunch:dist2gev} to \subref{fig:performance:desy2bunch:dist5gev}) show that an interval in the order of \SI{0.1}{\milli\second} or \SI{100}{bunch cycles} is most likely with nearly \SI{10}{\percent} occurrence,
and that the minimum occurs for one \desyii bunch cycle with below \SI{1}{\percent} occurrence.
By calculating the distribution's mean of the particle interval, the dependence on the particle momentum is in accordance to the effective rate measurements (see Sec.~\ref{sec:performance:rate_momentum}).
For example, test beam particles at \SI{2}{\GeV/c} are measured every \SI{0.3}{\milli\second} or every \SI{300}{bunch cycles} on average (Fig.~\ref{fig:performance:desy2bunch:interval_mean}).

In addition, the distributions in Figures~\ref{fig:performance:desy2bunch:dist2gev} to \subref{fig:performance:desy2bunch:dist5gev} resolve the \desyii cycle structure at time intervals between 20-\SI{80}{\milli\second} (Sec.~\ref{sec:performance:d2cycle}),
and the \petraiii top-ups below \SI{160}{\milli\second} which happened during this measurement (Sec.~\ref{sec:performance:p3topup}).

From this distribution, the probability of multiple test beam particles per bunch cycle can be directly estimated.
The measured distribution of particle intervals can be well described by
assuming that the probability for particles per bunch underlies a Poisson distribution
$$f(k;\lambda ){\frac {\lambda ^{k}e^{-\lambda }}{k!}}$$
with $k = {\rm bunch\, cycle}$ and $\lambda = \left(max({\rm bunch\, cycle}) < 10^4\right)^{-1}$ 
which is the maximum bin content ignoring the \desyii cycle structure and \petraiii top-ups described in the previous sections. 
For intervals below the maximum the measured distributions are well described by time intervals underlying this model which is depicted by the gray area in Figs.~\ref{fig:performance:desy2bunch:dist2gev} to \subref{fig:performance:desy2bunch:dist5gev}.
As an example the probability for more than one particle per bunch is $\lambda_{\rm model} = (6.08 \pm 0.36) \cdot 10^{-3}$
at \SI{2}{\GeV/c} applying this estimate. 

Measurements were performed by using the beam monitor system as the trigger signal and using this as the trigger input for the AIDA TLU (trigger logic unit).
The time resolution of the measurement setup is in the order of the bunch length of $f_{\rm HF\, cavities}^{-1} \approx$ \SI{2}{\nano\second}.
Thus, a bunch structure cannot be resolved by this measurement.
Multiple measurements were performed by selecting different particle momenta and each measurement includes approx. \SI{300}{k trigger events}.
The time differences of consecutive trigger signals representing consecutive particles were determined by the corresponding timestamps of the \SI{160}{MHz} clock of the TLU, 
and were used for calculating the distributions. 
The total counts of each distribution were normalized to one.

\subsection{Rate Dependencies}\label{sec:performance:rate}

The test beam rates measured in the individual  beam areas depend on the \desyii beam intensity (Sec.~\ref{sec:performance:d2intensity}) and 
on the setting of each  beam line components like 
\begin{itemize}
    \item the magnet current and therefore the selected momentum (Sec.~\ref{sec:performance:rate_momentum}),
    \item the primary target position (Sec.~\ref{sec:performance:target1}), 
    \item the material and thickness of the secondary target (Sec.~\ref{sec:performance:target2}), or
    \item the opening of the primary collimator (Sec.~\ref{sec:performance:primarycollimator}).
\end{itemize}
They are equally dependent on the measurement setup including active areas, thresholds or on the position in the beam area.
Thus, relative rates are primarily presented in this section, in order to demonstrate the influence of each component on the particle rate.

Since the rate strongly depends on the \desyii timing structure (Sec.~\ref{sec:performance:timing}) and therefore on the time window of the measurement, 
additionally all of the following results are based on effective rate measurements integrated over \SI{10}{\second} excluding periods with \petraiii top-up operations 
unless indicated otherwise.
By counting the coincidence signal of two scintillators the rate is calculated and 
normalized to the corresponding value of the \desyii beam intensity in order to take into account smaller fluctuations.
For each individual setting, the effective rate was measured three times in order to derive the mean and the RMS.

\subsubsection{Rate Dependence on the \desyii Beam Intensity}\label{sec:performance:d2intensity}

\begin{figure}[htbp]
\begin{center}
    \includegraphics[width=0.5\textwidth]{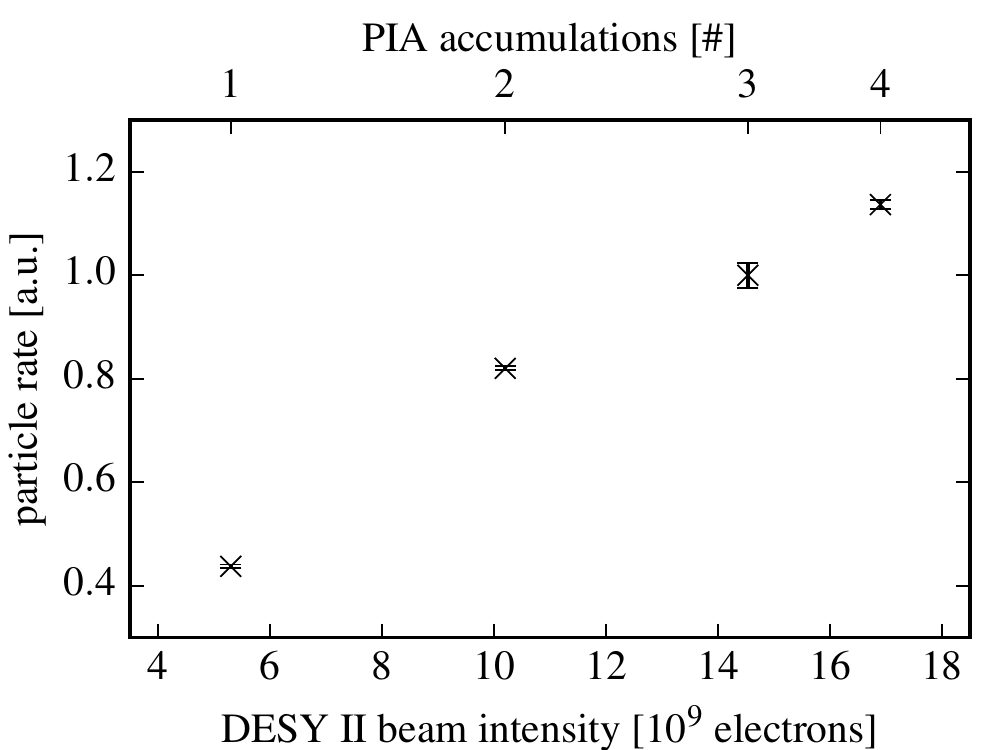}

\caption{
The test beam particle rate against the \desyii beam intensity: 
The rate strongly depends on the \desyii beam intensity.
}
    \label{fig:performance:d2beamintensity}
\end{center}
\end{figure}

As already indicated in Section~\ref{sec:performance:p3topup}, the test beam rate strongly depends on the \desyii beam intensity.
Figure~\ref{fig:performance:d2beamintensity} shows that the dependence is nearly linear.

In this measurement, the bunch size, and therefore the \desyii beam intensity, was varied by the different number of bunch accumulations in \pia before the injection to \desyii (see Sec.~\ref{sec:desy2}).
The test beam rate was measured by using the first two telescope scintillators in coincidence in TB21 and the \desyii beam intensity was measured by the intensity monitors.
The rate was normalized to the corresponding \desyii beam intensity value of $14.53 \times 10^{9}$~electrons or three accumulations in \pia for this measurement.
Typically, three \pia accumulations are used for routine operation.

\subsubsection{Rate Dependence on the Particle Momentum}\label{sec:performance:rate_momentum}

\begin{figure}[htbp]
\centering
\begin{subfigure}[b]{0.49\textwidth}
    \includegraphics[width=1.0\textwidth]{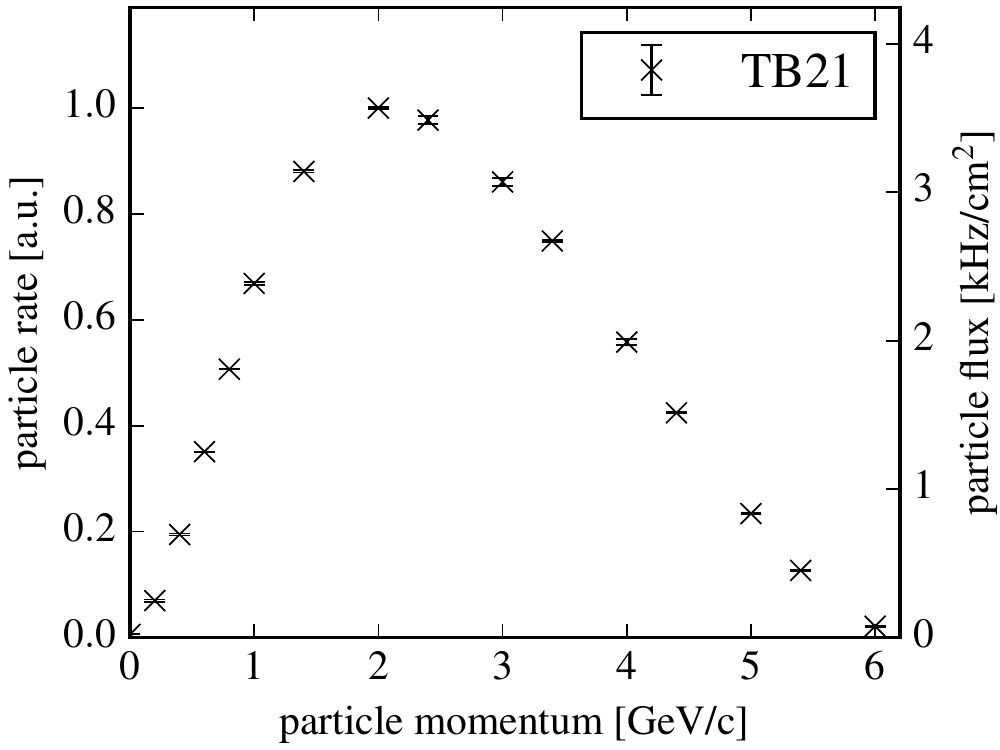}
    \subcaption{at TB21}
    \label{subfig:rate_momentum_T21}
\end{subfigure}    
\begin{subfigure}[b]{0.49\textwidth}
    \includegraphics[width=1.0\textwidth]{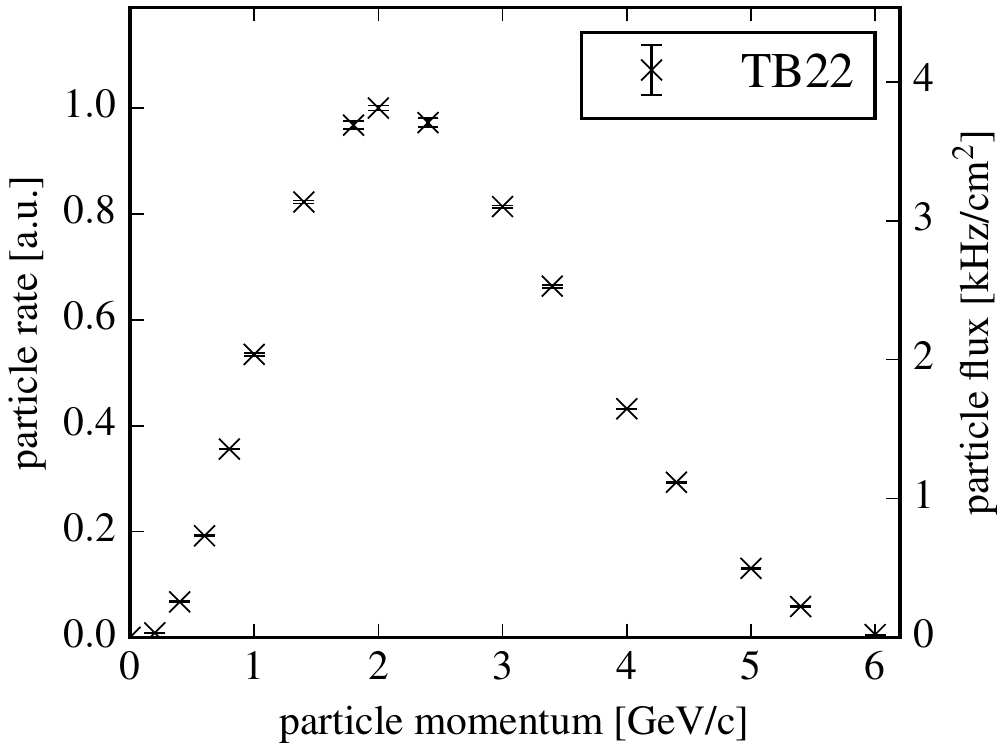}
    \subcaption{at TB22}
    \label{subfig:rate_momentum_T22}
\end{subfigure}
\begin{subfigure}[c]{0.49\textwidth} 
    \centering
    \includegraphics[width=1.0\textwidth]{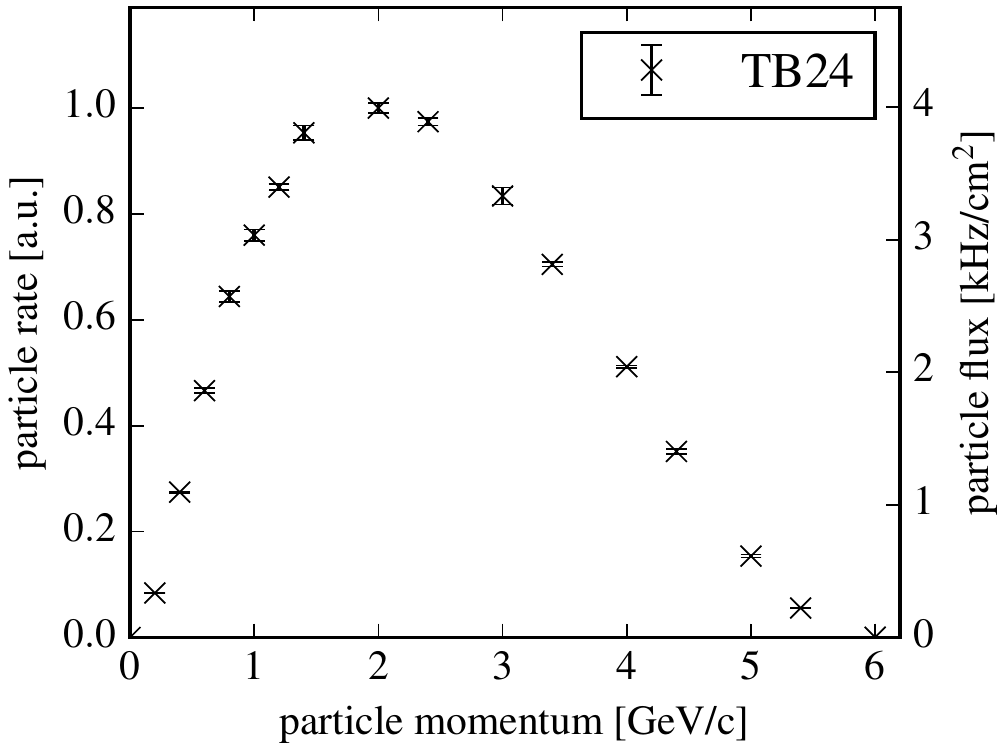}
    \subcaption{at TB24}
    \label{subfig:rate_momentum_T24}
\end{subfigure}
\caption{The test beam particle rate against the selected momentum for all three beam lines. }
\label{fig:performance:rate_momentum}
\end{figure}

The most important result for experimental groups using the test beam is the rate dependence on the particle momentum.
Thus, for this measurement an absolute particle flux ---rate per detection area--- is given on the right $y$-axis (Fig.~\ref{fig:performance:rate_momentum}).
The highest rate is measured for \SI{2}{\GeV/c} at each beam line.
The shape of the curve is mainly caused by the beam line geometry and the bremsstrahlung spectrum after the primary target (see Fig.~\ref{fig:bremsstrahlung_spectrum}):
The rising slope for momenta smaller than \SI{2}{\GeV/c} can be explained by the fixed geometry of the beam line and increasing scattering for low momenta. 
The falling slope for momenta larger than \SI{2}{\GeV/c} is explained by the decreasing bremsstrahlung spectrum for higher momenta with the upper limit of the maximum \desyii energy $E_{\rm max}$.
As already mentioned above, this measurement is directly correlated to the result shown in Fig.~\ref{fig:performance:desy2bunch:interval_mean}.
The momentum spread of each selected momentum is discussed in Sec.~\ref{sec:performance:primarycollimator}.

For these measurements, the selected particle momentum was varied by the corresponding magnet currents (see Tab.~\ref{tab:desy2beamgen:magnets:specs}).
For each measurement the corresponding primary collimator was symmetrically opened in each direction by \SI{5.0}{\mm}. 
The detailed measurement settings for all three beam lines can be found in Tab.~\ref{table:perfromance:rate_settings} and were chosen as similar as possible.

\begin{table}[htbp]
\begin{center}
    \begin{tabular}{c c c c}
    \hline
    \arraybackslash{\textbf{Setting}} & {\textbf{TB21}} & {\textbf{TB22}} & {\textbf{TB24}}\\ 
    \toprule
    Fiber position [mm] & -14.0 / 16.0 & -23.0 / 7.0  & -19.8 / 10.2 \\
    Copper target [mm] & 5.0 &  5.0  &  4.0 \\
    Lead collimator [mm$^2$] & $15 \times 10$  &  $10 \times 10$  &  (not needed) \\
    Scintillator area [mm$^2$] & $20 \times 10$ & $20 \times 10$ & $10 \times 10$ \\
    Scintillator position & DATURA & DURANTA & beam monitor \\
    \bottomrule
    \end{tabular}
\end{center}
\caption{Table of the used settings for the rate dependency on the particle momentum.
    This numbers are given here since an absolute rate normalized on the detector area is given in
    Fig.~\ref{subfig:rate_momentum_T21} to \subref{subfig:rate_momentum_T24}.
    }
\label{table:perfromance:rate_settings}
\end{table}

\subsubsection{Rate Dependence on the Primary Target Position}\label{sec:performance:target1}

As discussed in Sections~\ref{sec:performance:p3topup} and \ref{sec:performance:d2intensity} the beam rate strongly depends 
on the \desyii beam intensity or more precisely on the number of \desyii beam particles hitting the \SI{7}{\micro\meter} 
thick fiber positioned in the beam orbit (see Sec.~\ref{sec:desy2beamgen:primarytarget}).
This is given by the beam profile of each bunch traversing the fiber target. 
By measuring the rate for various positions on the axis being perpendicular to the \desyii beam axis the beam profile can be determined.
By performing such a horizontal position scan the measurement provides a one-dimensional vertically integrated beam profile.
Due to expected fluctuations of the absolute bunch position between different cycles the measurement indicates an effective beam profile 
including all statistical uncertainties for the time of the measurement date.  

\begin{figure}[htbp]
\centering
\begin{subfigure}[b]{0.49\textwidth}
    \includegraphics[width=1.0\textwidth]{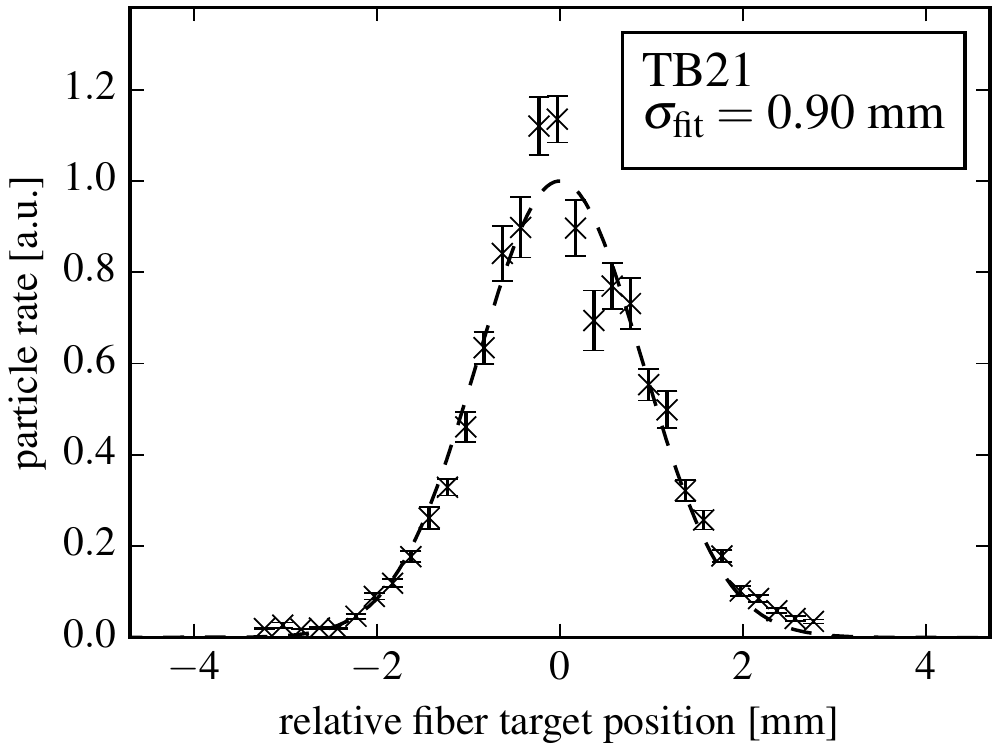}
    \subcaption{at TB21}
    \label{subfig:profileTB21}
\end{subfigure}    
\begin{subfigure}[b]{0.49\textwidth}
    \includegraphics[width=1.0\textwidth]{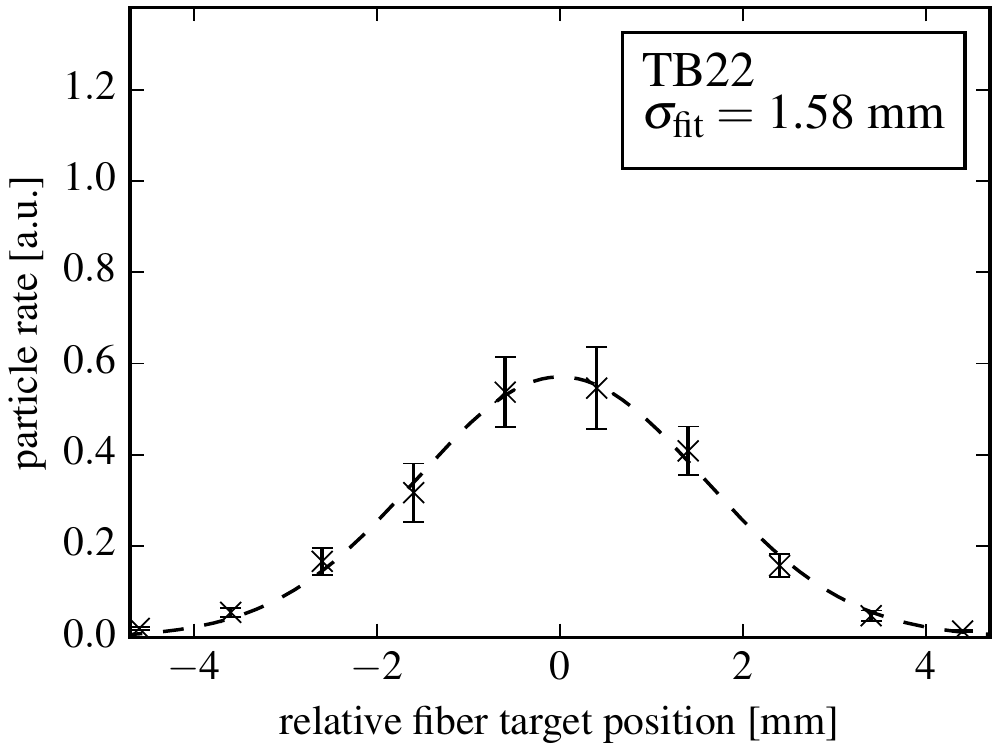} 
    \subcaption{at TB22}
    \label{subfig:profileTB22}
\end{subfigure}
\begin{subfigure}[c]{0.49\textwidth} 
    \centering
    \includegraphics[width=1.0\textwidth]{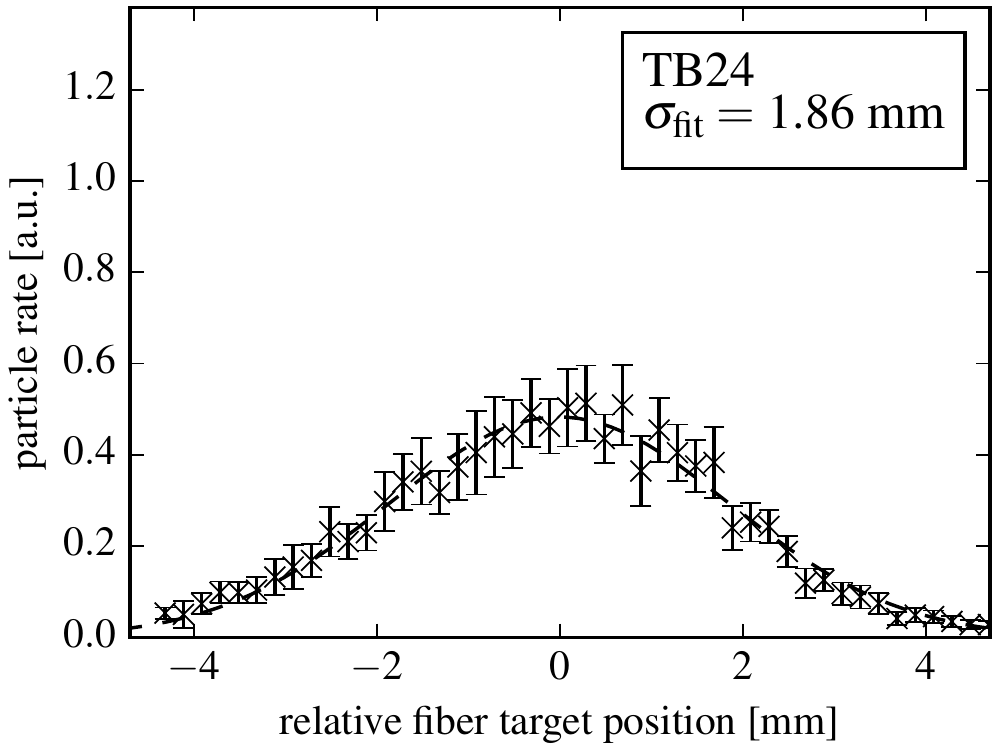}
    \subcaption{at TB24}
    \label{subfig:profileTB24}
\end{subfigure}
    \caption{
The rate dependence on the fiber target position.
This corresponds to an effective, vertically integrated profile of the \desyii beam at the fiber position of the three beam lines.
    }
\label{fig:performance:tbprofiles}
\end{figure}

Figure~\ref{fig:performance:tbprofiles} shows the effective vertically integrated beam profiles at the primary target positions of TB21, TB22 and TB24.
The particle rate was measured by using the beam monitor data from the machine control system (Sec.~\ref{sec:beaminstr:rate:beammonitor} and \ref{sec:beaminstr:tine}) and a test beam magnet current corresponding \SI{3}{\GeV/c}.

The effective horizontal beam width at these positions is defined as two times of the Gaussian standard deviation after performing a fit to the data. 
Statistical uncertainties are conservatively estimated to be $\pm$\SI{0.1}{\mm} since the RMS shows the same value in this interval which indicates a normal distribution.
For this measurement, this results in the following effective horizontal beam widths of \desyii at the corresponding primary target positions:
\begin{itemize}
    \item for TB21: \SI[separate-uncertainty = true]{1.80(10)}{\mm}
    \item for TB22: \SI[separate-uncertainty = true]{3.16(10)}{\mm}
    \item for TB24: \SI[separate-uncertainty = true]{3.72(10)}{\mm}
\end{itemize}
In addition, each profile is normalized by the total integral of the TB21 profile which indicates the relative rate differences by the fiber target position of the individual beam lines.
The difference of the profile shape or its Gaussian width is explained by the shape of the \desyii electron bunch at the locations of the fiber targets:
The bunch is broadened in the horizontal direction after the fiber position of TB21.

\subsubsection{Rate Dependence on the Secondary Target}\label{sec:performance:target2}

The conversion from GeV photons to leptons is dominated by the electron-positron pair production in the nuclear field.
The conversion probability depends on the material which can be scaled by the radiation length $X_0$ \cite[e.g.]{Tavernier2010}.
Thus, the beam rate is higher for copper targets than for aluminum ones (Fig.~\ref{fig:performance:rate_secondary_target} and see Sec.~\ref{sec:simulations:target2}).

Without any target inside the beam line the test beam particle are generated 
by a \SI{0.5}{\mm} thick aluminum window which is installed right behind the primary target at the \desyii beam line.
This explains the non-zero rate with no secondary target inside the beam axis (Fig.~\ref{fig:performance:rate_secondary_target}).
The saturation is mainly explained by scattering processes which increase the radiation angle with thicker material.
For this measurement, the secondary target was varied by the possible selections for TB22 (see Tab.~\ref{tab:desy2beamgen:secondarytargets:specs}).


\subsubsection{Rate Dependence on the Primary Collimator}\label{sec:performance:primarycollimator}

\begin{figure}[htbp]
\begin{subfigure}[t]{0.49\textwidth}
    \includegraphics[width=1.0\textwidth]{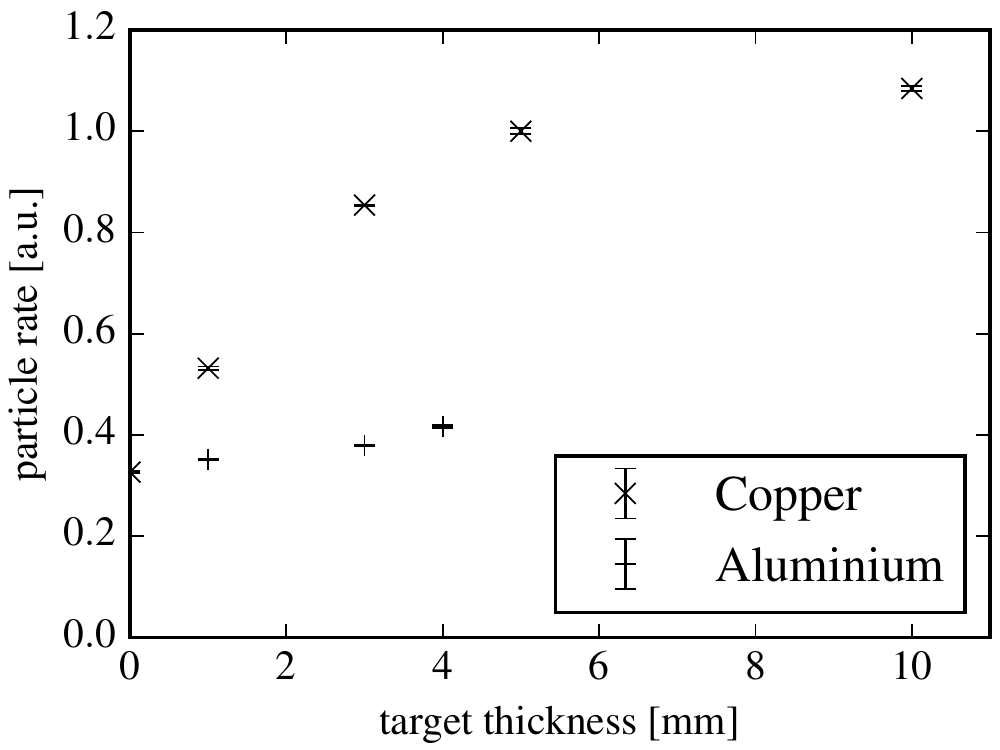}
    \subcaption{Secondary Target}
    \label{fig:performance:rate_secondary_target}
\end{subfigure}
\hfill
\begin{subfigure}[t]{0.49\textwidth}
    \includegraphics[width=1.0\textwidth]{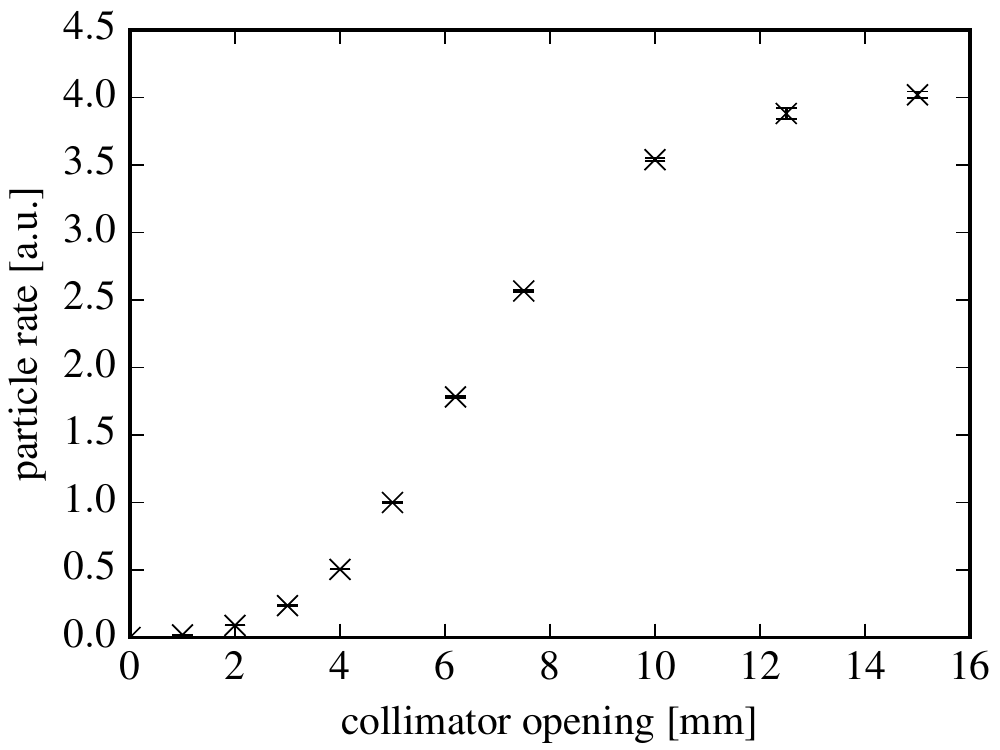}
    \subcaption{Primary Collimator}
    \label{fig:performance:rate_collimator}
\end{subfigure}
\label{fig:performance:ratedep}
    \caption{The rate dependence on the secondary target (\subref{fig:performance:rate_secondary_target}) and 
    on a symmetrically opening in each direction of the primary collimator (\subref{fig:performance:rate_collimator}).
    Both measurements were performed in TB22 at a selected momentum of \SI{2}{\GeV/c}.
    }
\end{figure}

The primary collimator (see Sec.~\ref{sec:desy2beamgen:primarycoll}) is the first beam line component 
shaping the spatial test beam profile behind the magnet for the momentum selection.
The wider the opening in one direction is, the higher the rate, 
and the saturation at a wide opening indicates the shape of the test beam profile (Fig.~\ref{fig:performance:rate_collimator}).  
For this measurement, the opening of the first collimator at TB22 was symmetrically varied in
all four directions from 0 to \SI{15}{\mm} (Sec.~\ref{sec:desy2beamgen:primarycoll}).


\subsection{Particle Momentum}
\label{sec:performance:momentum}
The particle momentum is set by the user via the corresponding current in the 
test beam magnet (Sec.~\ref{sec:desy2beamgen:magnets}). Detailed knowledge 
of the particle momentum and spread is essential for the analysis of the test beam data. 
In this section measurements of the particle momentum without using the information 
from the test beam magnet are shown.

\subsubsection{Calibration}
\label{sec:performance:momentum:calibration}
The verification of the momentum calibration for TB21 was performed both in 2013 and in 2016 \cite{schuetze2013, lange2016}. 
Measurements were performed to investigate the momentum spread and its spatial dependence.

For this measurement, the Big Red Magnet in TB21 (Sec.~\ref{sec:addinf:brm}) was used as a spectrometer magnet, deflecting the particles dependent on their momentum. 
The DATURA beam telescope (Sec.~\ref{sec:addinf:telescopes}) was split into two sets of three sensor planes each, one measuring the particle angle in front of and one behind the magnet. 
With this technique, the deflection angle for each particle could be measured. 
From the measured deflection angles of multiple particles, the mean and the width of the angular distribution is extracted.

To reconstruct the momentum, a simulation of the propagation of the positrons inside the magnet was developed, based on the field map of the Big Red Magnet (Fig.~\ref{fig:addinf:brmfieldmap}).
Hence, for any measured angular distribution, the momentum input of the simulation is varied such that the simulated deflection angle matches the mean of the measured distribution.
Similarly, the momentum spread is derived from the width of the deflection angle distribution.

Due to the good resolution of the MIMOSA26 sensors (Sec.~\ref{sec:addinf:telescopes}), the uncertainty of the angle 
measurement of about \SI{0.023}{\milli\radian} is lower than the nominal angular width induced by multiple Coulomb 
scattering in air of about \SI{0.37}{\milli\radian} for $p=\SI{3}{\GeV/c}$. 
The width measured for a non-deflected beam, representing the multiple scattering and measurement uncertainties, 
is quadratically subtracted from the measured widths for deflected beams.

\begin{figure}[htb!]
\begin{center}
\begin{subfigure}[b]{0.49\textwidth}
\includegraphics[width=\textwidth]{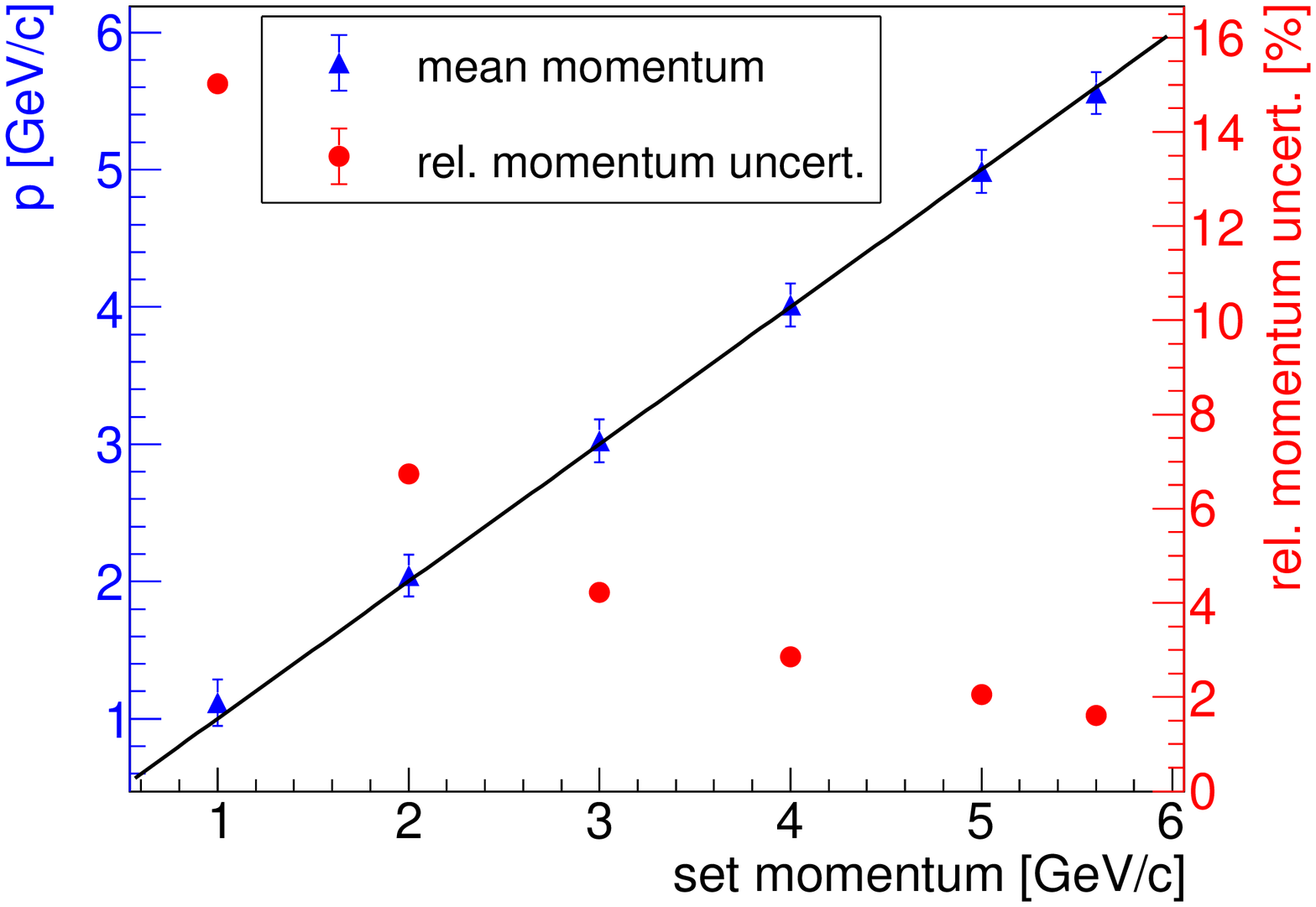}
\subcaption{Calibration}
\label{fig:performance:momentumMeasurement:Spread}
\end{subfigure}
\begin{subfigure}[b]{0.49\textwidth}
\includegraphics[width=\textwidth]{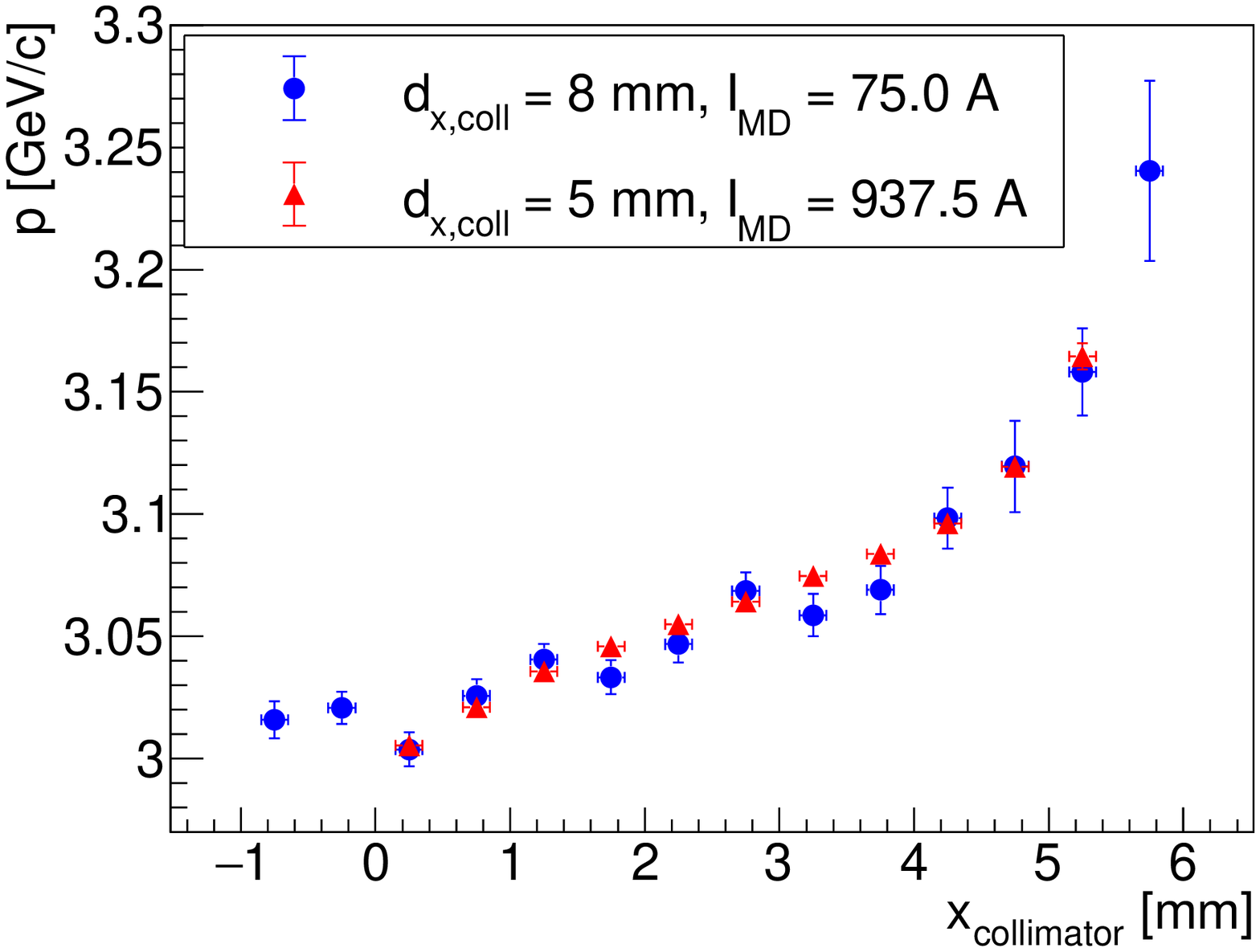}
\subcaption{Spread}
\label{fig:performance:momentumMeasurement:Collimator}
\end{subfigure}
\caption{Particle momenta: Measuring the deflection angle of the particles traversing the dipole magnet in TB21, 
the momentum distribution of the particle beam is derived. The mean particle momentum is shown in blue 
(\subref{fig:performance:momentumMeasurement:Spread}), with the black line marking the expectation values. The momentum spread is shown in absolute values 
as error bars and in relative values as red data points. The mean momentum shows a dependence on the transverse 
particle position, as shown for measurements at two  different openings of the secondary collimator and different 
dipole currents (\subref{fig:performance:momentumMeasurement:Collimator}). For this, the particle position was extrapolated to the position of the secondary collimator. 
Here the error bars represent the statistical uncertainty on the mean measured momentum.}
\label{fig:performance:momentumMeasurement}
\end{center}
\end{figure}

Figure \ref{fig:performance:momentumMeasurement:Spread} shows the results of the momentum calibration, with the blue markers showing the 
measured momentum of the particle beam and its momentum spread. 
The black line represents the set values for the momentum according to Table~\ref{tab:desy2beamgen:magnets:specs}. 
It is clearly visible, that the measured momentum distributions fit well to the nominal values. 

\subsubsection{Spread}
\label{sec:performance:momentum:scattering}
For many measurements a detailed knowledge of the momentum spread is a key ingredient to achieve the best possible track resolution.
The absolute momentum spread is found to be constant at \SI[separate-uncertainty = true]{158(6)}{\MeV/c} over the 
full momentum range for horizontal collimator openings of $\pm$\SI{3.2}{\mm} (primary) and \SI{8}{\mm} (secondary). 
Hence, the relative momentum spread decreases for larger momenta as shown in Figure~\ref{fig:performance:momentumMeasurement:Spread}. 

Since the particle momentum is selected by the test beam magnets with a vertical field (Sec.~\ref{sec:desy2beamgen:magnets}),
a dependence of the particle momentum on the horizontal position inside the beam is also expected.
In order to understand the horizontal momentum distribution, 
the momentum spread can be spatially resolved by including the particle position determined by the beam telescope (Sec.~\ref{sec:addinf:telescopes}).
Especially for larger devices this can be necessary to apply the appropriate corrections. 
This measurement is performed by extrapolating the measured track in the upstream telescope arm to the position of the secondary collimator and then combining the deflection angle and the particle position.
Evaluating the mean particle momentum as described above, 
the expected behavior of the momentum as a function of the position can be extracted, as shown in Figure~\ref{fig:performance:momentumMeasurement:Collimator}. 
The two measurements show the dependency for two different openings of the secondary collimator $d_{x, {\rm coll}}$ at different currents of the Big Red Magnet $I_{\rm MD}$ and demonstrate its reproducibility.

\newpage

\section{Simulations}\label{sec:simulations}

To further understand and optimize the beam lines of the \diitbf, a simulation package has been developed which 
simulates the entire test beam line from the beam generation in the \desyii synchrotron to the test beam area.
The simulation is based on the \slic simulation package~\cite{Graf:2006ei}, which is a \geant ~\cite{Agostinelli:2002hh,Allison:2006ve,Allison:2016lfl} 
simulation toolkit using the LCIO file format~\cite{Gaede:2003ip}.
Although the simulation package can be adapted to simulate all three test beam lines, so far only TB21 has been simulated in detail.

In the following, the simulation results for the bremsstrahlung photon generation in the primary target, the pair production 
in the secondary target, and the final momentum selection and collimation of the test beam are shown. 
The details on the simulation tool as well as on the full simulation study of TB21 can be found in \cite{Schutz:2015iya}.

\subsection{Test Beam Generation in the Primary Target}
\begin{figure}[htbp]
  \centering
    \includegraphics[width=0.7\textwidth]{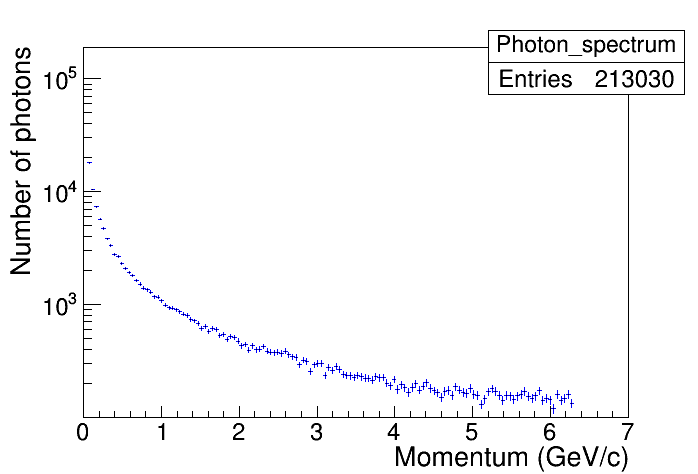}
  \caption[Simulated bremsstrahlung spectrum emitted from the primary target.]{(Colour online) 
  The \desyii beam bunch of 10$^{10}$ electrons with an energy of \SI{6.3}{\GeV} generates in total about \num{2e5} bremsstrahlung photons by hitting the primary target. 
  The bremsstrahlung spectrum reaches about \SI{6.3}{\GeV/c}, which is the maximum momentum of the photons, as they cannot exceed the maximum momentum of the respective electrons.}
  \label{fig:bremsstrahlung_spectrum}
\end{figure}

The first simulation step is the \desyii beam bunch consisting of 10$^{10}$ electrons or positrons hitting a primary target placed directly in the \desyii beam pipe. 
When the simulated \desyii beam bunch hits the carbon fiber target, bremsstrahlung photons are emitted.
Their momentum distribution, which follows the characteristic $\frac{1}{E}$ dependency of bremsstrahlung spectra, is shown in Figure~\ref{fig:bremsstrahlung_spectrum}.
The generated photons exit the beam pipe tangentially to \desyii.
Further test beam line components are placed alongside the beam path of the bremsstrahlung photons.
It is important to note that the \desyii beam energy was fixed to \SI{6.3}{\GeV} and 
does not account for the sinusoidal \desyii magnet cycle (see Sec.~\ref{fig:desy2:cycle}).

\subsection{Secondary Target}\label{sec:simulations:target2}
In the secondary target, the bremsstrahlung photons are converted to electron-positron pairs.  
As the cross section for pair production is dependent on the material and the target thickness~\cite{Tavernier2010}, the rate of the test beam depends on the choice of the secondary target.
The test beam users can choose from a set of different converter plates consisting of aluminum and copper of various thicknesses. 
The most commonly used converter target is a copper plate with \SI{5}{\milli\meter} thickness, as it yields the highest particle rate.
The test beam simulation presented in the following sections assumes this target choice.

\begin{figure}[htbp]
  \centering
  \includegraphics[width=0.7\textwidth]{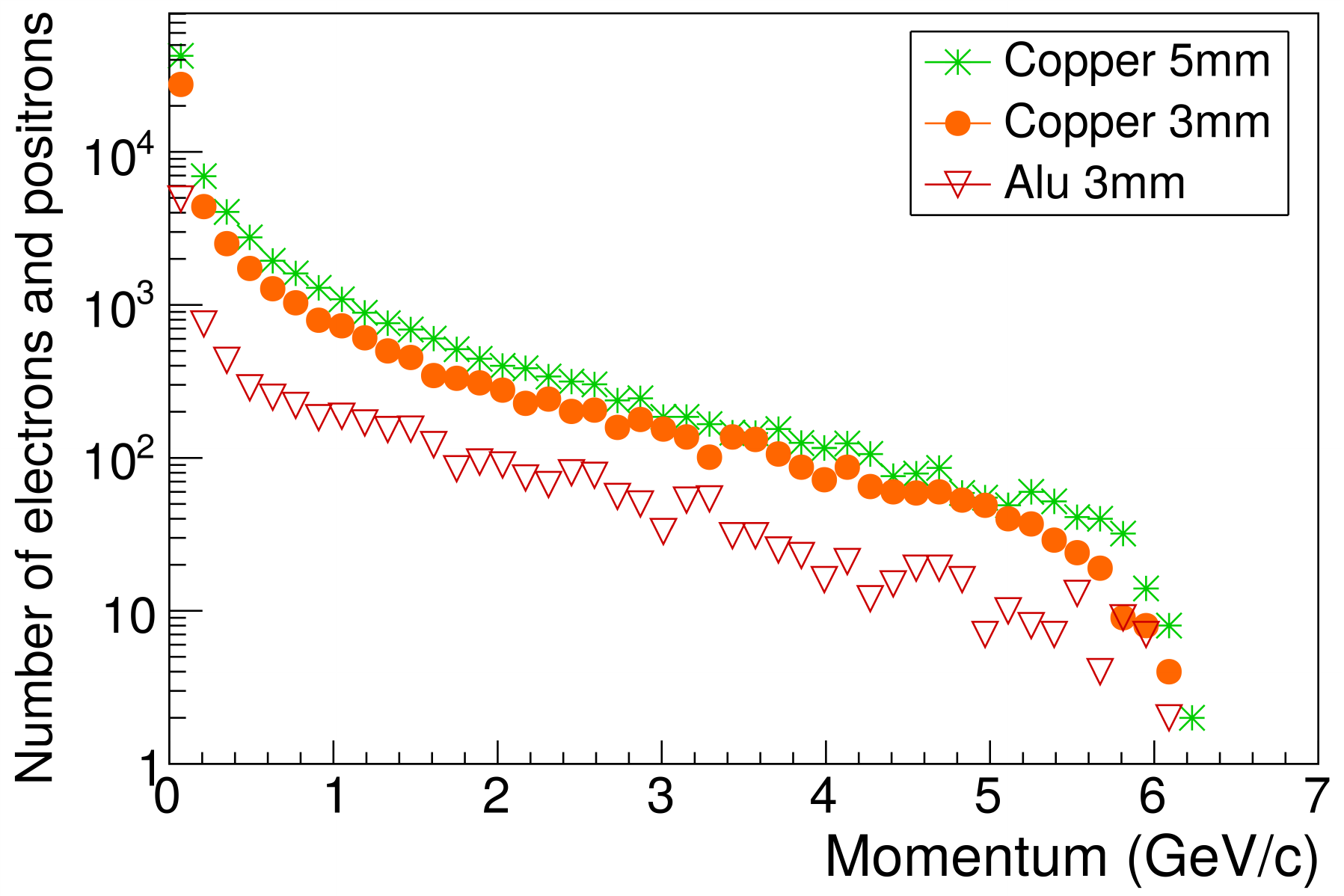}
  \caption[Momentum distribution of electrons and positrons after photon conversion in different secondary targets of TB21.]{(Colour online) 
  The histogram shows the momentum distribution of electrons and positrons behind the secondary target of TB21. 
  With the comparison between the different secondary targets in respect to their material and thickness, it is a direct comparison of their photon conversion efficiencies.}
  \label{fig:converter_rates}
\end{figure}

Figure~\ref{fig:converter_rates} shows the comparison between the numbers of electrons and positrons simulated after the conversion of \num{2e5} photons in the different converters. 
Copper targets yield higher pair production rates than the aluminum targets, and the thicker the material, 
the more photons are converted into electron-positron pairs (see also Sec.~\ref{sec:performance:target2}).

\subsection{Test Beam Energy Selection and Collimation}
The electron-positron pairs (converted from the bremsstrahlung photons in the secondary target) enter the test beam magnet through an evacuated beam pipe, 
about \SI{60}{\centi\meter} behind the converter plate.
In the magnetic field of the dipole magnet, the electrons and positrons are separated by charge and momentum such that the beam is spread into a particle fan in the $xz$-plane.
Neutral particles, such as unconverted photons, are not deflected in the magnetic field and stay on their initial path (in the case that they are not scattered or stopped).
They leave the test beam magnet centrally through the exit beam pipe.

\begin{figure}[htbp]
  \centering
  \includegraphics[width=0.9\textwidth]{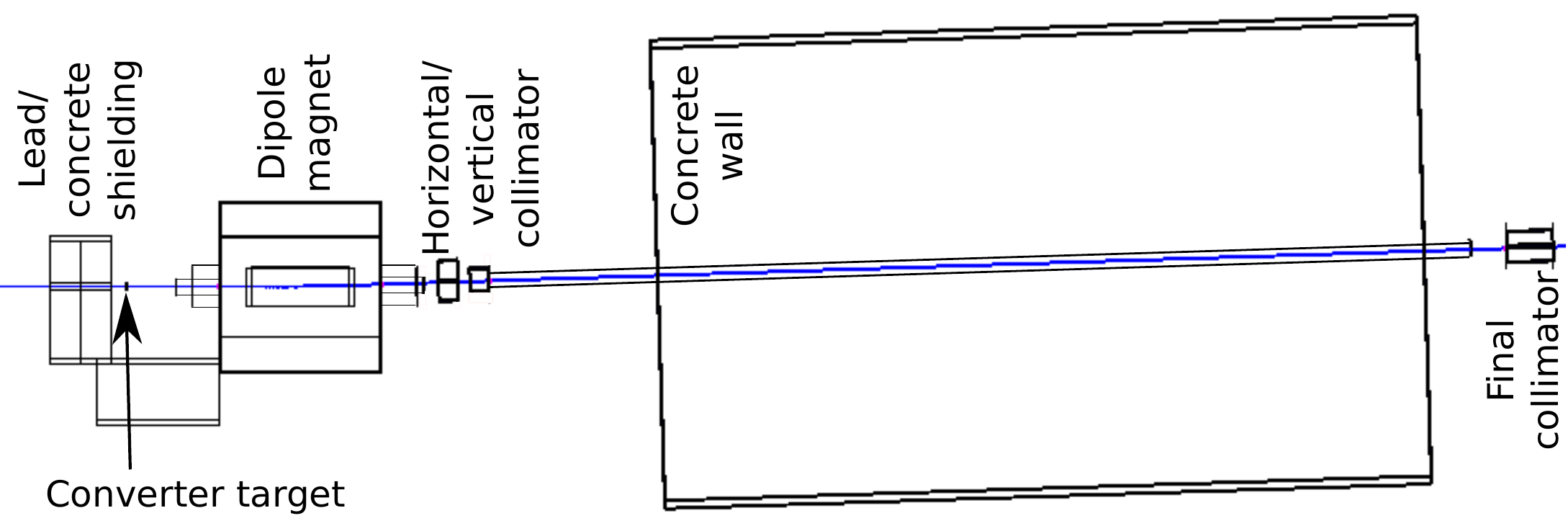}
  \caption[Visualization of a geantino trajectory in TB21.]{(Colour online) Visualization of TB21 with a geantino trajectory. 
  As the $xz$-plane is shown with a view on top of the geometry, the geantino passes through the test beam line from left to right.
  The origin of the test beam (and in that effect the start of the test beam simulation) is the carbon fiber in the \desyii beam pipe.
  Due to the scale, it is not shown here.}
    \label{fig:TBline_trajectory}
\end{figure}

To separate the neutral particles from the desired electrons and positrons of the final test beam, the beam pipe has a small kink shortly after the test beam magnet. 
All subsequent components of the test beam line are positioned along this new beam path. 
Only charged particles with the desired momentum enter the subsequent beam line components.
The particle momentum is Gaussian distributed, where the mean value belongs to the particles following the ideal beam path through the center of the pipe. 
Due to the large beam pipe aperture, the width of the momentum distribution of particles continuing on the test beam line is large.

To illustrate the trajectory of a beam particle, the simulation of a geantino trajectory in the test beam line is shown in Figure~\ref{fig:TBline_trajectory}. 
A geantino is a virtual particle for transportation processes in \geant simulations~\cite{Geantino}.

\begin{figure}[htbp]
  \centering
  \begin{subfigure}[t]{0.49\textwidth}

    \includegraphics[width=\textwidth]{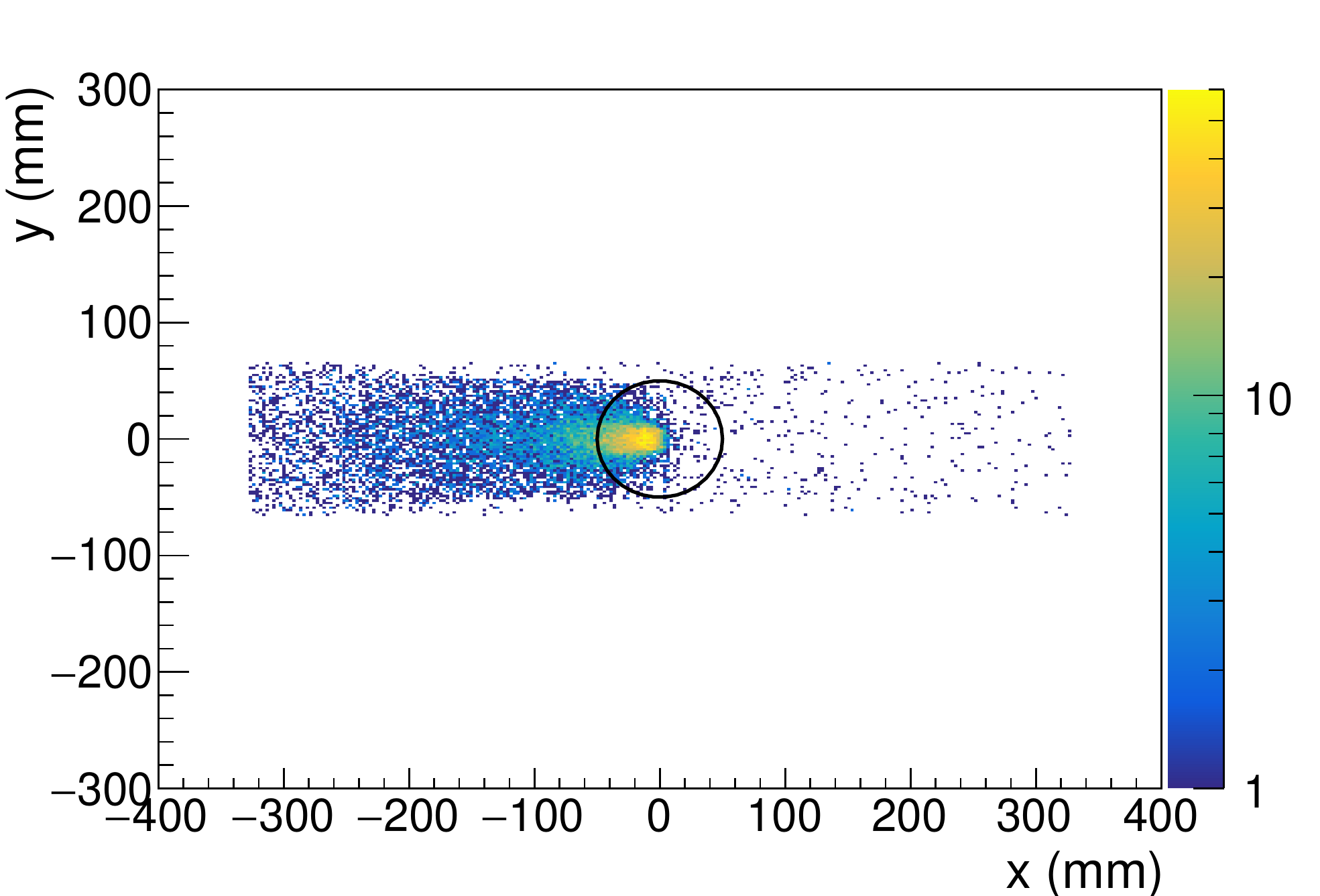}
    \subcaption{Negatively charged particles}
  \end{subfigure}
\hfill
  \begin{subfigure}[t]{0.49\textwidth}
       
    \includegraphics[width=\textwidth]{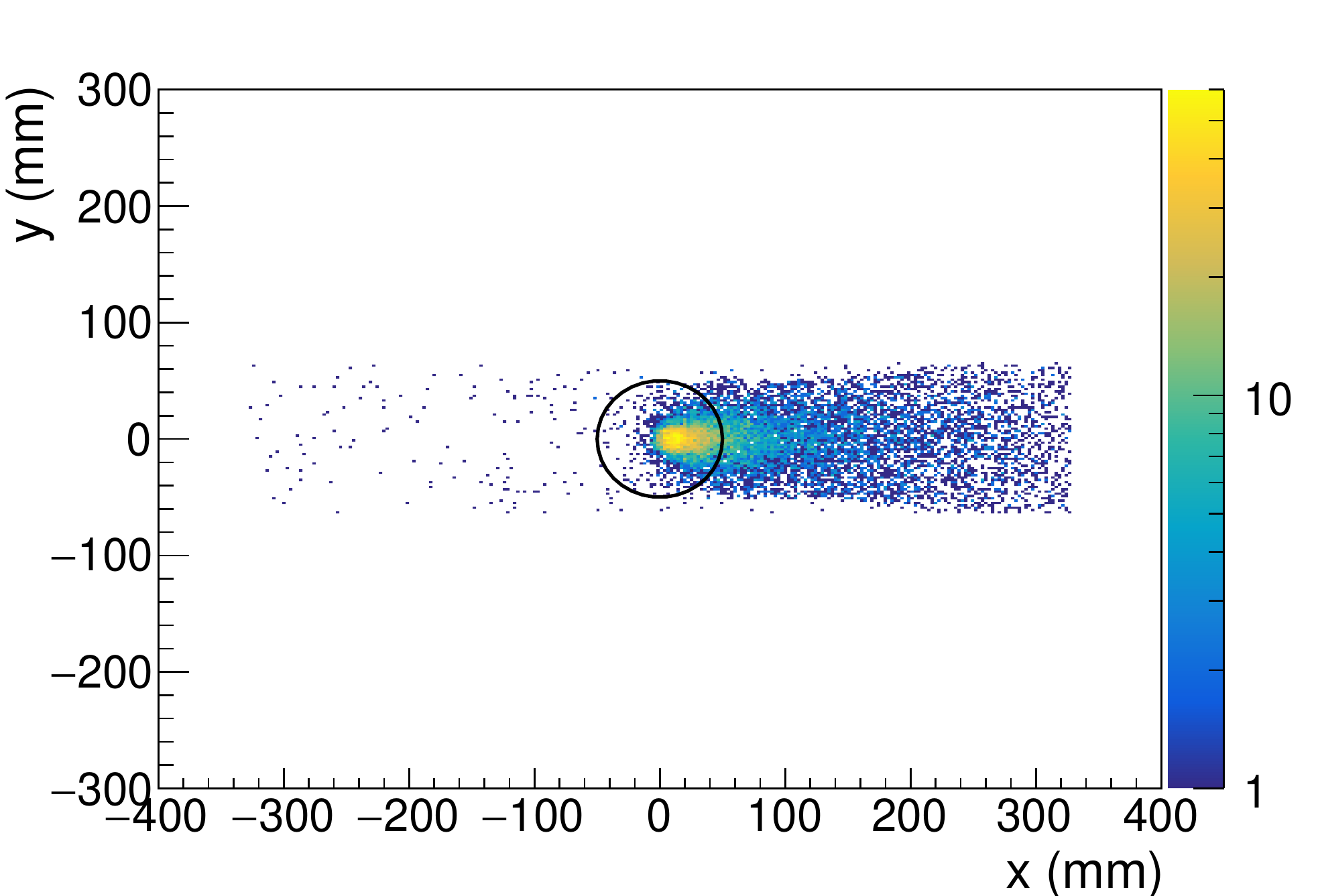}
 \subcaption{Positively charged particles}
  \end{subfigure}
 \caption[Position plots for negatively/positively charged particles behind the test beam magnet.]
 {(Colour online) The points represent the $xy$-positions of the particles directly behind the magnet, e.g. the positions where the particles leave the magnet. 
 The black circle illustrates the beam pipe coming out of the magnet. 
 For the simulation, a magnetic field strength of B\,=\,\SI{-0.3}{\tesla} was chosen. 
 The particles around the particle fan (particles with \SI{0}{\milli\meter}\,$<$\,x\,$<$\,\SI{400}{\milli\meter} 
 in Fig. (a), and \SI{-400}{\milli\meter}\,$<$\,x\,$<$\SI{0}{\milli\meter} in Fig. (b)) are scattered particles. }
  \label{fig:deflection_plot_neg/pos}
\end{figure}

The deflection of charged particles in the $xy$-plane behind the dipole magnet is shown in Figure~\ref{fig:deflection_plot_neg/pos}. 
The two separate particle fans for positively and negatively charged particles are clearly visible. 
The black circle indicates the contour of the beam pipe, and only particles within enter the subsequent beam line components. 
The other particles hit the iron surrounding of the magnet and do not proceed along the beam path. 
By changing the field strength of the magnet, the deflection of the particles is changed, and the desired beam momentum can be selected.
In the case of reversing the magnet polarization, particles with the opposite charge enter the test beam area.
Hence, the final test beam can consist of either electrons or positrons.

By collimating the beam using two separate collimators (for vertical and horizontal collimation), the Gaussian momentum distribution of the final test beam is narrowed.
The stronger the collimation is, the smaller the width of the distribution is.
After the first collimation, the beam continues along the evacuated beam pipe through the concrete wall into the test beam area. 
The concrete wall separates the test beam areas from the \desyii synchrotron tunnel.

In the test beam area, the particles are again collimated by a lead collimator with various available diameters. 
The size of the collimator hole affects the number of particles and the width of the momentum distribution of the final test beam.
For presented simulation, a collimator aperture of \SI{1}{\centi\meter} x \SI{1}{\centi\meter} was assumed.

\subsection{Momentum Distribution and Comparison to Measurements}
\label{sec:simulations:momentum_dist}

In order to gain information about the momentum distribution of the final test beam, the simulation was repeated for 
six different magnetic field strengths between \SI{0.1}{\tesla} and \SI{0.9}{\tesla}.
The statistics were increased by repeating the full simulation 200 times for each magnetic field strength.
For all six magnetic field strengths, the mean and the width of the momentum distributions were determined using a Gaussian fit. 
The results, listed in Table~\ref{table:momentum_spreads}, show a momentum spread of between \SI{1.5}{\percent} and \SI{12.8}{\percent}.

\begin{figure}[htbp]
\centering
    \begin{subfigure}[t]{0.49\textwidth}
        \includegraphics[width=1.0\textwidth]{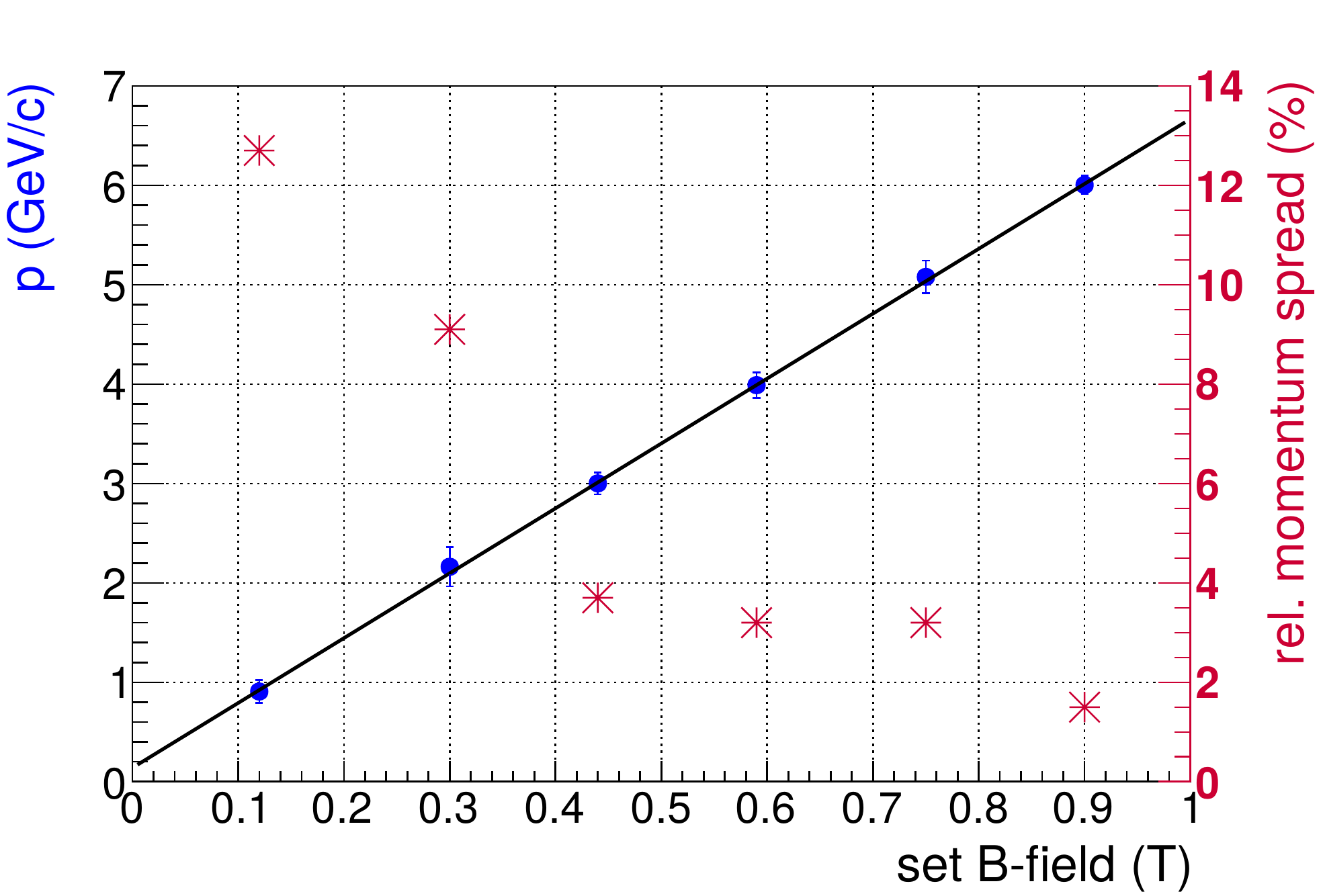}
	\subcaption{Momentum vs. $B$-field}
        \label{fig:simresults:E_over_B}
    \end{subfigure}
    \begin{subfigure}[t]{0.49\textwidth}

        \includegraphics[width=1.0\textwidth]{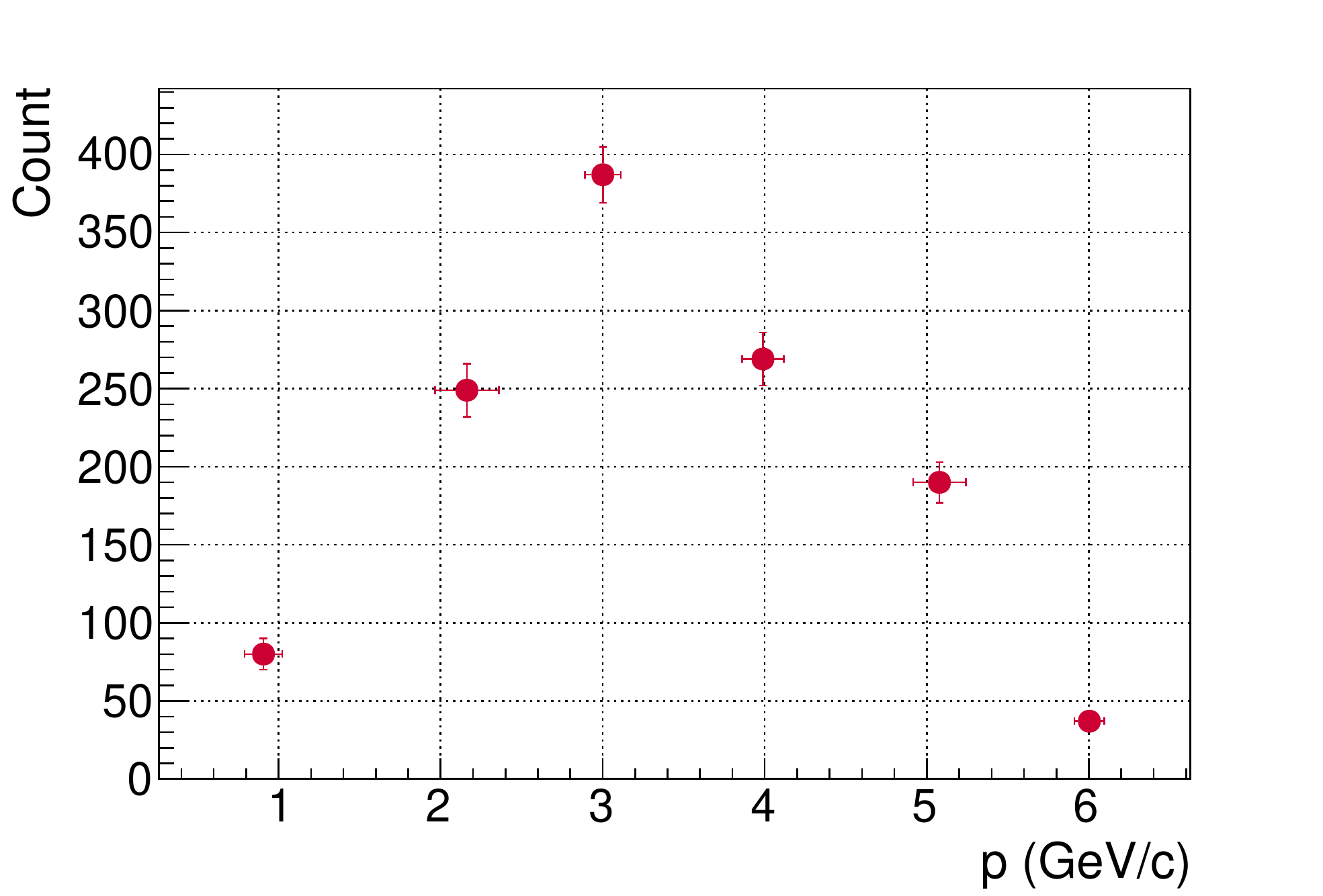}
	\subcaption{Counts vs. momentum}
        \label{fig:simresults:TBCollimator_rate}
    \end{subfigure}
\caption{(Colour online) 
(\subref{fig:simresults:E_over_B}): 
    The particle momentum of the simulated test beam as a function of the magnetic field strength of the dipole magnet of TB21. 
    The plot shows the mean momenta for corresponding magnetic field strengths.
(\subref{fig:simresults:TBCollimator_rate}): 
    The simulated particle counts in area of TB21 after the final collimation with respect to the electron momentum. 
    This assumes a fixed \desyii beam energy of \SI{6.3}{\GeV}.
    The non-constant \desyii beam intensity as well as the time structure given by the sinusoidal \desyii magnet cycle (see Sec.~\ref{fig:desy2:cycle}) are not considered in this simulation. 
}
\label{fig:simresults}
\end{figure}

The calibration and momentum spread measurements performed in Section~\ref{sec:performance:momentum} show a linear dependency between the particle momentum and the magnetic field strength.
This linear dependency is confirmed in Figure~\ref{fig:simresults:E_over_B}, in which the particle momentum determined from the simulation is plotted as a function of the selected magnetic field strengths.
The simulation results are in good agreement with the measurements shown in Figure~\ref{fig:performance:momentumMeasurement:Spread}.

Additionally, the accumulated counts are plotted against the determined particle momentum (Fig.~\ref{fig:simresults:TBCollimator_rate}).
The shape of this distribution including a peak at momenta of 2-\SI{3}{\GeV/c} is similar to the measured integrated rate (see Sec.~\ref{sec:performance:rate_momentum}).
Deviations are expected since the \desyii cycle structure including a non-constant \desyii beam intensity as well as the 
time structure given by the sinusoidal \desyii magnet cycle (see Sec.~\ref{fig:desy2:cycle}) is not considered in this simulation, which was using a fixed \desyii beam energy of \SI{6.3}{\GeV}.
Due to this behavior of the \desyii beam intensity, the peak is expected to move to lower particle momenta.
These deficiencies present possible future improvements to the simulation setup.

\begin{table}[htbp]
  \begin{center}
    \begin{tabular}{c c c S[table-format=2.3]}
    \hline
        \arraybackslash{\boldmath$B$~\textbf{(T)}} & \multicolumn{1}{>{\centering\arraybackslash}m{2.0cm}}{\boldmath$\bar{p}$~\textbf{(GeV/c)}} & \multicolumn{1}{>{\centering\arraybackslash}m{2.0cm}}{\boldmath$\sigma$~\textbf{(GeV/c)}} & \multicolumn{1}{>{\centering\arraybackslash}m{2.0cm}}{\boldmath$\frac{\sigma}{\bar{p}}$~\textbf{(\%)}}\\ 
\toprule
    0.12 & 0.907 & 0.116 & 12.7\\ 
    0.30 & 2.162 & 0.197 & 9.1\\ 
    0.44 & 3.001 & 0.111 & 3.7 \\
    0.59 & 3.989 & 0.128 & 3.2\\
    0.75 & 5.078 & 0.163 & 3.2\\
    0.90 & 6.003 & 0.093 & 1.5\\
\bottomrule
    \end{tabular}
  \end{center}
  \caption{Table of the particle momentum and its spread for the final test beam of TB21. The results are gained from 200 full simulations for each magnetic field strength.}
  \label{table:momentum_spreads}
\end{table}

\subsection{Conclusion}
For gaining a better understanding of the \desyii test beam generation, a \geant Monte Carlo simulation of TB21 was set up.
The presented simulation efforts have already provided valuable insights in test beam generation and its dependency on the beam attributes.
Improvements on the simulation setup, as well as the implementation of the remaining test beam lines and the \desyii time structure
would allow thorough studies which can serve as a key input for future test beam line improvements.

\newpage

\section{Access \& User Statistics}\label{sec:userstats}

In this section a summary of the usage of the \diitbf is mainly given for the years from 2013 till 2017.
First, the beam availability of \diitbf and the usage are considered (Sec.~\ref{sec:userstats:usage}).
Then details are given on the research fields and projects and the usage of the infrastructure (Sec.~\ref{sec:userstats:projects})
and on the individual aspects (Sec.~\ref{sec:userstats:individual}).
Finally, the support for European users is explained (Sec.~\ref{sec:userstats:transnational})
and the educational programmes at the \diitbf (Sec.~\ref{sec:userstats:education}).

\subsection{Availability and Usage}\label{sec:userstats:usage}

The \diitbf usually operates between 35 and 40 weeks a year which corresponds to 120 user slots for the three beam lines. 
Each slot lasts one week from Monday 8~am to Monday 8~am.
Slots are allocated to the users by 
the test beam coordinators based on the scientific merit of the individual proposals. 
Each fall, a call is made for beam time requests for the upcoming year. Based on the response, the user schedule is compiled. 
Over the remainder of the year, still available slots are allocated to requests on a first come first serve basis by the test beam coordinators.

\begin{figure}[!htb]
\begin{subfigure}[b]{0.49\textwidth}
\includegraphics[width=\textwidth]{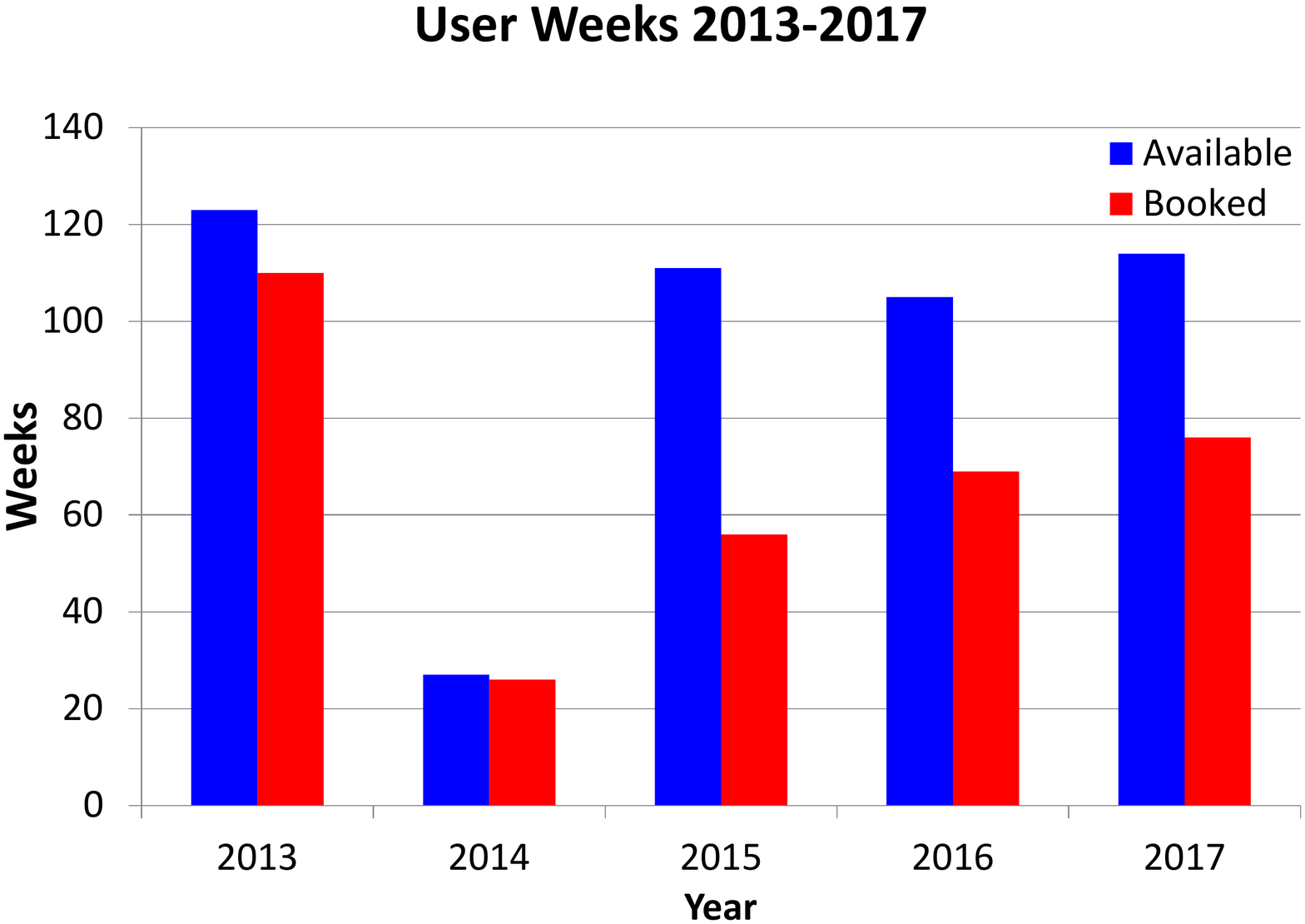}
\caption{}
\label{sfig:tbusrweeks2015-17}
\end{subfigure}
\begin{subfigure}[b]{0.49\textwidth}
\includegraphics[width=\textwidth]{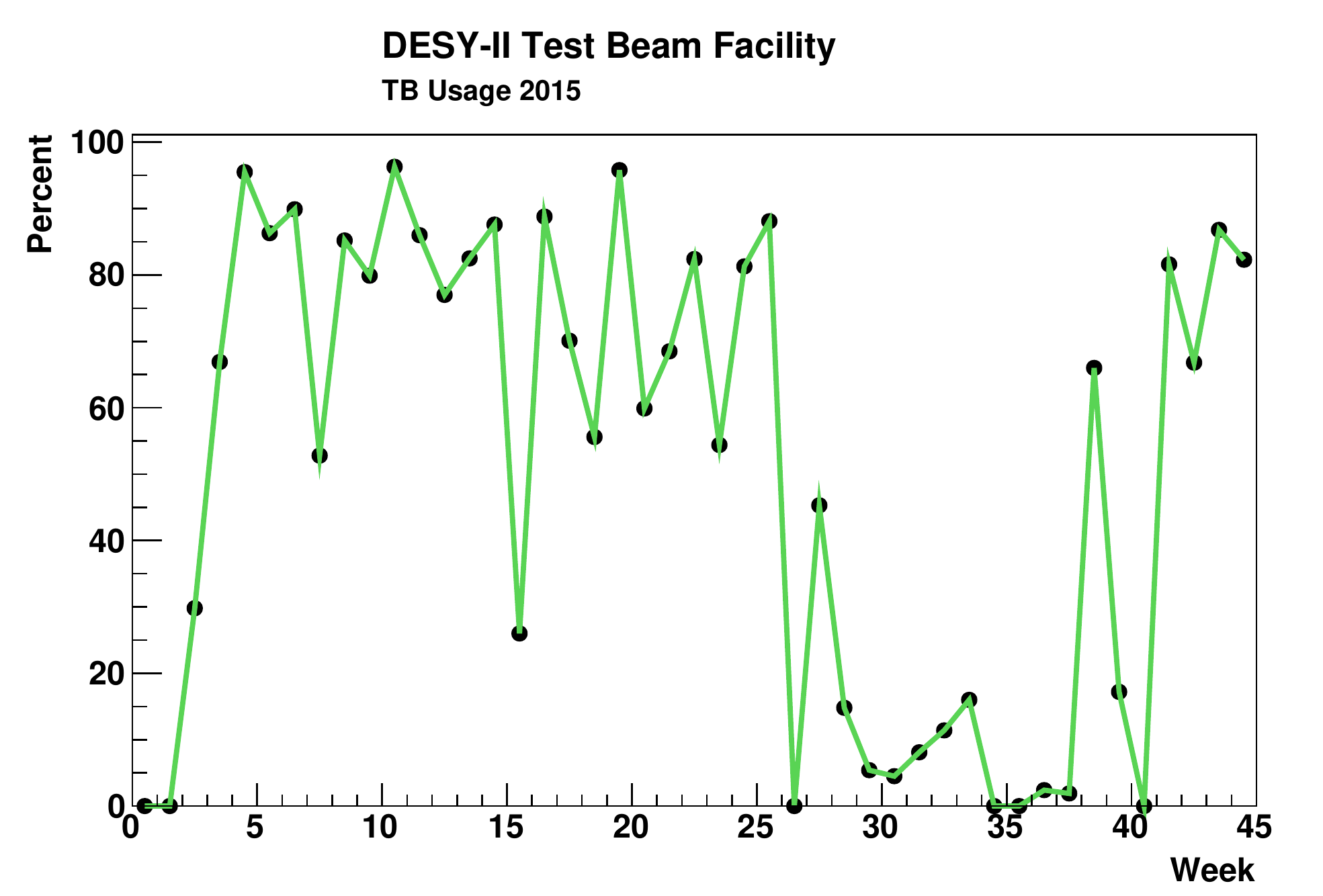}
\caption{}
\label{sfig:tbefficiencystats}
\end{subfigure}
\caption{(Colour online) \protect\subref{sfig:tbusrweeks2015-17})~The number of available (in blue) and booked weeks (in red) per year for the period from 2013 and 2017.  
\protect\subref{sfig:tbefficiencystats})~The average up-time of the \diitbf for the 2015 run period.}
\label{fig:userstats:usereff}
\end{figure}

Figure \ref{sfig:tbusrweeks2015-17} shows that the \diitbf has been well-used. 
In particular, 2013 was a very busy year 
due to the shutdown of the CERN test beams. This led to an increased demand, 
since the \diitbf was at this time the only facility operating in Europe 
providing multi-GeV particles. The usage number of 2014 is rather low, as 
\desyii was only running for nine weeks in 2014 due to a shutdown of the DESY 
accelerator complex from mid-March to December.

The usage statistics from 2015 to 2017 show a clear trend that the demand is 
continuously increasing. The average up-time ---defined as the time any beam 
line is taking data divided by total available beam time--- is with 
\SI{60}{\percent} to \SI{70}{\percent} rather high, given that this includes 
also the times for dis-/assembly of the user setups. As an example, the average 
up-time for 2015 is shown in Figure~\ref{sfig:tbefficiencystats}. Here, the 
average up-time was \SI{64.5}{\percent}, excluding the weeks where no users 
where allocated to any of the beam lines.

\subsection{Test Beam Projects and Usage of Infrastructures}\label{sec:userstats:projects}

\begin{figure}[htbp]
\begin{subfigure}[b]{0.49\textwidth}
\includegraphics[width=\textwidth]{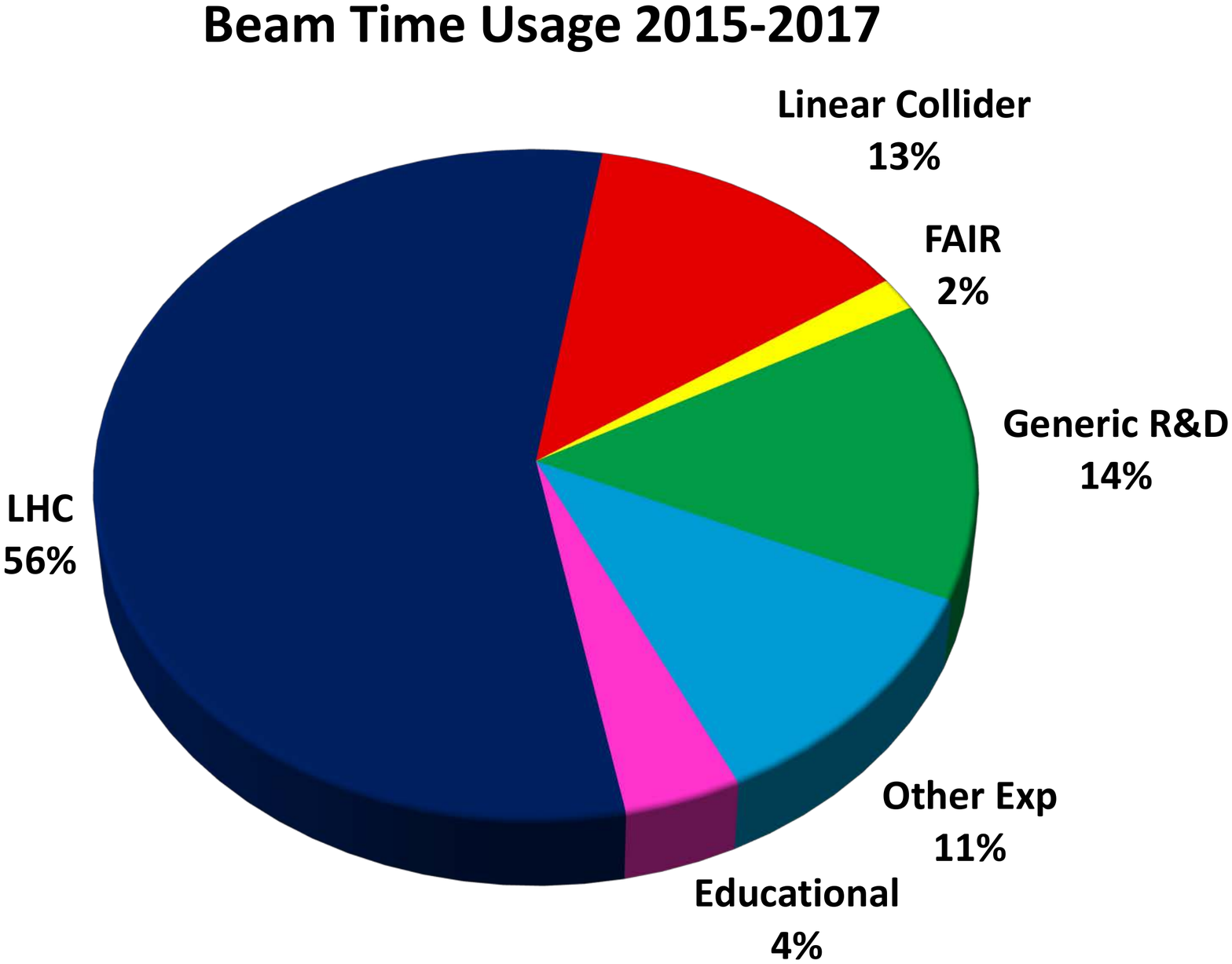}
\caption{}
\label{sfig:tbusrstatsexpcake}
\end{subfigure}
\begin{subfigure}[b]{0.49\textwidth}
\includegraphics[width=\textwidth]{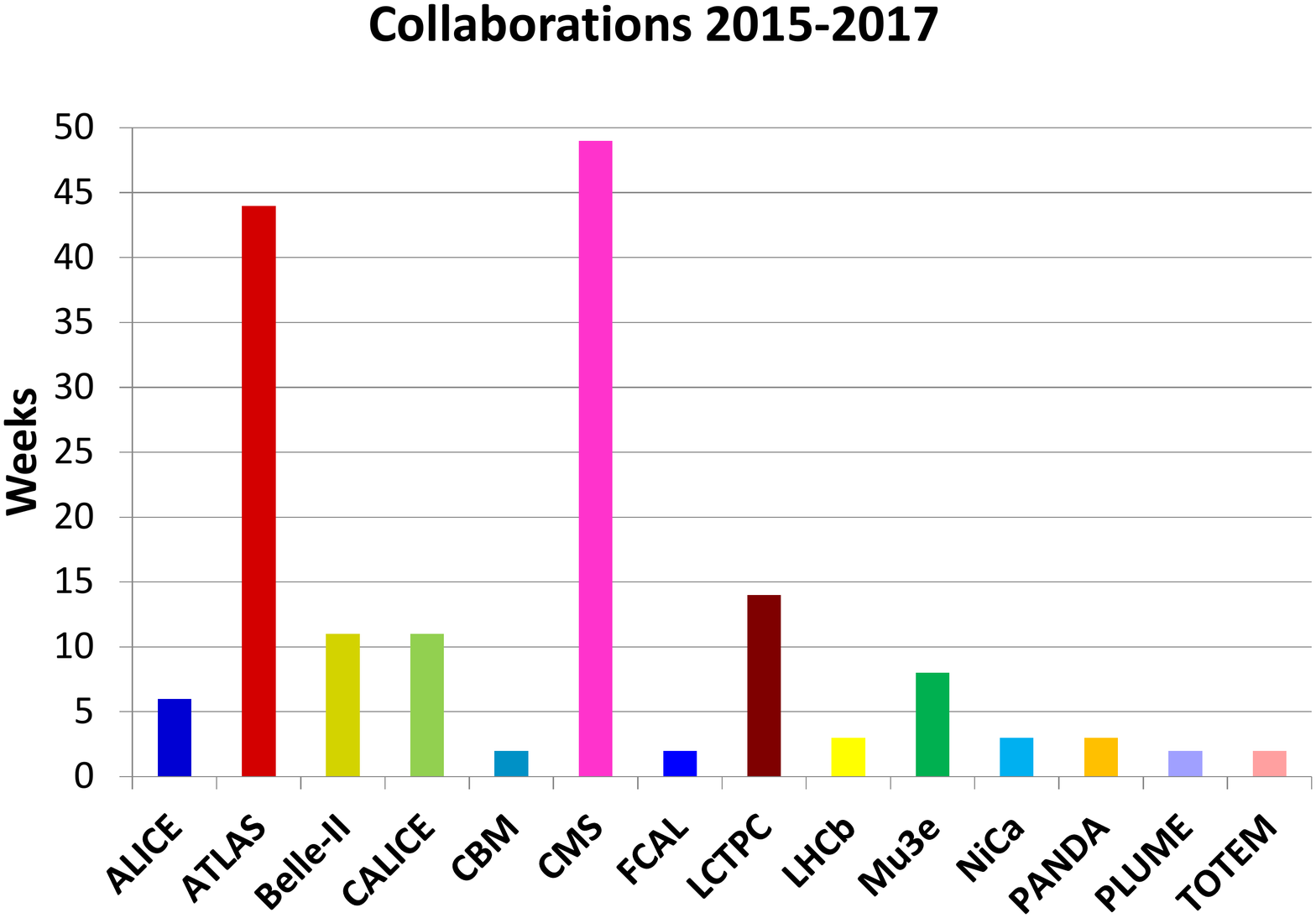}
\caption{}
\label{sfig:tbusrstatscollabbars}
\end{subfigure}
\caption{(Colour online) \protect\subref{sfig:tbusrstatsexpcake})~Composition of user groups from 2015-2017 and \protect\subref{sfig:tbusrstatscollabbars})~ number of weeks used by individual collaborations.}
\label{fig:userstats:groups}
\end{figure}

The core of the users at the \diitbf originates from particle physics, but the facility also has an increased use by adjacent fields like nuclear physics,
accelerator physics and beyond. In addition, many generic projects of detector R\&D for particle detection make use of the facility. 
Figure~\ref{sfig:tbusrstatsexpcake} shows the different user communities, with LHC groups making up more than half the groups. The category "Linear Collider" summarizes R\&D focused on 
detectors for ILC and CLIC, "FAIR" the nuclear physics groups developing detectors for FAIR at Darmstadt. "Other Experiments" includes all groups 
from Experiments/Collaborations including e.g. Belle II and Mu3e. Generic R\&D is used for all groups, whose developments are not or not yet 
geared at a specific experiment. The category "Educational" included all the activities described in more detail in Section~\ref{sec:userstats:edu}.
Figure~\ref{sfig:tbusrstatscollabbars} shows the percentage of weeks allocated to individual collaborations, with ATLAS and CMS being the dominant users.

\begin{figure} [htbp]
\begin{subfigure}[b]{0.49\textwidth}
\includegraphics[width=\textwidth]{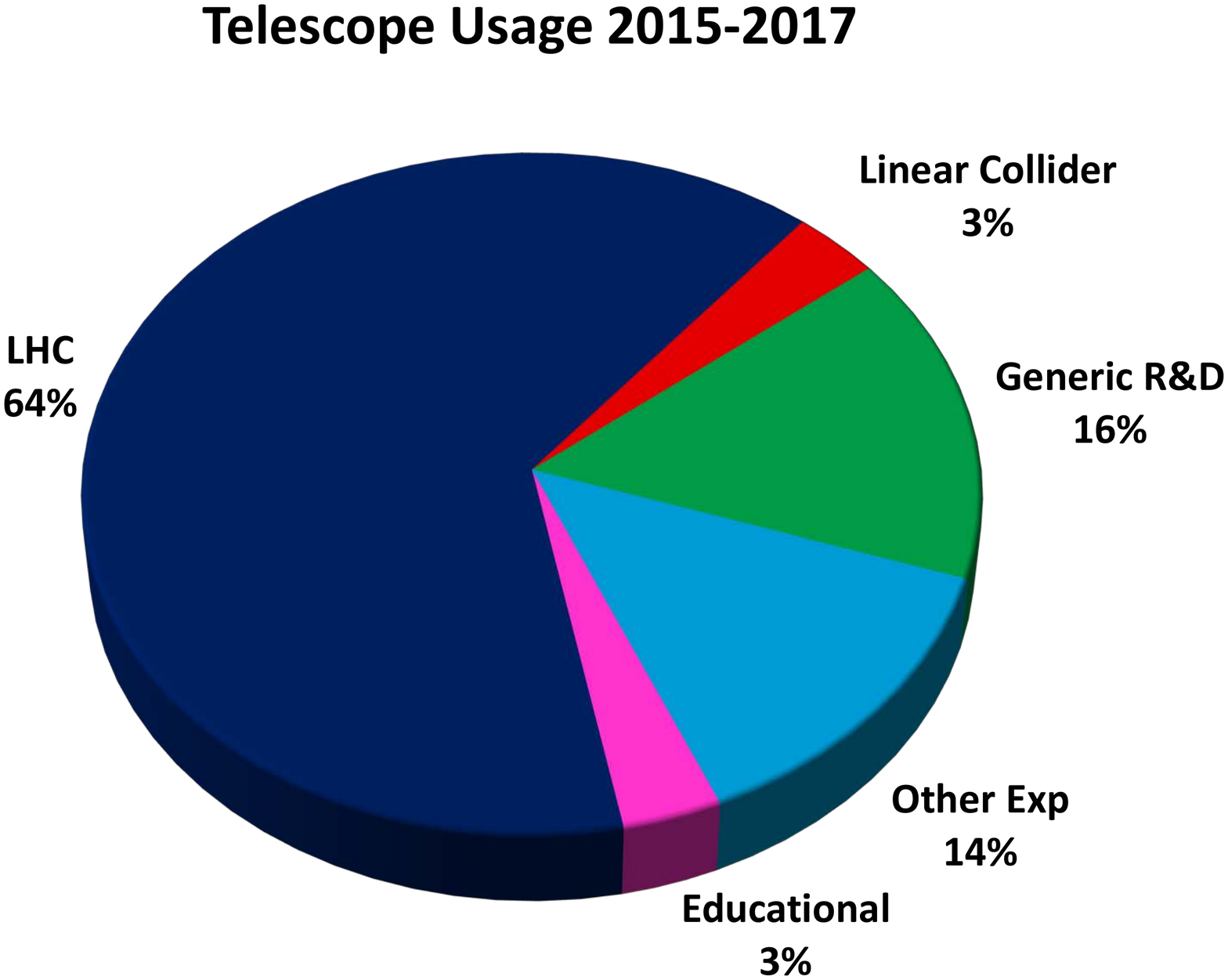}
\caption{}
\label{sfig:tbtelestats}
\end{subfigure}
\begin{subfigure}[b]{0.49\textwidth}
\includegraphics[width=\textwidth]{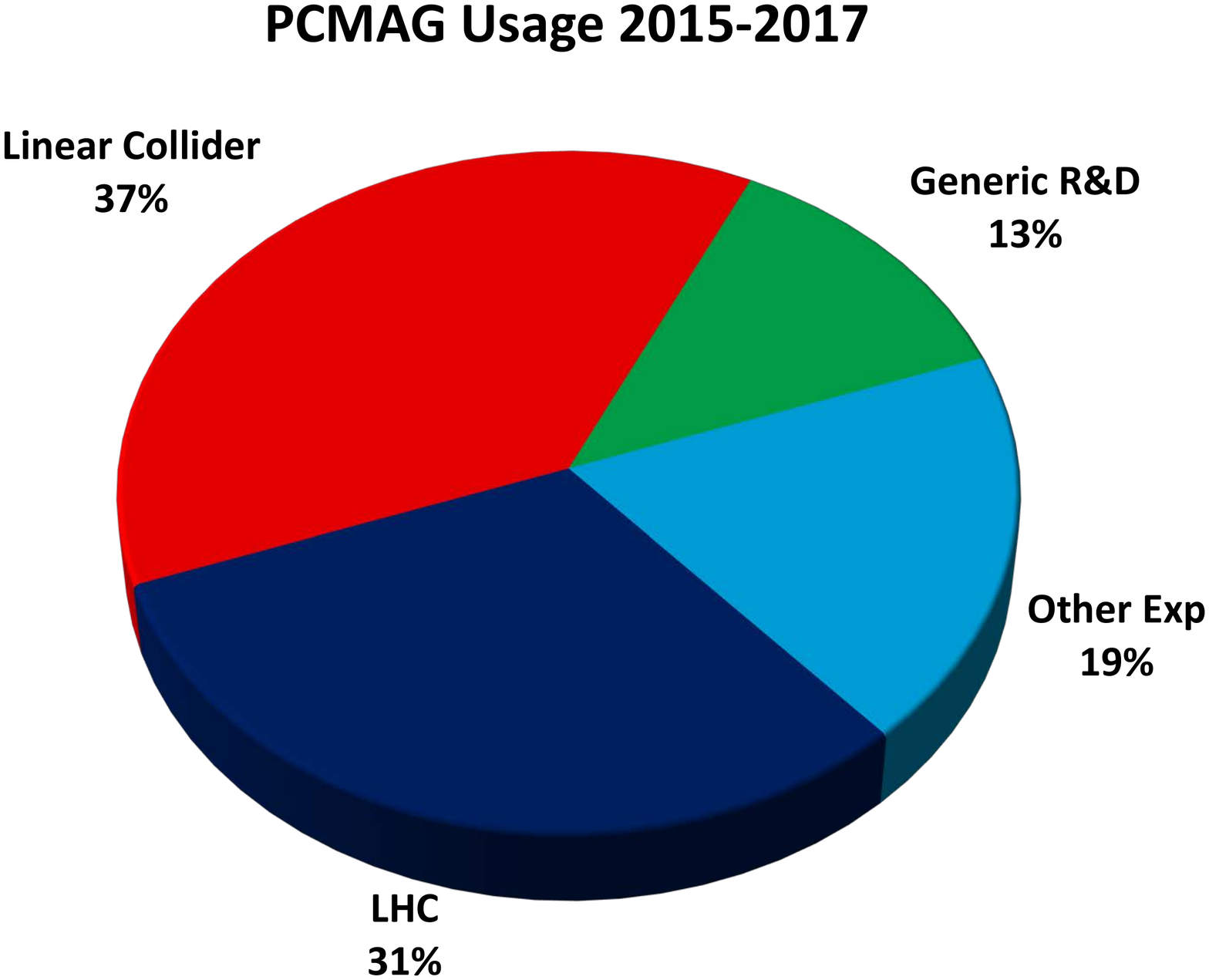}
\caption{}
\label{sfig:tbpcmagstats}
\end{subfigure}
\caption{(Colour online) \protect\subref{sfig:tbtelestats})~Use of the EUDET-type Pixel Beam Telescopes and \protect\subref{sfig:tbpcmagstats})~the PCMAG solenoid 
for the period from 2015 to 2017. Overall, about \SI{72}{\percent} of the groups requested the use of a telescope and \SI{13}{\percent} requested the use of the PCMAG.}
\label{fig:userstats:infrastructures}
\end{figure}

The usage of additional infrastructures provided at DESY like the EUDET-type Pixel Beam Telescopes (see Sec.~\ref{sec:addinf:telescopes}) and 
\SI{1}{\tesla} PCMAG Solenoid (see Sec.~\ref{sec:addinf:pcmag}) is shown in Figure~\ref{fig:userstats:infrastructures} with the user group categories described in detail above. 
For the telescopes, about \SI{72}{\percent} and for the PCMAG, about \SI{13}{\percent} of the groups requested their use.
Especially for the telescopes, the LHC groups are the predominant user, while the use of the PCMAG is more evenly spread between the different categories.
For the telescopes it should be noted that, from 2016 onward, the two beam lines TB21 and TB22 were equipped a telescope, so the usage went up significantly from \SI{68}{\percent} 
to almost \SI{80}{\percent}. The usage of the PCMAG has remained quite constant over this period. 

\subsection{Individual Users Statistics}\label{sec:userstats:individual}

\begin{figure}[htb]
\begin{center}
\includegraphics[width=0.7\textwidth]{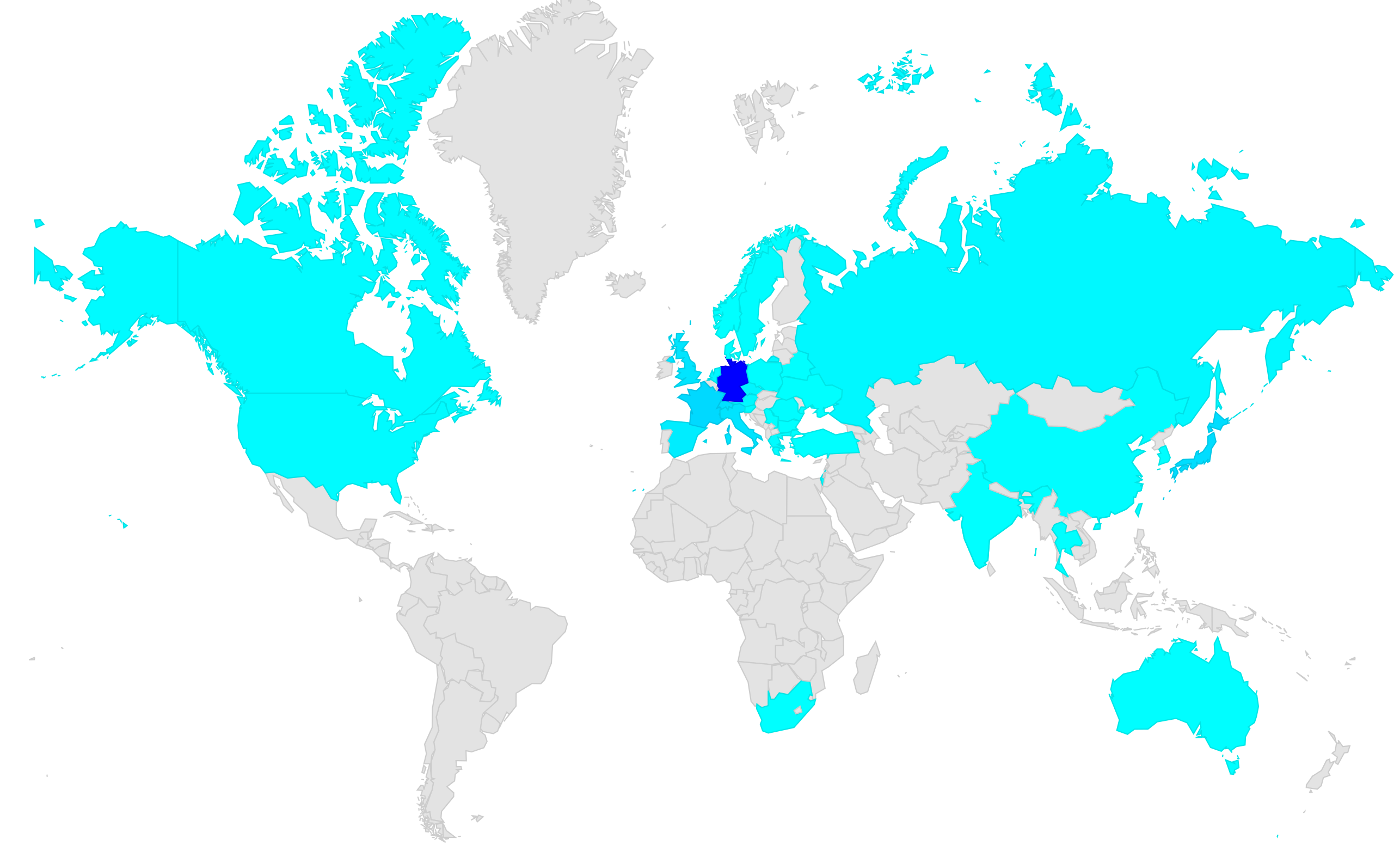}
\caption{(Colour online) Origin of the users at the \diitbf from 2013-2017}
\label{fig:userstats:worldmap}
\end{center}
\end{figure}

In the period 2013-2017, a total of 1266 registered user, originating from 34 different countries (Fig.~\ref{fig:userstats:worldmap}), visited the \diitbf. 
Users from Germany contribute \SI{51}{\percent}, users from the EU to \SI{30}{\percent} and non-EU users \SI{19}{\percent}. 
The users from Germany include local users from DESY, which for many projects act as local liaison. That about half of the users come from outside of Germany 
illustrates that the test beam at DESY is an international facility even though it is operated at and funded by a national laboratory.

It is important to note, that the facility is a key infrastructure for training the next generation detector experts, as about
\SI{50}{\percent} of the users are students and postdoctoral researchers.

\subsection{Transnational Access Support}\label{sec:userstats:transnational}

The EU projects EUDET~\cite{EUDET}, AIDA~\cite{AIDA} and AIDA2020~\cite{AIDA2020} have been, respectively are, supporting a transnational access scheme to large facilities in Europe for users outside of the host country. 
This scheme has been extensively used by many groups and enables them to conduct their 
experiments at the test beam facility. From 2013 to 2017, on average \SI{20}{\percent} of the users have received support from the Transnational 
Access scheme.

\subsection{Educational Use}\label{sec:userstats:education}

\label{sec:userstats:edu}
DESY has a very successful Summer Student Programme~\cite{DESYSSP} and since a few years, a few of the summer 
students are given the opportunity to conduct measurements at the \diitbf. 
This has been proven to be very popular since students can experience high energy physics from the experimental setup over test beam measurements to analysis and results.

Besides, DESY is offering since 2016 a particle physics training course for high school teachers. This includes a hands-on experiment at the test beam.
Further outreach and training activities targeted at high school students are being planned.

\newpage

\section{Summary and Outlook}\label{sec:summaryoutlook}

The \diitbf has been in operation since the early nineties and has since then continuously been improved.
Today it successfully operates as a well-established international facility for detector R\&D for particle physics 
and adjacent fields. The reliable beam and the infrastructures provided ranging from patch panels to 
telescopes and high-field magnets are a key ingredient to its success.

The facility is nowadays a key infrastructure for detector R\&D worldwide and its usage keeps increasing.

A Test Beam User Workshop was held at DESY in October 2017 to get feedback first-hand from the user community 
and to discuss possible future improvements and extensions to the facility~\cite{Arling:2018hca}. Key points made were 
to increase the particle rates in general and to have high rates available at the highest possible 
energy. Overall however, the user community was pleased with the status of the facility.
The current facility will continue to run until at least to 2025, which is the timescale for the start of an 
upgrade of \petraiii to \petraiv. This upgrade could imply changes to the injector chain, which then would 
have an impact on the \diitbf. However, the intention is to provide test beams at DESY for as long as there 
is a user need for multi-GeV electron test beams.

\section*{Acknowledgments}

The authors are greatly indebted to our colleagues from the DESY M and FH divisions, 
which are essential to successfully operate, maintain and extend this facility.

The described infrastructure has received funding from the European Commission 
under the 6$^{\textrm{th}}$ Framework Programme ``Structuring the European Research Area'' 
under contract RII3-026126 and under the FP7 Research Infrastructures project 
AIDA, grant agreement no. 262025. This project has received funding from the 
European Union's  Horizon 2020 Research and Innovation programme under Grant 
Agreement no.~654168. The PCMAG infrastructure was supported by JSPS KAKENHI 
Grant No.~23000002. 
The information herein only reflects the views of its authors and not those of 
the European Commission and no warranty expressed or implied is made with regard 
to such information or its use.

\newpage





\bibliographystyle{elsarticle-num}
\bibliography{bibliography.bib}

\begin{thebibliography}{10}
\expandafter\ifx\csname url\endcsname\relax
  \def\url#1{\texttt{#1}}\fi
\expandafter\ifx\csname urlprefix\endcsname\relax\def\urlprefix{URL }\fi
\expandafter\ifx\csname href\endcsname\relax
  \def\href#1#2{#2} \def\path#1{#1}\fi

\bibitem{EUDET}
\href{http://www.eudet.org/}{{Homepage of the EUDET project}}, [Online;
  accessed 23-January-2018] (2018).
\newline\urlprefix\url{http://www.eudet.org/}

\bibitem{AIDA}
\href{http://aida-old.web.cern.ch/aida-old/index.html}{{Homepage of the AIDA
  project}}, [Online; accessed 23-January-2018] (2018).
\newline\urlprefix\url{http://aida-old.web.cern.ch/aida-old/index.html}

\bibitem{AIDA2020}
\href{http://aida2020.web.cern.ch/}{{Homepage of the AIDA-2020 project}},
  [Online; accessed 23-January-2018] (2018).
\newline\urlprefix\url{http://aida2020.web.cern.ch/}

\bibitem{Hemmie:1983et}
G.~Hemmie, {DESY II, a New Injector for the Desy Storage Rings PETRA and DORIS
  II}, IEEE Transactions on Nuclear Science 30~(4) (1983) 2028--2030.
\newblock \href {http://dx.doi.org/10.1109/TNS.1983.4332706}
  {\path{doi:10.1109/TNS.1983.4332706}}.

\bibitem{Hemmie:1985uw}
G.~Hemmie, U.~Berghaus, H.~Bottcher, E.~Dabkowski, H.~R. Heller, G.~Meyer,
  G.~Nawrath, F.~Schwickert, K.~Sinram, G.~Wobke, H.~Wumpelmann, {Design,
  Construction and Performance of the DESY II Magnets}, IEEE Transactions on
  Nuclear Science 32~(5) (1985) 3625--3627.
\newblock \href {http://dx.doi.org/10.1109/TNS.1985.4334448}
  {\path{doi:10.1109/TNS.1985.4334448}}.

\bibitem{bielski2012}
R.~Bielski, {DESY Summer Student Project} (2012).

\bibitem{tlu}
D.~Cussans,
  \href{http://www.eudet.org/e26/e28/e42441/e57298/EUDET-MEMO-2009-04.pdf}{{Description
  of the JRA1 Trigger Logic Unit (TLU), v0.2c}}, {EUDET-MEMO} {2009-04}.
\newline\urlprefix\url{http://www.eudet.org/e26/e28/e42441/e57298/EUDET-MEMO-2009-04.pdf}

\bibitem{tine-0957-0233-18-8-012}
P.~Bartkiewicz, P.~Duval, {TINE} as an accelerator control system at {DESY},
  Measurement Science and Technology 18~(8) (2007) 2379.
\newblock \href {http://dx.doi.org/10.1088/0957-0233/18/8/012}
  {\path{doi:10.1088/0957-0233/18/8/012}}.

\bibitem{Agostinelli:2002hh}
S.~Agostinelli, J.~Allison, K.~Amako, J.~Apostolakis, H.~Araujo, P.~Arce,
  M.~Asai, D.~Axen, S.~Banerjee, G.~Barrand, F.~Behner, L.~Bellagamba,
  J.~Boudreau, L.~Broglia, A.~Brunengo, H.~Burkhardt, S.~Chauvie, J.~Chuma,
  R.~Chytracek, G.~Cooperman, G.~Cosmo, P.~Degtyarenko, A.~Dell'Acqua,
  G.~Depaola, D.~Dietrich, R.~Enami, A.~Feliciello, C.~Ferguson, H.~Fesefeldt,
  G.~Folger, F.~Foppiano, A.~Forti, S.~Garelli, S.~Giani, R.~Giannitrapani,
  D.~Gibin, J.~G. Cadenas, I.~González, G.~G. Abril, G.~Greeniaus, W.~Greiner,
  V.~Grichine, A.~Grossheim, S.~Guatelli, P.~Gumplinger, R.~Hamatsu,
  K.~Hashimoto, H.~Hasui, A.~Heikkinen, A.~Howard, V.~Ivanchenko, A.~Johnson,
  F.~Jones, J.~Kallenbach, N.~Kanaya, M.~Kawabata, Y.~Kawabata, M.~Kawaguti,
  S.~Kelner, P.~Kent, A.~Kimura, T.~Kodama, R.~Kokoulin, M.~Kossov,
  H.~Kurashige, E.~Lamanna, T.~Lampén, V.~Lara, V.~Lefebure, F.~Lei,
  M.~Liendl, W.~Lockman, F.~Longo, S.~Magni, M.~Maire, E.~Medernach,
  K.~Minamimoto, P.~M. de~Freitas, Y.~Morita, K.~Murakami, M.~Nagamatu,
  R.~Nartallo, P.~Nieminen, T.~Nishimura, K.~Ohtsubo, M.~Okamura, S.~O'Neale,
  Y.~Oohata, K.~Paech, J.~Perl, A.~Pfeiffer, M.~Pia, F.~Ranjard, A.~Rybin,
  S.~Sadilov, E.~D. Salvo, G.~Santin, T.~Sasaki, N.~Savvas, Y.~Sawada,
  S.~Scherer, S.~Sei, V.~Sirotenko, D.~Smith, N.~Starkov, H.~Stoecker,
  J.~Sulkimo, M.~Takahata, S.~Tanaka, E.~Tcherniaev, E.~S. Tehrani,
  M.~Tropeano, P.~Truscott, H.~Uno, L.~Urban, P.~Urban, M.~Verderi, A.~Walkden,
  W.~Wander, H.~Weber, J.~Wellisch, T.~Wenaus, D.~Williams, D.~Wright,
  T.~Yamada, H.~Yoshida, D.~Zschiesche, Geant4: a simulation toolkit, Nuclear
  Instruments and Methods in Physics Research Section A: Accelerators,
  Spectrometers, Detectors and Associated Equipment 506~(3) (2003) 250 -- 303.
\newblock \href {http://dx.doi.org/10.1016/S0168-9002(03)01368-8}
  {\path{doi:10.1016/S0168-9002(03)01368-8}}.

\bibitem{Allison:2006ve}
J.~Allison, K.~Amako, J.~Apostolakis, H.~Araujo, P.~A. Dubois, M.~Asai,
  G.~Barrand, R.~Capra, S.~Chauvie, R.~Chytracek, G.~A.~P. Cirrone,
  G.~Cooperman, G.~Cosmo, G.~Cuttone, G.~G. Daquino, M.~Donszelmann,
  M.~Dressel, G.~Folger, F.~Foppiano, J.~Generowicz, V.~Grichine, S.~Guatelli,
  P.~Gumplinger, A.~Heikkinen, I.~Hrivnacova, A.~Howard, S.~Incerti,
  V.~Ivanchenko, T.~Johnson, F.~Jones, T.~Koi, R.~Kokoulin, M.~Kossov,
  H.~Kurashige, V.~Lara, S.~Larsson, F.~Lei, O.~Link, F.~Longo, M.~Maire,
  A.~Mantero, B.~Mascialino, I.~McLaren, P.~M. Lorenzo, K.~Minamimoto,
  K.~Murakami, P.~Nieminen, L.~Pandola, S.~Parlati, L.~Peralta, J.~Perl,
  A.~Pfeiffer, M.~G. Pia, A.~Ribon, P.~Rodrigues, G.~Russo, S.~Sadilov,
  G.~Santin, T.~Sasaki, D.~Smith, N.~Starkov, S.~Tanaka, E.~Tcherniaev,
  B.~Tome, A.~Trindade, P.~Truscott, L.~Urban, M.~Verderi, A.~Walkden, J.~P.
  Wellisch, D.~C. Williams, D.~Wright, H.~Yoshida, Geant4 developments and
  applications, IEEE Transactions on Nuclear Science 53~(1) (2006) 270--278.
\newblock \href {http://dx.doi.org/10.1109/TNS.2006.869826}
  {\path{doi:10.1109/TNS.2006.869826}}.

\bibitem{Allison:2016lfl}
J.~Allison, K.~Amako, J.~Apostolakis, P.~Arce, M.~Asai, T.~Aso, E.~Bagli,
  A.~Bagulya, S.~Banerjee, G.~Barrand, B.~Beck, A.~Bogdanov, D.~Brandt,
  J.~Brown, H.~Burkhardt, P.~Canal, D.~Cano-Ott, S.~Chauvie, K.~Cho,
  G.~Cirrone, G.~Cooperman, M.~Cortés-Giraldo, G.~Cosmo, G.~Cuttone,
  G.~Depaola, L.~Desorgher, X.~Dong, A.~Dotti, V.~Elvira, G.~Folger,
  Z.~Francis, A.~Galoyan, L.~Garnier, M.~Gayer, K.~Genser, V.~Grichine,
  S.~Guatelli, P.~Guèye, P.~Gumplinger, A.~Howard,
  I.~H\v{r}ivn\'{a}\v{c}ov\'{a}, S.~Hwang, S.~Incerti, A.~Ivanchenko,
  V.~Ivanchenko, F.~Jones, S.~Jun, P.~Kaitaniemi, N.~Karakatsanis,
  M.~Karamitros, M.~Kelsey, A.~Kimura, T.~Koi, H.~Kurashige, A.~Lechner,
  S.~Lee, F.~Longo, M.~Maire, D.~Mancusi, A.~Mantero, E.~Mendoza, B.~Morgan,
  K.~Murakami, T.~Nikitina, L.~Pandola, P.~Paprocki, J.~Perl, I.~Petrovi\'{c},
  M.~Pia, W.~Pokorski, J.~Quesada, M.~Raine, M.~Reis, A.~Ribon, A.~R. Fira,
  F.~Romano, G.~Russo, G.~Santin, T.~Sasaki, D.~Sawkey, J.~Shin, I.~Strakovsky,
  A.~Taborda, S.~Tanaka, B.~Tom\'{e}, T.~Toshito, H.~Tran, P.~Truscott,
  L.~Urban, V.~Uzhinsky, J.~Verbeke, M.~Verderi, B.~Wendt, H.~Wenzel,
  D.~Wright, D.~Wright, T.~Yamashita, J.~Yarba, H.~Yoshida, {Recent
  developments in {Geant4}}, Nuclear Instruments and Methods in Physics
  Research Section A: Accelerators, Spectrometers, Detectors and Associated
  Equipment 835 (2016) 186 -- 225.
\newblock \href {http://dx.doi.org/10.1016/j.nima.2016.06.125}
  {\path{doi:10.1016/j.nima.2016.06.125}}.

\bibitem{BOHLEN2014211}
T.~B{\"o}hlen, F.~Cerutti, M.~Chin, A.~Fass{\`o}, A.~Ferrari, P.~Ortega,
  A.~Mairani, P.~Sala, G.~Smirnov, V.~Vlachoudis, {{The FLUKA Code:
  Developments and Challenges for High Energy and Medical Applications}},
  Nuclear Data Sheets 120 (2014) 211 -- 214.
\newblock \href {http://dx.doi.org/10.1016/j.nds.2014.07.049}
  {\path{doi:10.1016/j.nds.2014.07.049}}.

\bibitem{Ferrari_fluka:a}
A.~Ferrari, P.~R. Sala, A.~Fasso, J.~Ranft, O.~Europ{\'e}enne, P.~La,
  R.~Nucl{\'e}aire, A.~Ferrari, P.~R. Sala, A.~Fass{\`o}, J.~Ranft, {FLUKA: a
  multi-particle transport code}, in: CERN 2005-10 (2005), INFN/TC 05/11,
  SLAC-R-773, 2005.

\bibitem{DESYHandbuch}
I.~Borchert, K.~Holm, W.~Knaut, A.~Ladage, U.~M\"{u}ller, F.~Peters, H.~Pingel,
  H.~D. Schulz, K.~G. Steffen, H.~J. Stuckenberg, H.~W\"{u}mpelmann,
  {DESY-Handbuch}, Tech. rep., {Deutsches Elektronen Synchrotron (DESY)}
  ({1966}).

\bibitem{Wu:2290758}
M.~Wu, \href{http://cds.cern.ch/record/2290758}{{Environmental control system
  at DESY}} (Oct 2017).
\newline\urlprefix\url{http://cds.cern.ch/record/2290758}

\bibitem{pcmag:yamamoto}
A.~Yamamoto, K.~Anraku, R.~Golden, T.~Haga, Y.~{Higashi}, M.~Imori, S.~Inaba,
  B.~Kimbell, N.~{Kimura}, Y.~Makida, {Balloon-borne experiment with a
  superconducting solenoidal magnet spectrometer}, {Advances in Space Research}
  14 (1994) 2.
\newblock \href {http://dx.doi.org/10.1016/0273-1177(94)90071-X}
  {\path{doi:10.1016/0273-1177(94)90071-X}}.

\bibitem{pcmag:fieldmeas}
J.~Alozy, F.~Bergsma, F.~Formenti, et~al.,
  \href{http://www.eudet.org/e26/e28/e182/e599/eudet-memo-2007-51.pdf}{{First
  Version of the PCMAG Field Map}}, {EUDET-MEMO} {2007-51}.
\newline\urlprefix\url{http://www.eudet.org/e26/e28/e182/e599/eudet-memo-2007-51.pdf}

\bibitem{pcmag:fieldana}
C.~Grefe,
  \href{http://www-library.desy.de/cgi-bin/showprep.pl?desy-thesis-08-052}{{Magnetic
  Field Map for a Large TPC Prototype}}, Master's thesis, {Universit\"at
  Hamburg}, {DESY-THESIS-2008-052} ({June} 2008).
\newline\urlprefix\url{http://www-library.desy.de/cgi-bin/showprep.pl?desy-thesis-08-052}

\bibitem{ZenkerPHD}
K.~Zenker,
  \href{http://www-library.desy.de/cgi-bin/showprep.pl?desy-thesis-14-044}{Studies
  of field distortions in a time projection chamber for the international
  linear collider}, Ph.D. thesis, University of Hamburg, {DESY-THESIS-2014-044}
  (2014).
\newline\urlprefix\url{http://www-library.desy.de/cgi-bin/showprep.pl?desy-thesis-14-044}

\bibitem{Jansen:2016bkd}
H.~Jansen, S.~Spannagel, J.~Behr, A.~Bulgheroni, G.~Claus, E.~Corrin,
  D.~Cussans, J.~Dreyling-Eschweiler, D.~Eckstein, T.~Eichhorn, M.~Goffe, I.~M.
  Gregor, D.~Haas, C.~Muhl, H.~Perrey, R.~Peschke, P.~Roloff, I.~Rubinskiy,
  M.~Winter, {Performance of the EUDET-type beam telescopes}, EPJ Techniques
  and Instrumentation 3~(1) (2016) 7.
\newblock \href {http://dx.doi.org/10.1140/epjti/s40485-016-0033-2}
  {\path{doi:10.1140/epjti/s40485-016-0033-2}}.

\bibitem{www:telescope}
\href{https://telescopes.desy.de}{{Homepage of the EUDET-type beam
  telescopes}}, [Online; accessed 24-Januar-2018] (2018).
\newline\urlprefix\url{https://telescopes.desy.de}

\bibitem{HuGuo:2010zz}
C.~Hu-Guo, J.~Baudot, G.~Bertolone, A.~Besson, A.~Brogna, C.~Colledani,
  G.~Claus, R.~D. Masi, Y.~Degerli, A.~Dorokhov, G.~Doziere, W.~Dulinski,
  X.~Fang, M.~Gelin, M.~Goffe, F.~Guilloux, A.~Himmi, K.~Jaaskelainen,
  M.~Koziel, F.~Morel, F.~Orsini, M.~Specht, Q.~Sun, O.~Torheim, I.~Valin,
  M.~Winter, {First reticule size MAPS with digital output and integrated zero
  suppression for the EUDET-JRA1 beam telescope}, Nuclear Instruments and
  Methods in Physics Research Section A: Accelerators, Spectrometers, Detectors
  and Associated Equipment 623~(1) (2010) 480 -- 482, 1st International
  Conference on Technology and Instrumentation in Particle Physics.
\newblock \href {http://dx.doi.org/10.1016/j.nima.2010.03.043}
  {\path{doi:10.1016/j.nima.2010.03.043}}.

\bibitem{izzatullah2015}
M.~A. Izzatullah, {DESY Summer Student Project} (2015).

\bibitem{Tavernier2010}
S.~Tavernier, Interactions of Particles in Matter, Springer Berlin Heidelberg,
  Berlin, Heidelberg, 2010, pp. 23--53.
\newblock \href {http://dx.doi.org/10.1007/978-3-642-00829-0_2}
  {\path{doi:10.1007/978-3-642-00829-0_2}}.

\bibitem{schuetze2013}
P.~Sch{\"u}tze, {DESY Summer Student Project} (2013).

\bibitem{lange2016}
T.~Lange, {DESY Summer Student Project} (2016).

\bibitem{Graf:2006ei}
N.~Graf, J.~McCormick, {Simulator for the linear collider (SLIC): A tool for
  ILC detector simulations}, AIP Conf. Proc. 867 (2006) 503--512.
\newblock \href {http://dx.doi.org/10.1063/1.2396991}
  {\path{doi:10.1063/1.2396991}}.

\bibitem{Gaede:2003ip}
F.~Gaede, T.~Behnke, N.~Graf, T.~Johnson, {LCIO: A Persistency framework for
  linear collider simulation studies}, eConf C0303241 (2003) TUKT001,
  "SLAC-PUB-9992", "CHEP-2003-TUKT001".
\newblock \href {http://arxiv.org/abs/physics/0306114}
  {\path{arXiv:physics/0306114}}.

\bibitem{Schutz:2015iya}
A.~Sch{\"u}tz, \href{http://bib-pubdb1.desy.de/record/220092}{{Simulation of
  Particle Fluxes at the DESY-II Test Beam Facility}}, Master's thesis,
  {Karlsruhe Institute of Technology}, {DESY-THESIS-2015-017} (2015).
\newline\urlprefix\url{http://bib-pubdb1.desy.de/record/220092}

\bibitem{Geantino}
\href{https://geant4.web.cern.ch/support/user_documentation}{{Geant4 -- Users
  Documentation}}, [Online; accessed 20-July-2018].
\newline\urlprefix\url{https://geant4.web.cern.ch/support/user_documentation}

\bibitem{DESYSSP}
\href{https://summerstudents.desy.de/}{{DESY Summer Student Programme}},
  [Online; accessed 23-February-2018] (2018).
\newline\urlprefix\url{https://summerstudents.desy.de/}

\bibitem{Arling:2018hca}
J.-H. {Arling}, M.~{Sohail Amjad}, L.~{Bandiera}, T.~{Behnke}, D.~{Dannheim},
  R.~{Diener}, J.~{Dreyling-Eschweiler}, H.~{Ehrlichmann}, A.~{Gerbershagen},
  I.-M. {Gregor}, A.~{Hayrapetyan}, H.~{Jansen}, J.~{Kaminski}, J.~{Kroll},
  P.~{Martinengo}, N.~{Meyners}, C.~{M{\"u}ntz}, L.~{Poley}, B.~{Schwenker},
  M.~{Stanitzki}, {Summary and Conclusions of the First DESY Test Beam User
  Workshop} (2018).
\newblock \href {http://arxiv.org/abs/1802.00412} {\path{arXiv:1802.00412}}.

\end{thebibliography}

\end{document}